\pdfoutput=1
\documentclass[aps,prd,amsmath,floats,floatfix, twocolumn,
superscriptaddress,nofootinbib,showpacs,longbibliography]{revtex4-1}
\usepackage[T1]{fontenc}
\usepackage[utf8]{inputenc}
\usepackage{lmodern}
\usepackage{verbatim}
\usepackage[dvipsnames, usenames]{xcolor}
\definecolor{linkcolor}{rgb}{0.0,0.3,0.5}
\usepackage[hypertexnames=false, unicode, colorlinks=true, linkcolor=linkcolor,
citecolor=linkcolor, filecolor=linkcolor,urlcolor=linkcolor,
pdfusetitle]{hyperref}
\usepackage[all]{hypcap}
\usepackage{graphicx}
\usepackage{xspace}
\usepackage{amssymb}
\usepackage{microtype}

\usepackage[english]{babel}
\usepackage{blindtext}

\usepackage[normalem]{ulem} 
\usepackage{bm} 
\usepackage{mathrsfs}

\graphicspath{
	{plots/}
}

\DeclareMathAlphabet{\mathpzc}{OT1}{pzc}{m}{it}

\newcommand{\h}{\mathpzc{h}}

\newcommand{\model}{\texttt{BHPTNRSur1dq1e4}} 

\usepackage[nomessages]{fp}

\begin{document}
\title{Surrogate model for gravitational wave signals from non-spinning, comparable- to large-mass-ratio black hole binaries built on black hole perturbation theory waveforms calibrated to numerical relativity}
\newcommand{\UMassDMath}{\affiliation{Department of Mathematics,
		University of Massachusetts, Dartmouth, MA 02747, USA}}
\newcommand{\UMassDPhy}{\affiliation{Department of Physics,
		University of Massachusetts, Dartmouth, MA 02747, USA}}
\newcommand{\CSCVRUMass}{\affiliation{Center for Scientific Computing and Visualization Research, University of Massachusetts, Dartmouth, MA 02747, USA}}
\newcommand{\KITP}{\affiliation{Kavli Institute of Theoretical Physics, University of California Santa Barbara, Santa Barbara, CA 93106, USA}}
\newcommand{\URI}{\affiliation{Department of Physics, 
    University of Rhode Island, Kingston, RI 02881, USA}}    
\newcommand{\MIT}{\affiliation{Department of Physics and MIT Kavli Institute, Massachusetts Institute of Technology, Cambridge, MA 02139, USA}}   
\newcommand{\AEI}{\affiliation{Max Planck Institute for Gravitational Physics (Albert Einstein Institute), Am M\"uhlenberg 1, Potsdam 14476, Germany}}
\newcommand{\Caltech}{\affiliation{Theoretical Astrophysics, Walter Burke Institute for Theoretical Physics,
California Institute of Technology, Pasadena, California 91125, USA}}
\newcommand{\Cornell}{\affiliation{Cornell Center for Astrophysics and Planetary Science, Cornell University, Ithaca, New York 14853, USA}}

\author{Tousif Islam}
\email{tislam@umassd.edu}
\UMassDPhy
\UMassDMath
\CSCVRUMass
\KITP

\author{Scott E. Field}
\UMassDMath
\CSCVRUMass

\author{Scott A. Hughes}
\MIT

\author{\\Gaurav Khanna}
\URI
\UMassDPhy
\CSCVRUMass

\author{Vijay Varma}
\thanks{Marie Curie Fellow}
\AEI  

\author{Matthew Giesler}
\Cornell

\author{Mark A. Scheel}
\Caltech

\author{Lawrence E. Kidder}
\Cornell 

\author{Harald P. Pfeiffer}
\AEI

\hypersetup{pdfauthor={Islam et al.}}

\date{\today}

\begin{abstract}
We present a reduced-order surrogate
model of gravitational waveforms from non-spinning binary black hole systems
with comparable to large mass-ratio configurations. This surrogate
model, \model{}, is trained on waveform data generated by
point-particle black hole perturbation theory (ppBHPT) with mass
ratios varying from 2.5 to 10,000.
\model{} extends an earlier waveform model, \texttt{EMRISur1dq1e4}, by using an updated
transition-to-plunge model, 
covering longer durations up to 30,500 $m_1$  (where $m_1$ is the mass of the primary black hole),
includes several more spherical harmonic modes up to $\ell=10$,
and calibrates subdominant modes to numerical relativity (NR) data.
In the comparable mass-ratio regime, including mass ratios as low as $2.5$, the gravitational
waveforms generated through ppBHPT agree surprisingly well with those from
NR after this simple calibration step. 
We also compare our model to recent SXS and RIT NR simulations at mass ratios ranging from $15$ to $32$, and find the dominant quadrupolar modes agree to better than $\approx 10^{-3}$.
We expect our model to be useful to study intermediate-mass-ratio binary systems in current and future gravitational-wave detectors.
\end{abstract}

\maketitle
\section{Introduction}
\label{Sec:Introduction}

Detection of gravitational waves (GWs) \cite{LIGOScientific:2018mvr,Abbott:2020niy} from the coalescence of binary compact objects offer a new window to study black holes and fundamental physics. Most of the GW signals detected so far by the LIGO/Virgo collaboration are consistent with binary black holes (BBHs), with mass ratio $q=m_1/m_2 \le 10$ \footnote{We use the convention $q=m_1/m_2$, where $m_1$ and $m_2$ are the masses of the component black holes, with $m_1\ge m_2$.}.
Another interesting source of GWs are intermediate mass ratio inspirals (IMRIs) comprised of an intermediate-mass black hole (IMBH, mass$\sim10^2-10^4 M_\odot$) \cite{Feng:2011pc,Pasham:2015tca,Mezcua:2017npy} and a solar-mass black hole (mass$\sim3-20M_\odot)$.  
The resulting binaries will have a mass ratio in the range $10\le q \le10^4$. 
The existence of IMBHs on the low-end of this mass range has been confirmed
by the detection of GW190521 \cite{2017IJMPD..2630021M}, and electromagnetic evidence continues to mount indicating the likely existence of these objects across their possible mass range~\cite{Mezcua:2017npy}.
IMRIs are expected to form in dense globular clusters and galactic nuclei \cite{Leigh:2014oda,MacLeod:2015bpa}.
While these binaries are a prime source for future-generation detectors such as LISA~\cite{2017arXiv170200786A}, Cosmic Explorer (CE) 
or the Einstein Telescope (ET) \cite{AmaroSeoane:2007aw,Berry:2019wgg}, current detectors may also be able to detect IMRIs as their sensitivity improves. 
In particular, IMRIs with total mass $<2000M_\odot$ may be detected by the current generation of detectors \cite{Amaro-Seoane:2018gbb} while future space-based missions, \cite{Kawamura:2020pcg, Sedda:2019uro} such as LISA will observe heavier binaries.
LISA is also expected to detect GW signals from extreme mass ratio inspirals (EMRIs) (having a mass ratio $q \gtrsim 10^5$) that consists of a stellar-mass black hole paired with a supermassive black hole \cite{AmaroSeoane:2007aw,Berry:2019wgg}.

Detection of GWs from IMRIs would shed light on many issues in both astrophysics and fundamental
physics  \cite{AmaroSeoane:2007aw, Berry:2019wgg, Sedda:2019uro}.
IMRIs may form in dense globular clusters or galactic nuclei through multiple possible formation channels.
Detection and parameter inference with IMRIs~\cite{Gair:2010dx,Huerta:2010un,Huerta:2010tp,Haster:2015cnn} will help us probe formation channels and the evolutionary pathway to
supermassive black holes \cite{Bellovary:2019nib}.
As these systems form in dense environments, GW signals from IMRIs may carry an imprint of its surrounding environment and
are ideal sources to investigate possible environmental effects
\cite{AmaroSeoane:2007aw, Barausse:2006vt, Barausse:2007dy, Gair:2010iv, Yunes:2011ws, Barausse:2014tra, Barausse:2014pra, Derdzinski:2020wlw}.  
IMRI signals could also be used to test the nature of gravity in an unexplored strong-field regime  \cite{Gair:2012nm, Piovano:2020ooe,Yunes:2009ry, Canizares:2012ji, Canizares:2012is,Rodriguez:2011aa, Chua:2018yng}, complementing tests of general relativity (GR) performed with GW events detected so far~\cite{LIGOScientific:2020tif,LIGOScientific:2020tif}.

Carrying out accurate parameter inference and performing fundamental physics analysis with IMRI GW signals will require multi-modal waveform models that are fast and reliable~\cite{Islam:2021zee}.
While numerical relativity (NR) provides the most accurate waveforms from BBH mergers, it takes weeks to months to generate a single waveform, making them unfit to be directly used in multi-query studies. The availability of a large number of NR simulations in the comparable mass regime, however, has paved the way to build either NR-based reduced-order surrogate models \cite{Blackman:2015pia,Blackman:2017pcm,Blackman:2017dfb,Varma:2018mmi,Varma:2019csw,Islam:2021mha} or calibrated phenomenological or effective-one-body (EOB) models \cite{bohe2017improved,cotesta2018enriching,cotesta2020frequency,pan2014inspiral,babak2017validating,husa2016frequency,khan2016frequency,london2018first,khan2019phenomenological} to NR.
Only a few of these models \cite{Nagar:2022icd}, however, are tested or calibrated in the intermediate mass ratio regime, where only a few NR simulations are available \cite{Lousto:2020tnb,Lousto:2022hoq,Yoo:2022erv, Giesler:2022inPrep}. 
In lieu of NR-based calibration, some EOB models~\cite{Bohe:2016gbl} have been tuned to results from point particle black hole perturbation theory (ppBHPT), which provides an accurate waveform model as $q \rightarrow \infty$.
Due to computational cost, ppBHPT waveform models also cannot be directly used in multi-query data analysis.
This has been partly overcome by developing ``kludge'' models \cite{Barack:2003fp,Babak:2006uv,Chua:2017ujo,Chua:2015mua} that are fast and capture the qualitative features of an EMRI waveform in the inspiral regime using approximations for the amplitudes and phases.
Recently Refs.\ \cite{Chua:2020stf,Katz:2021yft} have introduced gravitational self-force-based waveform models that are as fast to compute as kludge models by using a combination of reduced order methods, deep-learning techniques and hardware acceleration. Also relevant is Ref.\ \cite{Wardell:2021fyy}, which presents a fully relativistic second-order self-force model that can generate first-principles inspiral waveforms in milliseconds, at least for the case of quasi-circular inspiral of non-spinning black holes.

To begin addressing these issues, Ref.~\cite{Rifat:2019ltp} built a proof-of-principle ppBHPT surrogate model \texttt{EMRISur1dq1e4} for non-spinning binaries that extends from mass ratio $q=3$ to 
$q=10,000$ and covers $\sim13,500m_1$ in duration.
The model's dominant mode has been tuned to NR in the comparable mass ratio regime ($q\leq10$), and it was shown that after this simple calibration step the ppBHPT and NR waveforms agreed to better than $\approx 1\%$ at mass ratios $q \gtrsim 8$.
These initial encouraging results suggest that suitably calibrated ppBHPT waveform data could provide for an accurate model of gravitational waves from IMRI systems. The agreement between NR and ppBHPT (with radiative corrections to the orbit) after a simple rescaling~\cite{Rifat:2019ltp} is a surprising observation on its own.

In this paper, we describe more fully the methods we have used to
build \texttt{EMRISur1dq1e4} as well as making numerous important improvements to the underlying model. 
The updated version of our surrogate model -- which we call \model{} -- 
is $\sim30,500m_1$ in duration and covers all phases of the system's evolution from inspiral through plunge and ringdown -- making it suitable to be used in
a wider range of data analysis studies.
It features a total of 50 important higher order modes up to $\ell=10$ thereby permitting studies to quantify
the effect of higher order modes in GW signals.
Furthermore, by applying a simple calibration, we find the NR-calibrated ppBHPT waveforms agree
remarkably well with NR for all of the higher order modes up to $\ell=5$ in the comparable mass ratio regime.

The rest of the paper is organized as follows. Sec.~\ref{imridata} describes our method for computing ppBHPT
waveforms by solving the Teukolsky equation.
We describe the surrogate-modelling framework, calibration to NR, and assess model accuracy in Sec.~\ref{Sec:modelling}. Sec.~\ref{sec:compare_to_nr} provides a more detailed comparison between ppBHPT waveforms and NR data in the comparable and intermediate mass
ratio regime with a focus on subdominant modes and new SXS and RIT simulations at mass ratios greater than 10.
Finally, we outline future directions in Sec.~\ref{Sec:Conclusion}.

\section{Waveform data using perturbation theory}
\label{imridata}

We generate the surrogate-model training data using point-particle black hole perturbation theory (ppBHPT).
First, we compute the trajectory taken by the point-particle and then we use that trajectory to compute the gravitational wave emission. The next three subsections summarize the  equations and algorithms for accomplishing this.

\subsection{Numerically solving the Teukolsky equation}

In the ppBHPT framework, the smaller black hole is modeled as a point-particle with no internal structure and a mass of $m_2$, moving in the spacetime of the larger Kerr black hole with mass $m_1$ 
and spin angular momentum per unit mass $a$.
Here, we provide an executive summary of this framework and refer to 
Refs. \cite{Sundararajan:2007jg,Sundararajan:2008zm,Sundararajan:2010sr,Zenginoglu:2011zz} for additional details.

Gravitational radiation is computed by first numerically solving the
Teukolsky equation
\begin{eqnarray}
\label{teuk0}
&&
-\left[\frac{(r^2 + a^2)^2 }{\Delta}-a^2\sin^2\theta\right]
\partial_{tt}\Psi
-\frac{4 m_1 a r}{\Delta}
\partial_{t\phi}\Psi \nonumber \\
&&- 2s\left[r-\frac{m_1(r^2-a^2)}{\Delta}+ia\cos\theta\right]
\partial_t\Psi\nonumber\\  
&&
+\,\Delta^{-s}\partial_r\left(\Delta^{s+1}\partial_r\Psi\right)
+\frac{1}{\sin\theta}\partial_\theta
\left(\sin\theta\partial_\theta\Psi\right)+\nonumber\\
&& \left[\frac{1}{\sin^2\theta}-\frac{a^2}{\Delta}\right] 
\partial_{\phi\phi}\Psi +\, 2s \left[\frac{a (r-m_1)}{\Delta} 
+ \frac{i \cos\theta}{\sin^2\theta}\right] \partial_\phi\Psi  \nonumber\\
&&- \left(s^2 \cot^2\theta - s \right) \Psi = -4\pi\left(r^2+a^2\cos^2\theta\right)T  \,  ,
\end{eqnarray}
sourced by the moving particle, where $\Delta = r^2 - 2 m_1 r + a^2$ and $s$ is the ``spin weight'' of the field. 
The $s=-2$ case for $\Psi$ describes the radiative degrees of freedom of the gravitational field, the Weyl scalar 
$\psi_4$, in the radiation zone, and is directly related to the Weyl curvature scalar as $\Psi = (r -ia\cos\theta)^4\psi_4$. 
The source term $T$ in Eq.~(\ref{teuk0}) for the  smaller compact object $m_2$ 
is related to the energy-momentum tensor $T_{\alpha\beta}$ of a point particle.
The Weyl scalar $\psi_4$ can then be integrated twice at future null infinity ${\mathscr{I}^+}$ to find the two polarization states $h_{+}$ and $h_{\times}$ of the transverse-traceless metric perturbations,
\begin{equation}
\psi_4 = \frac{1}{2}\left(\frac{\,\partial^2h_{+}}{\,\partial t^2}-i\frac{\,\partial^2h_{\times}}{\,\partial t^2}\right)\, .
\end{equation}
The complex gravitational wave strain
\begin{align}
h_{+}(t,\theta, \phi;q) &- {\mathrm i} h_{\times}(t,\theta, \phi; q) \nonumber \\
& =  \sum_{\ell=2}^{\infty} \sum_{m=-\ell}^{\ell} h^{\ell m}(t;q) {}_{-2}Y_{\ell m}(\theta, \phi) \, ,
\end{align}
can be formed from the two polarization states,
which is subsequently decomposed into a basis of spin-weighted spherical harmonics ${}_{-2}Y_{\ell m}$. We build models for the harmonic modes $h^{\ell m}(t;q)$.

Once the trajectory of the perturbing compact body is fully specified (cf. Sec.~\ref{sec:Trajectory}), we solve the inhomogeneous Teukolsky equation in the time-domain while feeding the trajectory information 
into the particle source-term of the equation \cite{Sundararajan:2007jg,Sundararajan:2008zm,Sundararajan:2010sr,Zenginoglu:2011zz,Field:2021}.
This involves a four step procedure:
(i) rewriting the Teukolsky equation using compactified hyperboloidal 
coordinates (Eq.~\ref{teuk0} is shown using standard Boyer-Lindquist coordinates) that allow us to extract the gravitational waveform directly at null infinity while also solving the 
issue of unphysical reflections from the artificial boundary of the finite computational domain; (ii) obtaining 
a set of (2+1) dimensional PDEs by using the axisymmetry of the background Kerr space-time, and separating the 
dependence on azimuthal coordinate; (iii) recasting these equations into a first-order, hyperbolic PDE system; 
and finally (iv) implementing a high-order WENO (3,5) finite-difference scheme with Shu-Osher (3,3) time-stepping~\cite{Field:2021}.  
The point-particle source term on the right-hand-side of the Teukolsky equation requires some specialized techniques 
for a finite-difference numerical implementation~\cite{Sundararajan:2007jg,Sundararajan:2008zm}. We set the 
spin of the central black hole to a value slightly away from zero, $a/m_1 = 10^{-8}$ for technical 
reasons\footnote{For example, to avoid a change in the definition of the coordinates from Kerr to Schwarzschild.}. 
Our numerical evolution scheme is implemented using OpenCL/CUDA-based GPGPU-computing which allows for
very long duration and high-accuracy computations within a reasonable time-frame. Numerical errors in the phase and amplitude are 
typically on the scale of a small fraction of a percent~\cite{McKennon:2012iq, Rifat:2019ltp}.

\subsection{Trajectory model}
\label{sec:Trajectory}

The particle's motion is characterized by three distinct regimes -- an initial adiabatic 
inspiral, a late-stage geodesic plunge into the horizon, and a transition regime between those two.

During the initial adiabatic inspiral, the particle follows a sequence of geodesic orbits driven by radiative 
energy and angular momentum losses.
The flux radiated to null infinity and through the event horizon are computed by solving the frequency-domain 
Teukolsky equation \cite{Fujita:2004rb,Fujita:2005kng,Mano:1996vt,throwe2010high} using the open-source code \texttt{GremlinEq} \cite{OSullivan:2014ywd,Drasco:2005kz} from the Black Hole Perturbation Toolkit \cite{BHPToolkit}.
The inspiral trajectory is then extended to include a plunge geodesic and a smooth transition region following a 
procedure similar to one proposed by Ori-Thorne \cite{Ori:2000zn}. We compute the transition between initial inspiral and 
the plunge using a generalized Ori-Thorne algorithm \cite{Hughes:2019zmt,Apte:2019txp} (hereafter, the ``GOT'' algorithm).
The GOT algorithm uses a parameterization of strong-field Kerr orbits based on {\em Mino time}, which separates the radial and polar motions of Kerr black hole orbits.  It also introduces a correction that smooths a rather sharp discontinuity in the evolution of an inspiral's integrals of motion as presented in the original Ori-Thorne model.  Detailed discussion of this point is given in Sec.\ IV A 2 of Ref.\ \cite{Apte:2019txp}.  The use of Mino time is not so critical for our analysis since the separation of radial and polar motions is not an issue for equatorial orbits, but smoothing of the integrals of the motion is of great importance.  Note that we use ``Model 2'' from Ref.\ \cite{Apte:2019txp} for this smoothing.

Our trajectory model does not include the effects of the conservative or second-order self-force~\cite{Hinderer:2008dm}, although once these post-adiabatic corrections are known they could be easily incorporated to improve the accuracy of the inspiral's phase. 

\begin{figure}[h!]
	\includegraphics[width=\columnwidth]{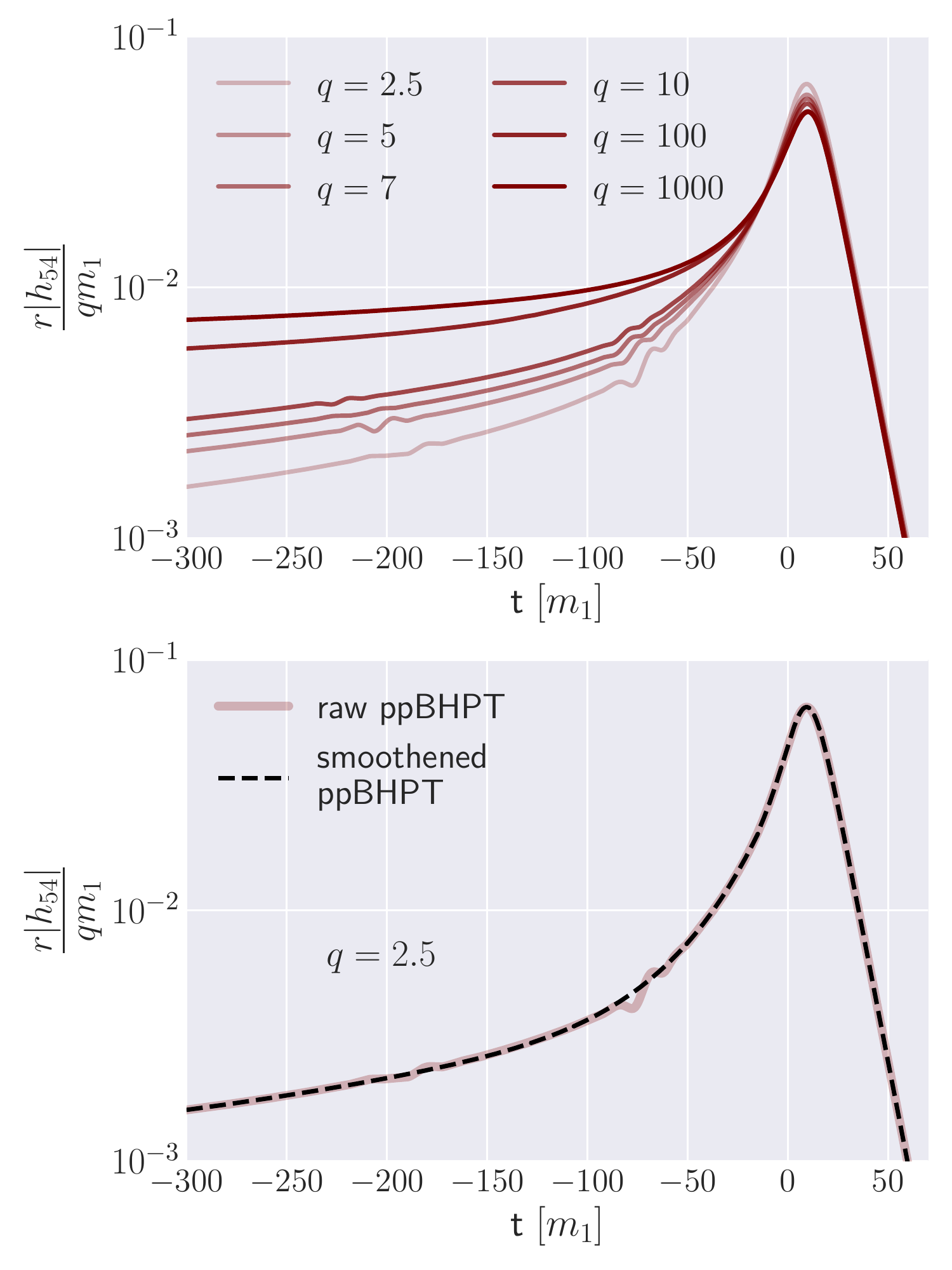}
	\caption{At mass ratios $q \lesssim 10$ the generalized Ori-Thorne transition trajectory
	causes small, nonphysical oscillations in some of the $\ell \neq m$ modes. The raw ppBHPT waveform amplitudes are shown for one of the representative modes $(\ell,m)=(5,4)$
    for increasing value of the mass ratio $q$ (upper panel) and one example waveform after performing the smoothing procedure (lower panel) described in Sec.~\ref{sec:fixer}.
	}
	\label{Fig:smoothing}
\end{figure}
\subsection{Waveform smoothing}
\label{sec:fixer}

At low mass ratios, the GOT transition trajectory produces small non-physical oscillations in high-order modes of the waveforms.  Because the GOT algorithm is designed for the regime $q \gg 1$, it is not surprising that some pathologies enter at low mass ratios.  
These non-physical oscillations are results of a small jump in the acceleration of the point-particle as it exits the  adiabatic inspiral and also when it begins the  plunge. 
It is interesting, however, that these oscillations are more apparent in our data for modes with $\ell \neq m$. 
It is also worth noting that the oscillations are larger in amplitude when we use Model 1 from Ref.\ \cite{Apte:2019txp} for smoothing the evolution of the integrals of motion.

In Fig. \ref{Fig:smoothing} (upper panel), we show the unphysical oscillations in the scaled amplitudes in one of the representative modes $(\ell,m)=(5,4)$ for increasing values of the mass ratios.
It is clear that while $q\le10$ shows unphysical oscillations in the transition regime, these features vanish for high mass ratio simulations. At lower mass ratios we remove these unwanted oscillations by using a ``smoothening'' procedure~\cite{Rifat:2019ltp}. 
To smooth the data, we (i) first remove the unphysical oscillatory portion of the waveform that we mention above, (ii) then use the rest of the waveform data to construct a polynomial fit of degree 7, and finally (iii) evaluate the polynomial to obtain smoother data in the problematic regions.
The lower panel of Fig. \ref{Fig:smoothing} shows the rescaled $(\ell,m)=(5,4)$ amplitude for $q=2.5$ before and after the `smoothing'. 
A similar smoothing procedure is applied to the phase data.
Our surrogate model is trained on --- and for validation purposes, compared to --- these smoothed waveform data.

To quantify the amount by which our smoothing procedure has modified the waveform, we compute a relative $L_{2}$-norm difference between the smoothed and original data. 
Normalized $L_{2}$-norm between two functions $h_1(t)$ and $h_2(t)$ is defined as a time-domain overlap integral with white-noise:
\begin{gather}
\label{Eq:cost_function}
\mathcal{E} [h_1 , h_2] =
\frac{1}{2} \frac{\sum_{(\ell,m)}\int_{t_1}^{t_2}|
	h_1(t) - h_2(t)|^2 dt}
{\sum_{(\ell,m)}\int_{t_1}^{t_2} |h_1(t)|^2 dt} \,,
\end{gather}
Here, $t_1$ and $t_2$ denote the start and end of the waveform data respectively whereas $h_1$ and $h_2$ denote the smoothed and original data respectively.
We find that the differences between the smoothed and original data for each mode is on average $8 \times 10^{-5}$ with a maximum $5 \times 10^{-4}$.  
To compute errors for individual modes, we restrict the sum in Eq.(\ref{Eq:cost_function}) to only the mode of interest.

\begin{table*}
	\centering
	\begin{tabular}{l|c|c|c|c}
		\toprule
		Model &Plunge Model &Available positive modes $(\ell,m)$ &Waveform length &NR-calibration\\
		\hline
		\texttt{EMRISur1dq1e4} &Ori-Thorne~\cite{Ori:2000zn} &$(2,\{1,2\})$, $(3,\{1,2,3\})$ & $13500m_1$ &$(2,2)$\\
		& &$(4,\{2,3,4\})$, $(5,\{3,4,5\})$ & &\\
		\hline
		\model{} & Generalized & $(2,\{1,2\})$, $(3,\{1,2,3\})$ & $30500m_1$ &$(2,\{1,2\})$\\
		&Ori-Thorne~\cite{Hughes:2019zmt,Apte:2019txp} & $(4,\{2,3,4\})$, $(5,\{3,4,5\})$  & &$(3,\{1,2,3\})$\\
		& &$(6,\{4,5,6\})$, $(7,\{5,6,7\})$  & &$(4,\{2,3,4\})$\\
		& &$(8,\{6,7,8\})$, $(9,\{7,8,9\})$  & &$(5,5)$\\
		& & $(10,\{8,9\})$ & &\\ 
		\botrule	
	\end{tabular}
	\caption{Overview of the \texttt{EMRISur1dq1e4} and \model{} models. Both models used the smoothing procedure described in Sec.~\ref{sec:fixer}.
	}
	\label{Tab:modes}
\end{table*}

\section{Surrogate modelling}
\label{Sec:modelling}
In this section, we briefly describe the framework used to build the surrogate model. Our framework is
constructed using a combination of methodologies proposed in earlier
works~\cite{Blackman:2015pia,Field:2013cfa,Purrer:2014fza}. 
\subsection{Building the surrogate}
\begin{figure}[h!]
	\includegraphics[width=.9\columnwidth]{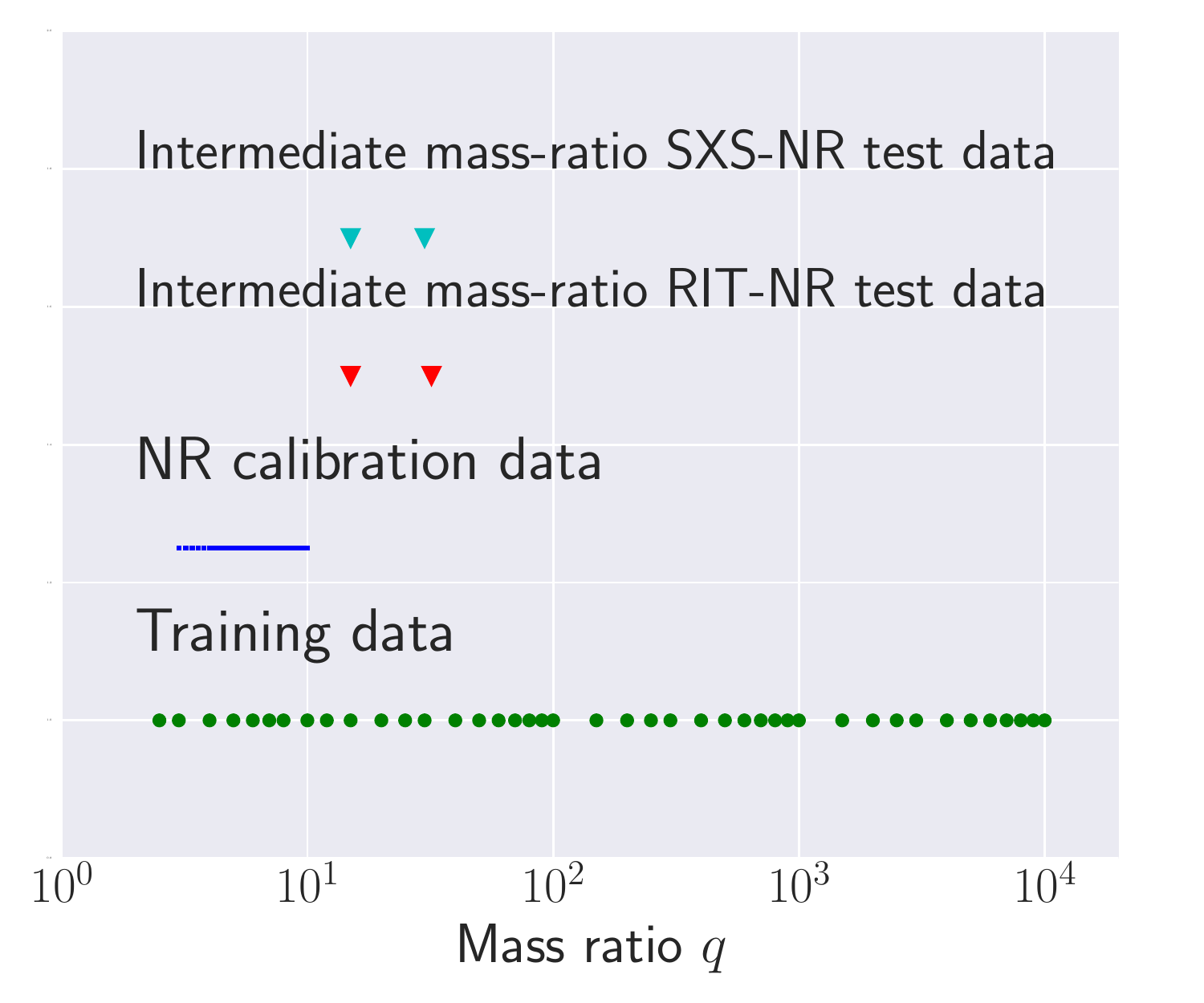}
	\caption{
		ppBHPT Waveform data points used in training \model{} (green circles). We also show points used in calibrating the ppBHPT waveforms to NR (blue squares) and data points used in validating NR calibration against high mass ratio NR simulation (red and cyan triangles).
	}
	\label{Fig:training_points}
\end{figure}
\paragraph{\textbf{Training data : }}
To train the model, we collect a total of 41 ppBHPT waveforms by numerically solving the inhomogeneous Teukolsky 
equation at different values of the mass ratio $q$.
These values of $q$ have been chosen in such a way that it populates the mass ratio axis from $q=2.5$ to $q=10,000$ in a logarithmic scale (green circles in Fig. \ref{Fig:training_points}).
For each value of $q$, we then extract the harmonic modes, $h^{\ell,m}(t;q)$. 
Note that we only model $m>0$ modes as the negative $m$ modes are computed from the positive $m$ modes using the 
symmetry of the nonprecessing system under reflections about the orbital plane: $h^{\ell, -m} = (-1)^{\ell} h^{\ell,m}{}^*$.
\paragraph{\textbf{Data alignment : }}
We first determine the peak of each waveform $\tau_{peak}$ to be the time when the quadrature sum,
\begin{equation}
A_{\rm tot}(\tau)=\sqrt{\sum_{\ell,m} |h^{\ell,m}(\tau)|^2} \,,
\end{equation}
reaches its maximum.
Here the summation is taken over all the modes being modeled.
In order to construct smooth parametric fits for the surrogate model, we align all the waveforms such 
that their peaks occur at the same time. This is done by choosing a new time coordinate,
\begin{equation}
t=\tau - \tau_{\rm peak} \,,
\end{equation}
such that $A_{\rm tot}(t)$ for each waveform peaks at $t=0$.
Next, we use cubic splines to interpolate the real and imaginary parts of the waveform modes onto a common time 
grid of [$-30500m_1$, $115m_1$] with a uniform time spacing of $dt=0.1 m_1$.
Once all the waveforms are interpolated onto a common time grid, we perform a 
rotation about the z-axis such that at the start of the waveform 
$\phi_{22}=0$
and $\phi_{21} \in [-\pi,0]$, where
$\phi_{22}$ and $\phi_{21}$ are the phases of the complex $(2,2)$ and $(2,2)$ modes, respectively. 
These pre-processing steps are necessary to ensure a smooth dependence of the training-set waveforms on mass ratio.

\paragraph{\textbf{Data decomposition : }}
After time and phase alignments, we decompose the inertial frame waveform modes into {\em waveform data pieces} that are slowly varying functions 
of time and are therefore simpler to model. 
We employ different decomposition strategy for the quadrupolar mode and the higher-order modes.
The complex $(2,2)$ waveform mode,
\begin{gather}
\label{Eq:AmpPhase_22}
h_{22} = A_{22} ~e^{-\mathrm{i} \phi_{22}} \,,
\end{gather}
is decomposed into an amplitude, $A_{22}$, and phase, $\phi_{22}$.
For the higher order modes, we first apply a time-dependent rotation given by the instantaneous orbital phase $\phi_\mathrm{orb}$ to transform the waveform into a co-orbital frame:
\begin{gather}
h_{\ell m}^C = h_{\ell m} ~e^{\mathrm{i} m \phi_\mathrm{orb}},
\label{Eq:coorb_frame}
\end{gather}
where $h_{lm}^C$ represents the complex modes in the co-orbital frame and the orbital phase is taken to be
\begin{gather}
\phi^{\rm orb} \equiv \frac{\phi_{22}}{2} \,.
\label{Eq:orb_phase}
\end{gather}
We then use the real and imaginary parts of $h_{\ell m}^C$ as our waveform data pieces for the non-quadrupole modes.
To summarize, the full set of waveform {\em data pieces} we model is as follows:  $A_{22}$, $\phi_{22}$ for the 
$(2,2)$ mode, and real and imaginary parts of $h_{\ell m}^C$ for the 24 higher order modes with $m>0$.
\paragraph{\textbf{Empirical interpolants : }}
The next step is to construct an empirical interpolant (EI) in time using a greedy algorithm that picks the most representative time nodes~\cite{Maday:2009,chaturantabut2010nonlinear, Field:2013cfa, Canizares:2014fya}. 
The number of the time (or EI) nodes for each data piece is equal to the number of basis functions used. 
The empirical interpolant gives a compact representation for each data piece (and hence the full waveform) in the training set by permitting the full time-series to be reconstructed through a significantly sparser sampling 
defined by the EI nodes.
We choose 7 basis functions for $A_{22}$ and $\phi_{22}$. 
For higher order modes, we use 13 basis functions for the real and imaginary parts if $\ell\le5$. 
Otherwise, 16 basis functions are used.
We inspect the basis functions visually to ensure they are free from noise.
Furthermore, unlike recently built surrogate models~\cite{Varma:2018mmi}, we put no restriction on the location of EI nodes as we did not find this to improve our model.
\paragraph{\textbf{Parameteric fits :}}
The final surrogate-building step is to construct parametric fits for each data piece at each of the EI nodes 
over the one-dimensional parameter space defined by ${q}$.
Following Ref. \cite{Rifat:2019ltp}, we fit the data-pieces using second degree interpolating splines (with smoothing factor $s=0.0005$) after performing a logarithmic transformation of $q$~\cite{Varma:2018aht, Varma:2018mmi}.

\begin{figure}[h!]
	\includegraphics[width=.9\columnwidth]{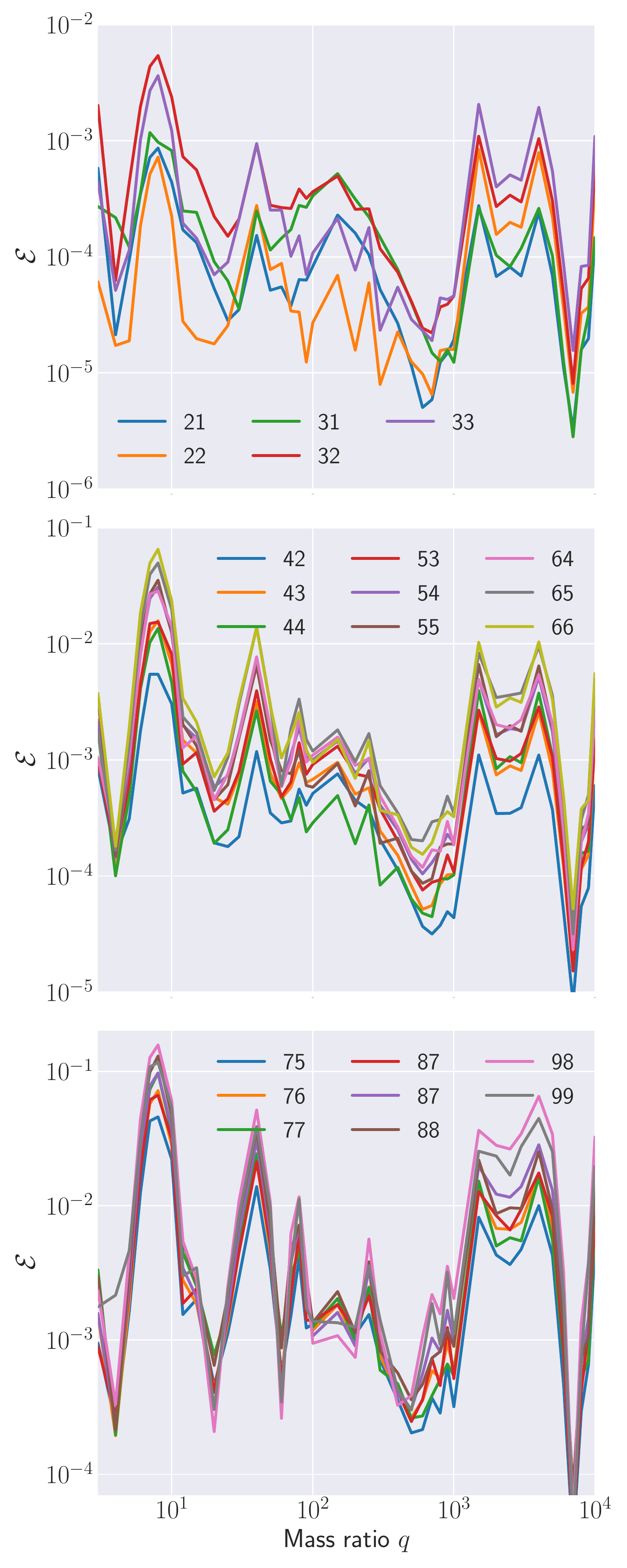}
	\caption{
		Validation errors, computed as $\mathcal{E} [h^{\ell m}, h^{\ell m}_{\tt S}]$, for the individual modes in our surrogate model as a function of mass ratio $q$.} 
	\label{Fig:L2_error}
\end{figure}

\subsection{Evaluating the surrogate}
To generate the \model{} surrogate model waveforms, we provide mass ratio $q$ as input.
We then evaluate the parametric fits for each waveform data pieces at each EI node at the requested value of $q$.
Next, the empirical interpolant is used to reconstruct the surrogate prediction of the waveform data pieces as a dense time-series.
We evaluate the surrogate models for the amplitude and phase of the (2,2) mode and combine them to get the complex strain as $h^S_{22} = A_{22}^S ~ e^{-\mathrm{i}\phi_{22}^S}$.
For the non-quadrupole modes, we first evaluate the surrogate models for the real and imaginary parts of the co-orbital frame waveform data pieces $h_{\ell m}^{C,S} \approx h_{\ell m}^C$ and treat it as $h_{\ell m}^C$.
Finally, we use Eqs.~(\ref{Eq:AmpPhase_22}), (\ref{Eq:coorb_frame}), and (\ref{Eq:orb_phase}) to obtain the surrogate prediction for the inertial frame strain $h^S_{\ell,m}$ for these modes.
The full surrogate, $h_{\tt S}$, is then written
\begin{align} \label{eq:full}
h^{\tt S}(t,\theta,\phi;q) = \sum_{\ell,m} h^{\tt S}_{\ell,m}(t;q) {}_{-2}Y_{\ell m} (\theta,\phi) \,,
\end{align}
where $h^{\tt S}_{\ell,m}$ is the
surrogate model prediction for each harmonic mode.

\subsection{Surrogate errors}
\label{Sec:error}
In this section we assess the accuracy of \model{} by performing some of the tests
described in Ref.~\cite{Blackman:2017dfb} using the relative $L_2$-type norm defined in Eq. (\ref{Eq:cost_function}). 
In our case, ppBHPT waveforms used in training are already aligned in time and phase, and the surrogate is expected to reproduce this alignment.  
Therefore, we compute the time-domain error $\mathcal{E}$ without any further time/phase shifts.

To assess the surrogate model's error, we compute three different types of errors.
First, we build the surrogate using all 41 ppBHPT training waveforms and calculate the {\em training error} between the training waveform and surrogate prediction. This checks whether the surrogate model can accurately reproduce the training waveforms.

Next we perform a \textit{leave-one-out cross-validation} study. 
In this study, we hold out one ppBHPT waveform from the training set and build a trial surrogate from the 
remaining 40 ppBHPT waveforms. 
We then evaluate the trial surrogate at the mass ratio corresponding to the held out data, and compare its 
prediction with the held-out ppBHPT waveform.
We refer to these errors as \textit{validation errors}. 
Validation errors represent conservative error estimates for the surrogate model's generalization error against ppBHPT.
Since we have 41 ppBHPT waveforms, we build 41 trial surrogates for each error study and assess the model's 
ability to predict new waveforms it was { not} trained on. For boundary cases (i.e. for $q=2.5$ and $q=10,000$), the test surrogate predictions are effectively extrapolation and therefore yield uninformative errors. We exclude these points from Fig. \ref{Fig:L2_error}.

We compare both of these errors to the numerical truncation error of the Teukolsky solver used to produce the ppBHPT training data. 
We refer to these errors as \textit{ppBHPT numerical errors}.

\begin{figure*}[h!]
	\includegraphics[width=1.0\textwidth]{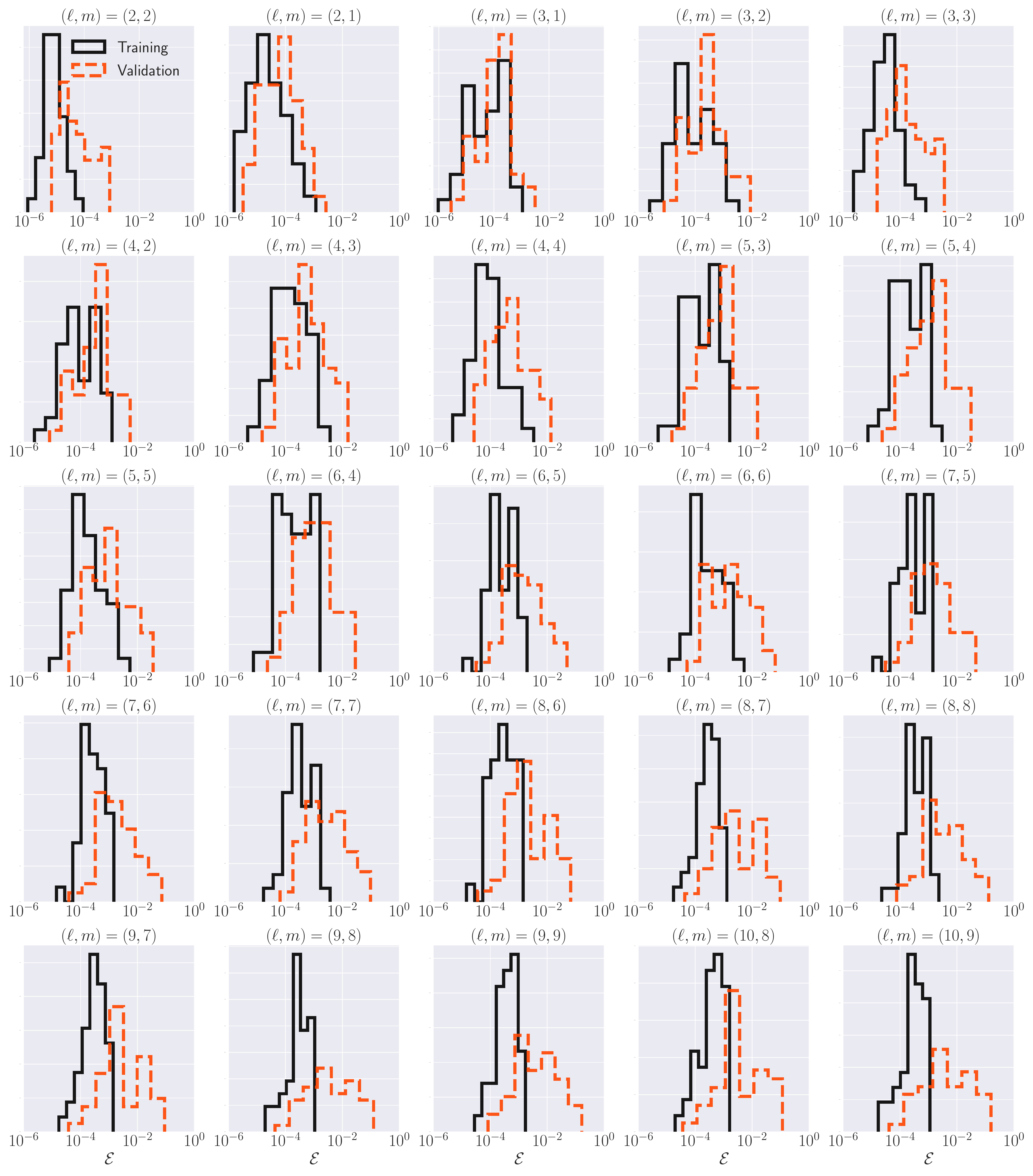}
	\caption{
		Time-domain errors $\mathcal{E}$, defined in Eq.\ref{Eq:cost_function}, for individual modes considered in \model{}. For comparison, we show both training (black solid lines) and validation (orange-red dashed lines) 
		errors.
	}
	\label{Fig:mode_by_mode}
\end{figure*}

In Fig.~\ref{Fig:L2_error}, we report the individual-mode validation errors for \model{} as a function of mass ratio $q$.
We find that the errors are mostly $\le10^{-3}$ for modes up to $\ell=3$, $\le10^{-2}$ for modes with $4 \le \ell\le6$ and $\le10^{-1}$ for modes with $7\le\ell\le10$.
We further note in Fig.~\ref{Fig:L2_error} that the highest errors in each mode corresponds to the same value of $q$. 
We also find that the zigzag structure of errors in Fig.~\ref{Fig:L2_error} is a result of the chosen $q$ values in the parameter space (blue circles in Fig. \ref{Fig:training_points}). 
We model the \textit{data pieces} as a logarithmic function of $q$. 
However, the mass ratio values are not spaced uniformly in logarithmic scale.
This results in repetitive patches of dense and sparsely spaced data points.
Modelling errors are smaller (larger) around the dense patches (sparsely spaced patches).

Despite the problematic parameter-sampling strategy, the final model should be sufficiently accurate for many data analysis studies in the large-mass ratio regime
(cf. Figs.~\ref{Fig:ligo_mismatch} and \ref{Fig:scaled_waveform_q30} for more details).  
In Fig.~\ref{Fig:mode_by_mode} we provide a mode-by-mode comparison of the 
training and validation errors.
For many of the modes considered in our model, both error measurements are consistent. For some of the modes,
and especially for modes with $\ell\ge8$, the larger validation errors indicate overfitting. Note that the high-error tails seen in Fig.~\ref{Fig:mode_by_mode} comes from the smallest and highest mass ratio boundaries.

\begin{figure}[h!]
	\includegraphics[width=\columnwidth]{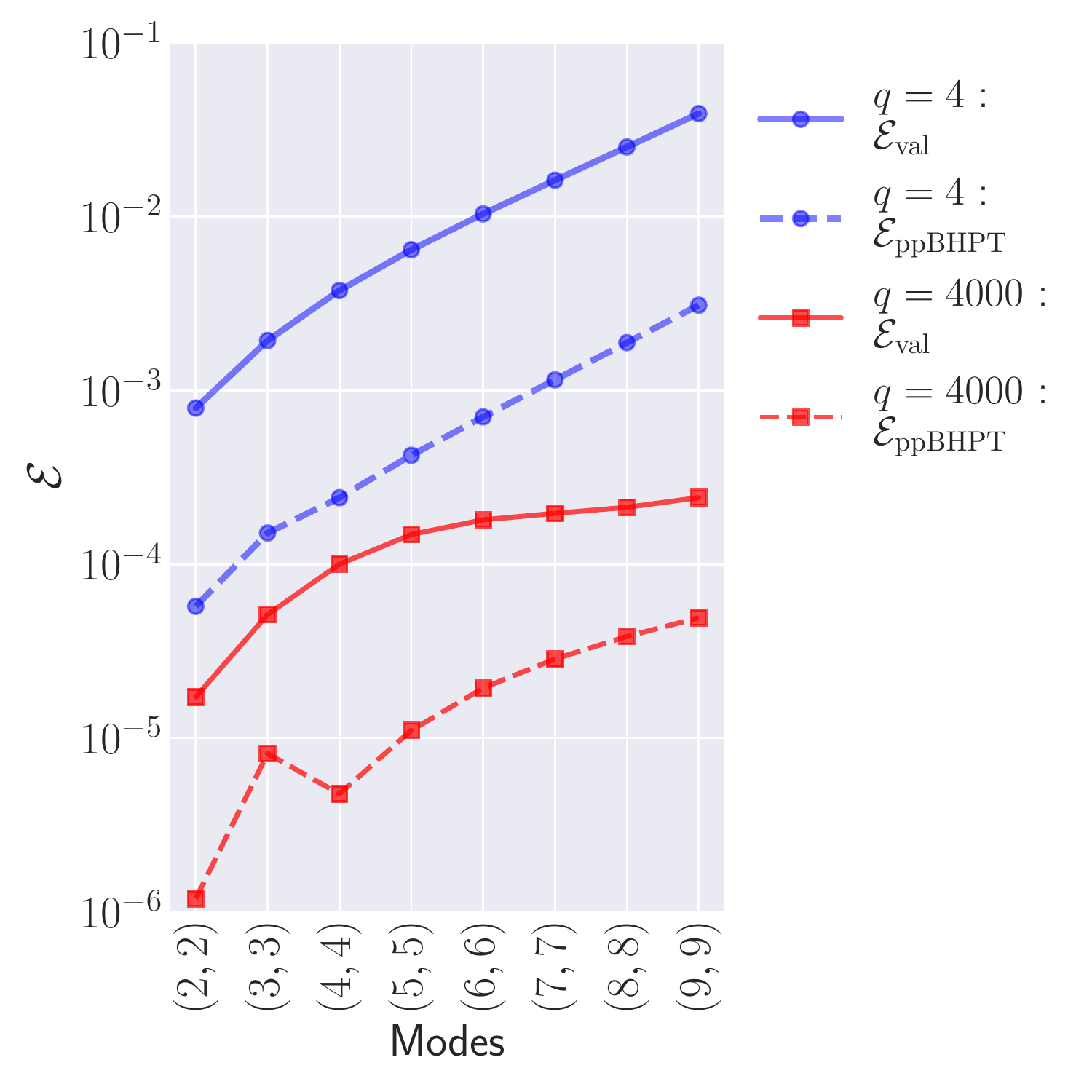}
	\caption{
		Validation errors (solid lines) and ppBHPT numerical errors (dashed lines) of select modes for two representative cases: $q=4$ (blue circles) and $q=4000$ (red squares). 
	}
	\label{Fig:ppBHPT_error}
\end{figure}

Finally, in Fig.~\ref{Fig:ppBHPT_error}, we compare validation error and the ppBHPT Teukolsky solver's numerical error for a select number of modes for two representative cases: $q=4$ (blue data) and $q=4000$ (red data). We compute ppBHPT numerical errors by comparing ppBHPT waveforms of different resolution. We find that validation errors are around one order of magnitude larger than ppBHPT numerical errors across all modes. 

\subsection{Calibating ppBHPT to numerical relativity in the comparable mass regime}
\label{Sec:nr_comparison}

In the previous subsections, we have built a surrogate model for gravitational waveforms computed within the ppBHPT framework. These waveforms faithfully approximate the physically-correct ones only in the limit $q\rightarrow \infty$. In order to construct an accurate model at moderate to large mass ratios, we introduce model calibration parameters and set their values by comparing to NR. These parameters modify each mode's amplitude and phase in a simple way and have the correct behavior in the limit of $q\rightarrow \infty$.

Before discussing the particulars of our calibration procedure, its important to 
consider how one should compare NR and ppBHPT waveforms.
For example, both ppBHPT and NR frameworks express dimensioned quantities in terms of a freely-specifiable mass scale, which is not the same in the two frameworks.  
For ppBHPT this scale is selected to be the background black hole spacetime's mass parameter, while in NR it is
the sum of the Christodoulou masses of each black hole~\cite{Boyle:2007ft,Boyle:2019kee}. 
Our Teukolsky solver sets the background black hole's mass to $m_1 = 1$ (the ppBHPT's mass scale), while 
the corresponding NR simulation for nonspinning black holes sets the total mass to $m_1 + m_2 = 1$ (the NR simulation's mass scale).
So before comparing, we should adjust the ppBHPT's mass-scale to use the NR convention of total mass, which in the 
ppBHPT's simulation would be $m_1 + m_2 = 1 + 1/q$.
This line of reasoning suggests that the ppBHPT modes should be adjusted according to the formula $h^{\ell m}(t) \rightarrow \beta h^{\ell m}(t \beta)$ before  comparing to NR, where $\beta = 1/(1+1/q)$. 
This straightforward identification works well when comparing post-Newtonian and NR waveforms~\cite{Boyle:2007ft} in the comparable mass ratio regime. In Ref.~\cite{Rifat:2019ltp}, it was found that (i) the naive value of $\beta$ accounts for much of the discrepancy between NR and ppBHPT waveforms and (ii) additional model improvements can be obtained by solving an optimization problem for its value. 

\subsubsection{Previous calibration of \texttt{EMRISur1dq1e4}}
\label{Sec:EMRISur1dq1e4}

\begin{figure}[h!]
	\includegraphics[width=\columnwidth]{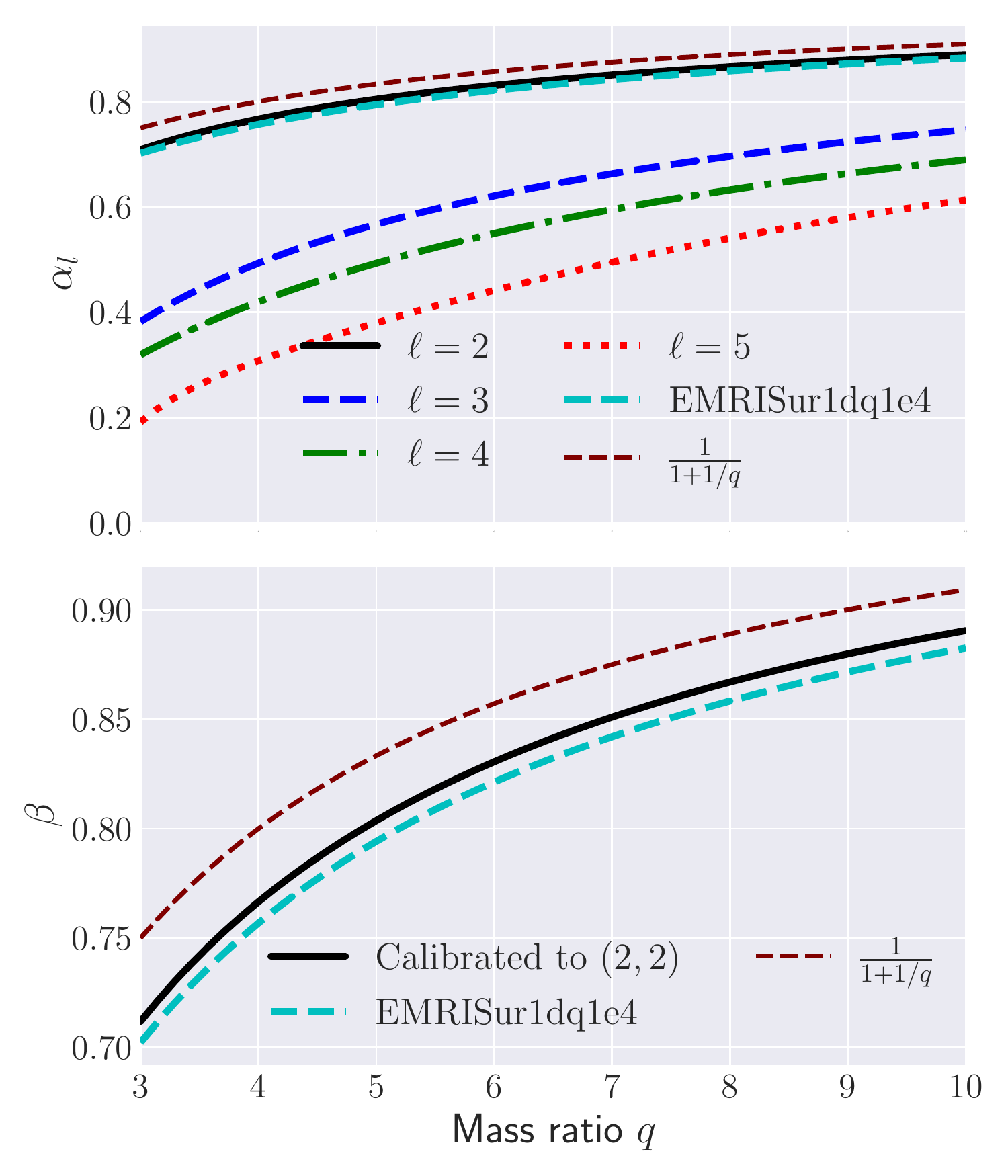}
	\caption{
		Scaling parameters $\alpha_{\ell}(q)$ and $\beta(q)$ as a function of mass ratio $q$. These parameters
		are obtained by minimizing the errors between ppBHPT and hybridized NR waveforms.
		In both panels, the dashed maroon line refers to a naive value  of $\alpha = \beta = 1/(1+1/q)$ set  by  including  the  mass  the  of smaller black hole as part of the background spacetime. 
		We also show the scaling used in the previous model \texttt{EMRISur1dq1e4} as a dashed cyan line in both panels.} 
	\label{Fig:scaling}
\end{figure}

Motivated by the mass scaling argument given above, Ref.~\cite{Rifat:2019ltp} proposed modifying the ppBHPT waveforms according to the formula
\begin{align} 
h^{\ell,m}_{\beta}(t ; q)= {\beta} h^{\ell,m} \left( t \beta;q \right) \,,
\end{align}
where $\beta$ was set by minimizing the difference
\begin{align} \label{eq:old_opt}
    \min_{\beta} \frac{\int \left| h^{22}_{\beta}(t ; q) - h^{22}_{NR}(t ; q) \right|^2 dt}{\int \left|  h^{22}_{NR}(t ; q) \right|^2 dt} \,,
\end{align}
between ppBHPT waveforms and 
nonspinning NR surrogate model~\cite{Blackman:2015pia}
trained on SXS simultation data~\cite{SXSCatalog,Mroue:2013xna,Boyle:2019kee}
for the $(2,2)$ harmonic mode
and mass ratios  $3 \leq q\leq 10$. The integral appearing in Eq.~\eqref{eq:old_opt}
was evaluated from -2,750M to 100M (where $M$ is the total mass of the binary), the duration of the NR surrogate model~\cite{Blackman:2015pia}.
The data $\beta(\nu)$
was then fit to a degree 4 polynomial in the symmetric mass $\nu = q / (1+q)^2$. The resulting 
function is shown as a dashed cyan line in Fig.~\ref{Fig:scaling}.

While the calibration choices and techniques of Ref.~\cite{Rifat:2019ltp} yielded 
surprisingly good agreement with NR, a number of deficiencies have been identified. These include
(i) the NR surrogate model~\cite{Blackman:2015pia} was built before center-of-mass (CoM) corrected waveform data was available and so the model inherited undesirable features due to CoM drifts,
(ii) the NR surrogate model~\cite{Blackman:2015pia} only included about 15 orbits before merger, and it was later found that the calibration parameter deduced on this short interval does not work adequately well on longer time intervals, and
(iii) it should be expected that, due to the point-particle approximation, the higher mode amplitudes computed within the ppBHPT framework will be overestimated as compared to NR. 

\subsubsection{Calibration of \model{}}
\label{sec:alpha-beta}

To overcome the limitations of the previous calibration method discussed in Sec.~\ref{Sec:EMRISur1dq1e4}, 
we present an updated set of choices that provide improvements over the original method. 
Instead of calibrating \model{} data directly to NR data, we use the
\texttt{NRHybSur3dq8} model~\cite{Varma:2018mmi} -- a surrogate model for hybridized non-precessing NR waveforms with the early inspiral waveform obtained using both PN and EOB waveforms. This model was trained on center-of-mass (CoM) corrected waveform data and is much longer in duration, thereby removing two of the three key limitations mentioned in Sec.~\ref{Sec:EMRISur1dq1e4}.

To calibrate ppBHPT waveforms, we propose modifying the \model{} model according to the formula
\begin{align} \label{eq:EMRI_rescale}
h^{\ell,m}_{\tt S, \alpha_{\ell}, \beta}(t ; q)= {\alpha_{\ell}} h^{\ell,m}_{\tt S}\left( t \beta;q \right) \,,
\end{align}
where $\alpha_{\ell}$ and $\beta$ are obtained by minimizing the difference between \texttt{NRHybSur3dq8} and rescaled ppBHPT waveforms 
\begin{align} \label{eq:alpha_lm}
\min_{\alpha_{\ell},\beta} \frac{\int \left| h^{\ell,m}_{\tt S, \alpha_{\ell}, \beta}(t ; q) - h^{\ell,m}_{\rm NRHyb}(t ; q) \right|^2 dt}{\int \left|  h^{\ell,m}_{\rm NRHyb}(t ; q) \right|^2 dt} \,,
\end{align}
between our model \model{} and hybridized NR surrogate waveform \texttt{NRHybSur3dq8} in 
its non-spinning limit for individual modes over the time window.
The integral appearing in Eq.~\eqref{eq:alpha_lm}
is evaluated from -5000M to 115M, which corresponds to the portion of the surrogate model described by NR simulations (i.e., after hybridization).
The motivation for the new parameters $\alpha_{\ell}$ can be seen as a correction to the point-particle approximation in the comparable mass regime i.e. it accounts for the larger relative size of the smaller black hole. We allow for $\ell$-dependent values of $\alpha$ while keeping $\beta$ fixed for all modes. By numerical computation we have checked that there is essentially no $m$-dependence  $\alpha_{\ell m} \approx \alpha_{\ell}$ on these amplitude corrections, which can also be motivated by noting that under rotations the harmonic modes mix in $m$ but not $\ell$.

\begin{figure}[hbt!]
	\includegraphics[width=\columnwidth]{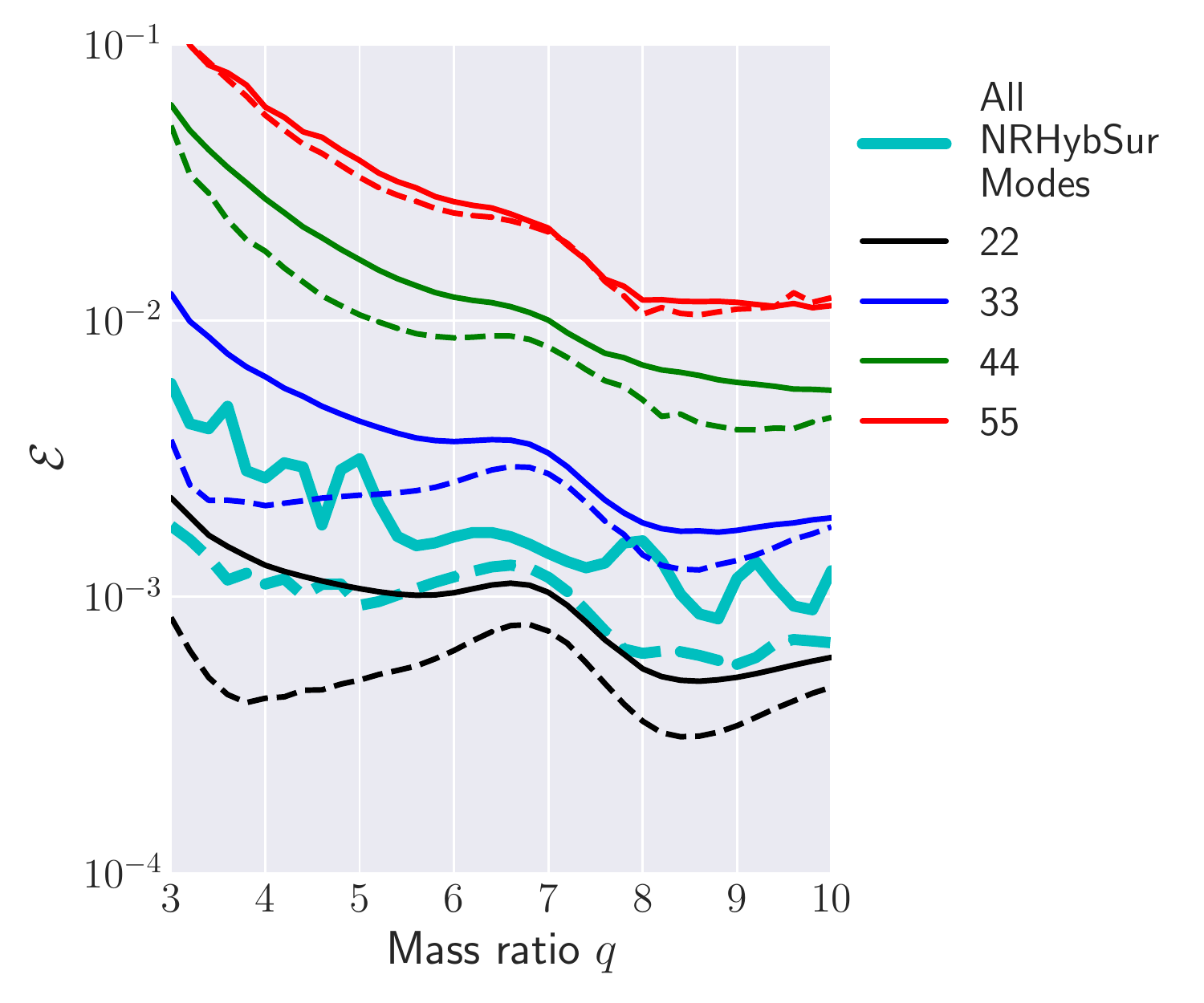}
	\caption{
            Time-domain error between the calibrated ppBHPT waveforms and the NR hybrid surrogate model as a function of mass ratio $q$. We show the errors computed over the full inspiral-merger-ringdown regimes (solid lines) and inspiral-only errors (dashed lines) restricting the waveform to  $t\leq-50M$. We show errors for select modes ($(2,2)$,$(3,3)$,$(4,4)$, and $(5,5)$) as well as for the case including all available \texttt{NRHybSur3dq8} modes including $\ell \neq m$ (\textit{referred as `All NRHybSur Modes'}).
		}
	\label{Fig:scaling_err_fig}
\end{figure}

\begin{table*}[htb]
	\centering
	\setlength{\tabcolsep}{10pt}
	\begin{tabular}{l|l|l|l|l}
		\toprule
		$\ell$ &$A_{\alpha}^{\ell}$ &$B_{\alpha}^{\ell}$ &$C_{\alpha}^{\ell}$ &$D_{\alpha}^{\ell}$\\
		\hline
		2 &-1.330$\pm$0.007 &2.720$\pm$0.116 &-5.904$\pm$0.556 &5.548$\pm$0.833\\
        3 &-3.067$\pm$0.017 &6.244$\pm$0.265 &-9.944$\pm$1.261 &6.437$\pm$1.894\\
        4 &-3.909$\pm$0.032 &9.431$\pm$0.498 &-14.734$\pm$2.367 &9.744$\pm$3.556\\
        5 &-4.509$\pm$0.102 &4.751$\pm$1.554 &21.959$\pm$7.381 &-52.350$\pm$11.085\\
		\botrule	
	\end{tabular}
	\caption{Fitting coefficients  for $\alpha_{\ell}$ parameters as defined in Eq.(\ref{alpha_fit}).}
	\label{Tab:alpha_values}
\end{table*}

\begin{table*}[htb]
	\centering
	\setlength{\tabcolsep}{10pt}
	\begin{tabular}{l|l|l|l}
		\toprule
		$A_{\beta}$ &$B_{\beta}$ &$C_{\beta}$ &$D_{\beta}$\\
		\hline
		-1.238$\pm$0.003 &1.596$\pm$0.049 &-1.776$\pm$0.237 &1.0577$\pm$0.356\\
		\botrule	
	\end{tabular}
	\caption{Fitting coefficients for $\beta$ parameters as defined in Eq.(\ref{beta_fit}).}
	\label{Tab:beta_values}
\end{table*}

\begin{figure*}[htb!]
	\includegraphics[width=1.0\textwidth]{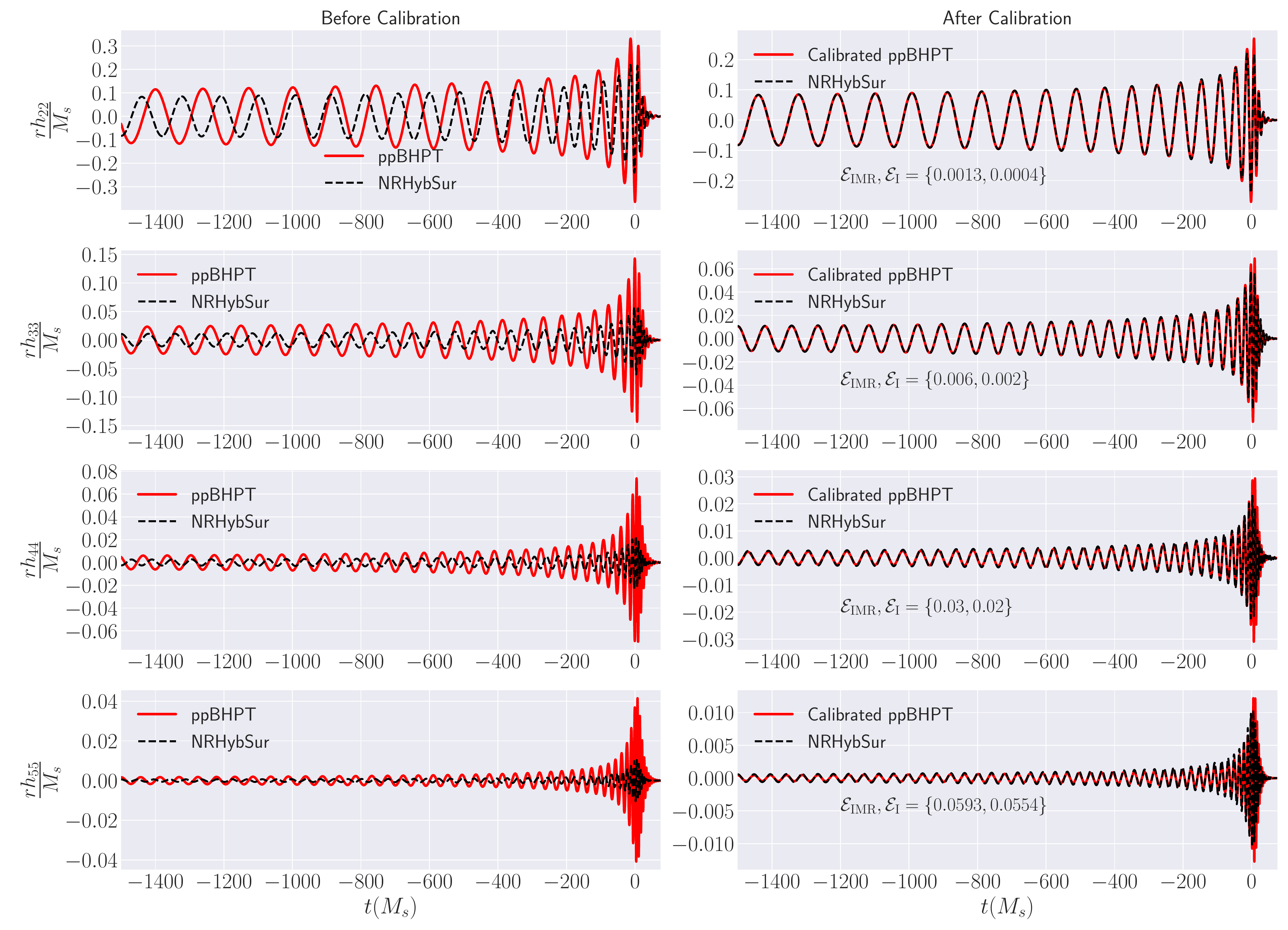}
	\caption{
            Waveform difference between \texttt{NRHybSur3dq8} (\textit{referred as `NRHybSur'}) and ppBHPT waveforms before and after calibration. We show the five different representative modes for $q=4$ to demonstrate the efficacy of the calibration used. Errors for the full inspiral-merger-ringdown waveform and only inspiral part are denoted by $\mathcal{E}_{\rm IMR}$ and $\mathcal{E}_{\rm I}$ respectively.
		The mass scale is denoted by $M_s$, which is either $m_1$ (ppBHPT) or $m_1 + m_2$ (NR) on the left column or either $m_1 \beta$ (calibrated ppBHPT) or $m_1 + m_2$ (NR) on the right column.
	}
	\label{Fig:scaled_waveform}
\end{figure*}

\begin{figure}[htb!]
	\includegraphics[width=\columnwidth]{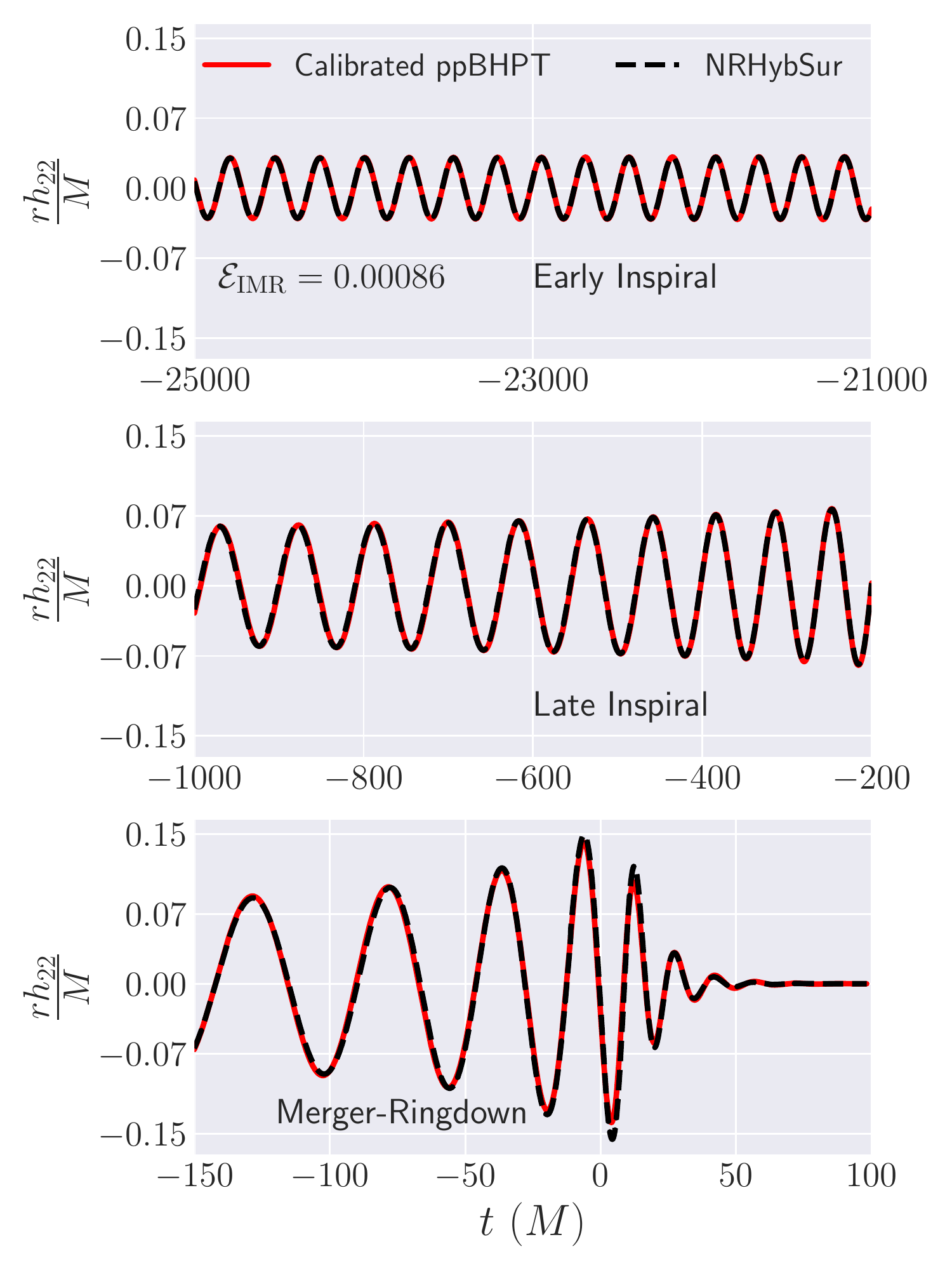}
	\caption{
		Effectiveness of the calibration obtained from a restricted time window $[-5000,115]M$ over the entire length of the waveform. Shown are the early inspiral (upper panel), late inspiral (middle panel), and merger-ringdown (lower panel) parts of the $(2,2)$ mode for $q=8$.
		The hybridized NR surrogate model \texttt{NRHybSur3dq8} (\textit{referred as `NRHybSur'}) and calibrated ppBHPT waveforms are shown in black dashed and solid red, respectively. $\mathcal{E}_{\rm IMR}$ is the $L_2$ norm error computed over the entire waveform.	
	}
	\label{Fig:q8_early_inspiral}
\end{figure}

\begin{figure}[h!]
	\includegraphics[width=0.5\textwidth]{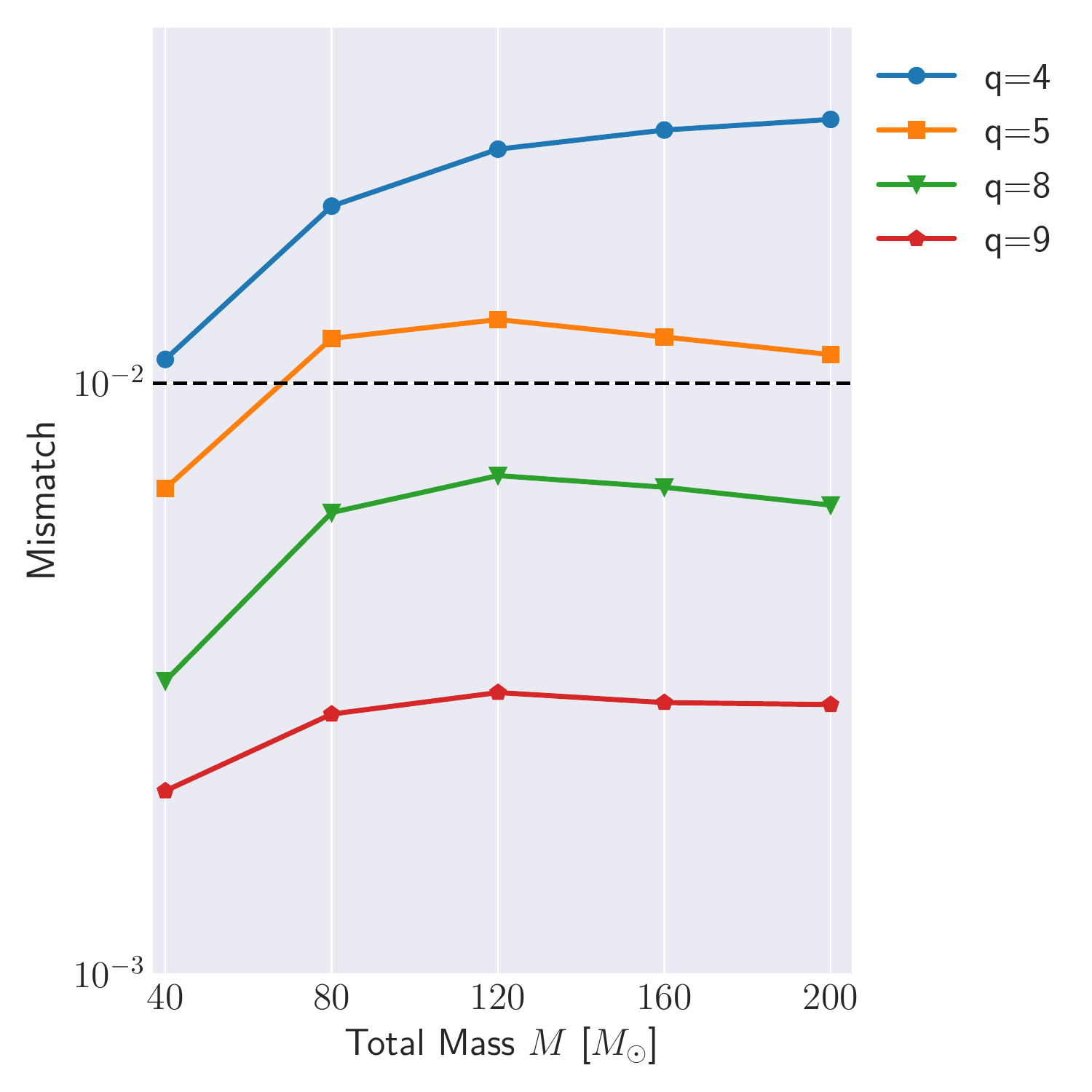}
	\caption{
		Frequency-domain mismatches between the rescaled-ppBHPT surrogate waveforms and \texttt{NRHybSur3dq8} model for different mass ratios. The mismatches are shown as a function of the binary total mass $M$ at inclination $\iota=0.0$ and orbital phase $\varphi=0$, and are computed using the advanced LIGO design sensitivity noise curve. 
		We set the minimum (maximum) frequency appearing in Eq.~\eqref{Eq:freq_domain_Mismatch} to be 20Hz (990Hz). The dashed horizontal line demarcates a mismatch of $0.01$, a commonly used threshold for sufficiently good model quality.
	}
	\label{Fig:ligo_mismatch}
\end{figure}

To obtain values for $\beta$ we minimize the cost function~\eqref{eq:alpha_lm} using the $(2,2)$ mode, while to find best-fit   $\alpha^{l}$ values we use $\ell=m$ modes (i.e. $(2,2)$, $(3,3)$, $(4,4)$ and $(5,5)$ modes). To discover each calibration parameters' $q$-dependence we sample from $q=3.0$ to $q=10.0$ with an increment of $0.2$, giving a total of $36$ data points. These data are then used to fit $\alpha_{\ell}$ and $\beta$ to polynomials in $1/q$:
\begin{align}
\begin{split}
\alpha_{\ell}(q) =  1 & + \frac{A_{\alpha}^{\ell}}{q} + \frac{B_{\alpha}^{\ell}}{q^2}  + \frac{C_{\alpha}^{\ell}}{q^3} + \frac{D_{\alpha}^{\ell}}{q^4}\;.
\end{split}
\label{alpha_fit}
\end{align}
\begin{align}
\begin{split}
\beta(q) =  1 & + \frac{A_{\beta}^{\ell}}{q} + \frac{B_{\beta}^{\ell}}{q^2} 
+ \frac{C_{\beta}^{\ell}}{q^3} + \frac{D_{\beta}^{\ell}}{q^4}\;.
\end{split}
\label{beta_fit}
\end{align}
The order of the polynomial is chosen in the following way: We first build fits for $\alpha^\ell$ and $\beta$ using degrees of polynomial order in $q^{-1}$ from one to four. Next, we select the fit order that minimizes the leave-one-out cross-validation error.
A similar strategy has helped us identify that polynomials in $q^{-1}$ lead to more stable fits to the data than polynomials in symmetric mass ratio $\nu$.
Values of the coefficients for the final fits are given in Table \ref{Tab:alpha_values} and Table \ref{Tab:beta_values} for $\alpha_{\ell}$ and  $\beta$, respectively.

Figure \ref{Fig:scaling} shows the scaling parameters $\alpha_{\ell}$ and  $\beta$ as a function
of mass ratio. 
We find that the $\ell=2$ calibration parameters $\alpha_{2}$ and $\beta$ closely match the analogous parameter used to calibrate the \texttt{EMRISur1dq1e4} model.
For higher order modes, $\alpha_{\ell}$ becomes smaller, which we interpret as a correction in the point-particle framework
to account for the extended size of the smaller black hole. 
Overall, $\alpha_{\ell}$ shows a monotonically increasing behavior with $q$ implying naive $\frac{1}{1+q}$
behavior will be recovered in the larger $q$ limit.
While we are using a simple $\beta$ parameter for all modes, we have experimented having $\beta^{\ell}$ for different modes. However, this did not appreciably change the final match between ppBHPT and NR hybrid waveforms. Furthermore, individual $\beta^{\ell}$ take almost the same values for different modes bolstering the claim that $\beta$, used to scale the time-axis, is related to the mass scaling and provides support for using one single $\beta$ for all modes.

\section{Comparison between the calibrated ppBHPT model and NR}
\label{sec:compare_to_nr}

\subsection{Time domain error in the comparable mass regime}

Figure ~\ref{Fig:scaling_err_fig} shows the time-domain error between
the calibrated \model{} and \texttt{NRHybSur3dq8} models. We show the error using all available \texttt{NRHybSur3dq8} modes~\footnote{The \texttt{NRHybSur3dq8} model includes the following harmonic modes: $\{\ell,m\}=\{(2,1),(2,2),(3,1),(3,2),(3,3),(4,2),(4,3),(4,4),(5,5)\}$.} as well as individual mode errors. Our comparison includes both the entire waveform over all inspiral-merger-ringdown (IMR) regimes (solid lines) and inspiral-only errors (dashed lines), restricting the waveform to  $t\leq-50M$.

In all cases, the differences before calibration are of order unity, while the agreement between the calibrated \model{}
and NR waveforms improves to $\mathcal{E}\sim10^{-3}$ for $(\ell,m)=(2,2)$ and $\mathcal{E}\sim10^{-2}$ for $(\ell,m)=\{(3,3),(4,4),(5,5)\}$.
Many of the $\ell\ne m$ modes have errors as high as $\mathcal{E}\sim 10^{-1}$ even after calibration (not shown).

We expect that in the merger and ringdown regimes, the calibrated ppBHPT waveforms that match so well in the inspiral will no longer serve as a faithful physical ansatz. Instead, we expect the ringdown signal to be described by perturbations of the remnant black hole whose mass and spin only agree with the initial background solution in the limit $q \rightarrow \infty$. This expectation is confirmed in Fig.~\ref{Fig:scaling_err_fig} as we see nearly an order-of-magnitude increase in the IMR error as compared to the inspiral-only error for mass ratios less than $q\approx 5$. By $q\approx 10$, however, the NR-calibrated \model{} does a reasonably good job even in the ringdown regime, which is also apparent in the bottom panel of Fig.~\ref{Fig:q8_early_inspiral}. This suggests that high-accuracy models based on calibrated ppBHPT waveforms may require special treatment in the merger-ringdown regime -- which is commonly employed in other waveform modeling efforts -- although at mass ratios beyond $q\approx 10$ the current approach already does a reasonably good job.

These results are shown in more detail for $q=4$, where Fig.~\ref{Fig:scaled_waveform} shows four of the most important harmonic modes before and after calibration. Before calibration the \model{} and \texttt{NRHybSur3dq8} waveforms differ visibly in both amplitude and phase evolution. The calibrated ppBHPT and \texttt{NRHybSur3dq8} waveforms, however, show surprisingly good agreement although some differences remain in the merger-ringdown part.

We note that these calibration parameters have been obtained by comparing the raw ppBHPT waveform to NR over a time window $[-5000,115]M$ that characterizes the late insprial through ringdown. To test whether the scaling works at earlier times too, we compare \texttt{NRHybSur3dq8} to the calibrated ppBHPT waveforms over the longest possible duration, which is $30,500 m_1$ using the ppBHPT's mass scale.  We show an example case in Fig.\ref{Fig:q8_early_inspiral}. We plot the dominant $(2,2)$ mode for both \texttt{NRHybSur3dq8} (dashed black line) and rescaled ppBHPT (solid red line) waveforms at $q=8$ in early inspiral as well as in the late inspiral and merger-ringdown parts. The waveforms are nearly indistinguishable for the entire duration, and we compute the error to be $\mathcal{E}_{\rm IMR}=0.00086$.

\subsection{Frequency domain mismatch between rescaled-ppBHPT surrogate and NR}

In Fig \ref{Fig:ligo_mismatch}, we show the frequency-domain mismatch between the calibrated ppBHPT surrogate waveforms and \texttt{NRHybSur3dq8} waveforms as a function of total mass for different mass ratios. Frequency domain mismatch between two waveforms $\h_1$ and $\h_2$ is
defined as:
\begin{gather}
	\left<\h_1, \h_2\right> = 4 \mathrm{Re}
	\int_{f_{\mathrm{min}}}^{f_{\mathrm{max}}}
	\frac{\tilde{\h}_1 (f) \tilde{\h}_2^* (f) }{S_n (f)} df,
	\label{Eq:freq_domain_Mismatch}
\end{gather}
where $\tilde{\h}(f)$ indicates the Fourier transform of the complex strain
$\h(t)$, $^*$ indicates complex conjugation, $\mathrm{Re}$ indicates the real
part, and $S_n(f)$ is the one-sided power spectral density of the Advanced LIGO 
detector at its design sensitivity. 
We set $f_{\rm min}$ to be 20Hz while $f_{\rm max}$ is set to be $990$Hz. We note that \model{} waveforms are long enough for all of the mass ratio and total mass configuratios considered in Fig. \ref{Fig:ligo_mismatch}, even the initial instantaneous frequency of the $(5,5)$ mode is below 20Hz.
Before transforming the time domain waveform to the frequency domain, we first
taper the time domain waveform using a Planck window~\cite{McKechan:2010kp},
and then zero-pad to the nearest power of two. The tapering at the start of the
waveform is done over $1.5$ cycles of the $(2,2)$ mode. The tapering at the end
is done over the last $20M$. The mismatches are always optimized over shifts in time,
polarization angle, and initial orbital phase. We find that as the mass ratio increases, 
agreement between the calibrated ppBHPT surrogate waveform and \texttt{NRHybSur3dq8} model
improves. For $q\geq5$, mismatches are below $0.01$ (Fig. \ref{Fig:ligo_mismatch}) indicating 
at least an order of magnitude improvement over our precursor \texttt{EMRISur1dq1e4} model.

\subsection{Extrapolating the model to $q \rightarrow 1$}
\label{Sec:nr_comparison_q1to3}

\begin{figure}[htb!]
	\includegraphics[width=\columnwidth]{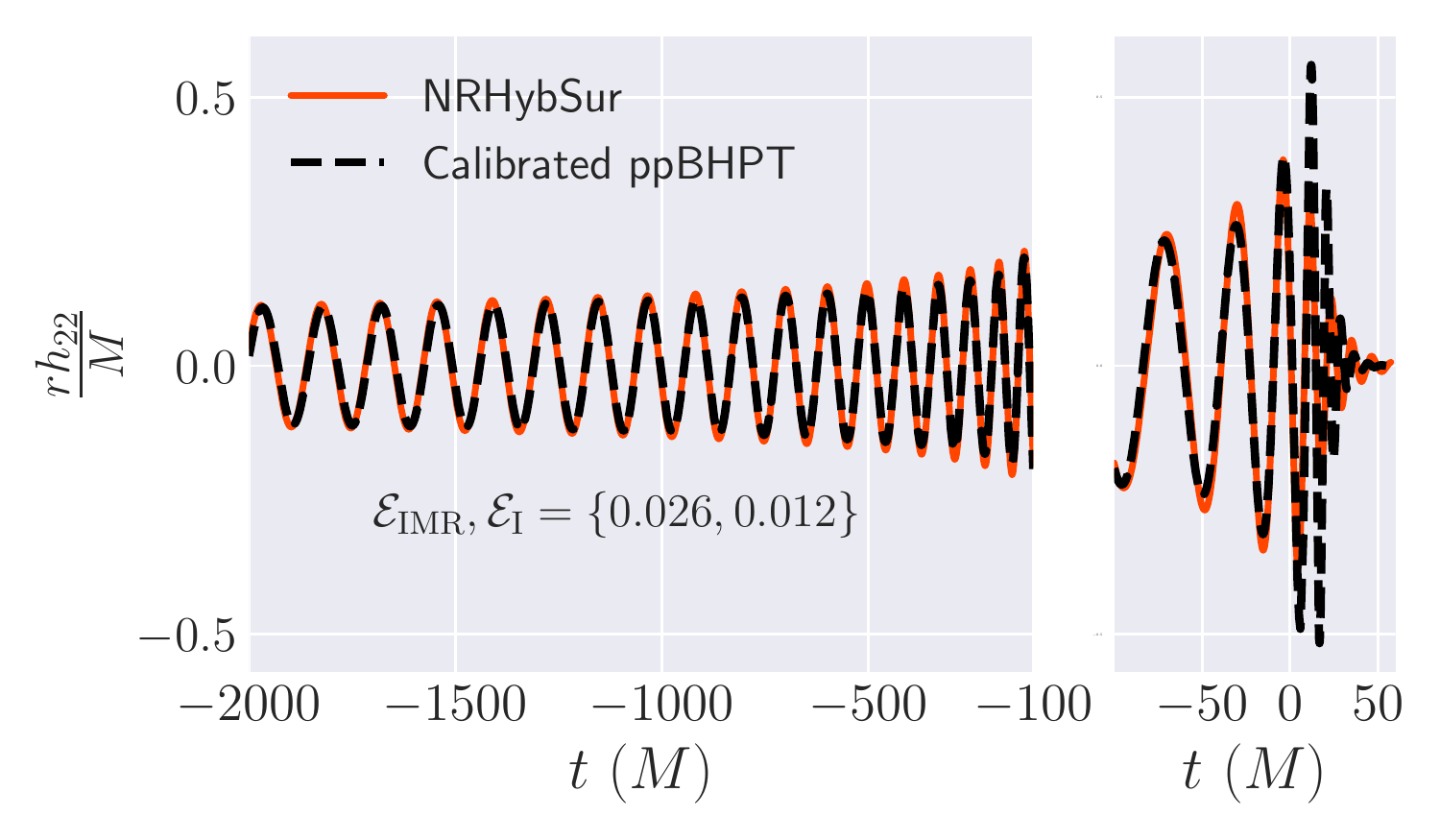}
	\caption{
		Comparison between the calibrated ppBHPT waveform and the hybridized NR surrogate model \texttt{NRHybSur3dq8} (\textit{referred as `NRHybSur'}) at $q=1.2$. We show the dominant $(\ell,m)=(2,2)$ mode. Errors for the full inspiral-merger-ringdown waveform and inspiral-only part are denoted by $\mathcal{E}_{\rm IMR}$ and $\mathcal{E}_{\rm I}$, respectively.
	}
	\label{Fig:scaled_waveform_q1.2}
\end{figure}

Even though \model{} is trained for $q \geq2.5$, we find that the model can be extrapolated to mass ratio $q=1.2$, although 
we advise caution with any extrapolation. This is particularly exciting as our time-domain Teukolsky solver struggles to generate waveforms below $q \approx 2.5$. The model can therefore be used to simulate ppBHPT waveforms for binaries with mass ratios $1.2\le q \le 2.5$. Although, as can be anticipated from Fig.~\ref{Fig:scaling_err_fig}, the calibrated \model{} is not a faithful approximation to GR as $q\rightarrow 1$.
Fig. \ref{Fig:scaled_waveform_q1.2} shows the dominant $(2,2)$ mode of these two models evaluated at $q=1.2$. At this mass ratio, we find the $(2,2)$ mode error of our model to be $\sim 0.026$ for the full inspiral-merger-ringdown waveform and $\sim 0.012$ for inspiral-only part.

\begin{figure*}[htb!]
	\includegraphics[width=1.0\textwidth]{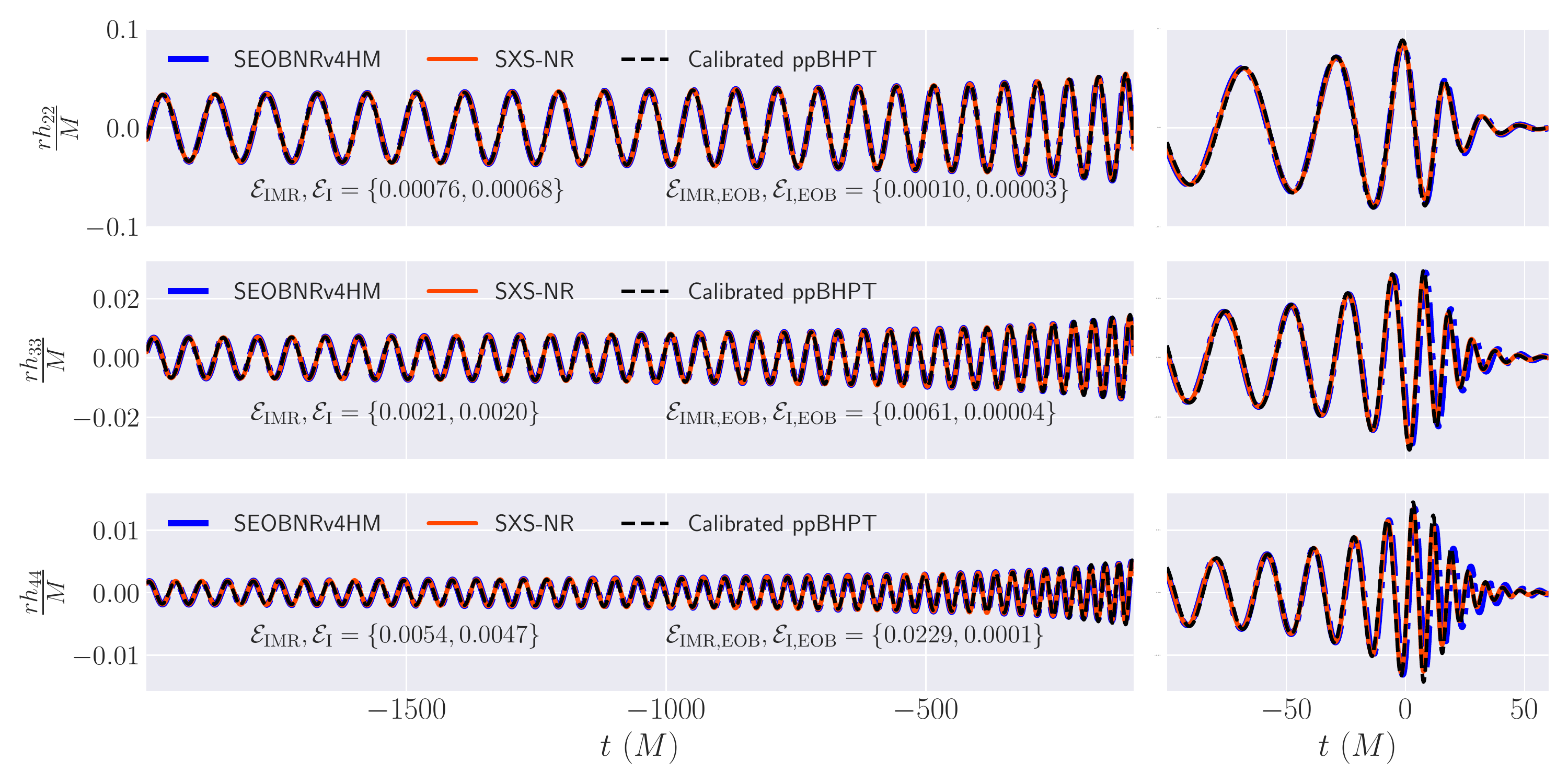}
	\caption{
		Waveform difference between the calibrated ppBHPT waveform (dashed black) and NR data (solid red) from the SXS collaboration (simulation ID SXS:BBH:2304) for $q=15$. We show three different representative modes to demonstrate the efficacy of our model. Errors for the full inspiral-merger-ringdown waveform and only inspiral part is denoted by $\mathcal{E}_{\rm IMR}$ and $\mathcal{E}_{\rm I}$ respectively. For comparison, we also show \texttt{SEOBNRv4HM} waveform modes (solid blue).
		}		
	\label{Fig:scaled_waveform_q15}
\end{figure*}

\begin{figure*}[htb!]
	\includegraphics[width=1.0\textwidth]{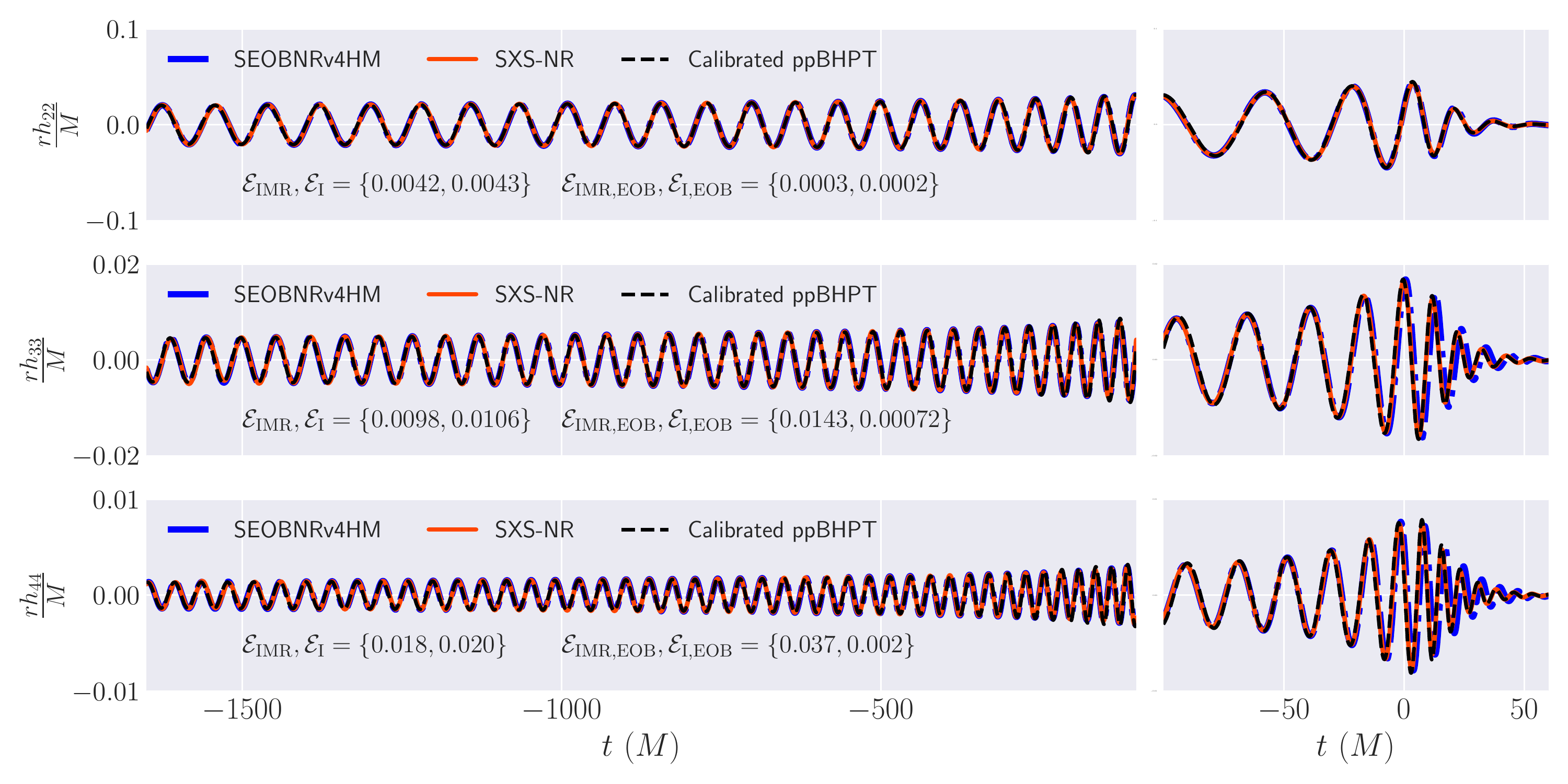}
	\caption{
		Waveform difference between the calibrated ppBHPT waveform (dashed black) and NR data (solid red) from the SXS collaboration for $q=30$~\cite{Giesler:2022inPrep}. We show three different representative modes to demonstrate the efficacy of our model. Errors for the full inspiral-merger-ringdown waveform and only inspiral part is denoted by $\mathcal{E}_{\rm IMR}$ and $\mathcal{E}_{\rm I}$ respectively. For comparison, we also show \texttt{SEOBNRv4HM} waveform modes (solid blue).
		}
	\label{Fig:scaled_waveform_q30}
\end{figure*}

\subsection{Numerical relativity in the intermediate mass ratio regime}
\label{Sec:nr_imri_comparison}

Due to a scarcity of NR data for intermediate-mass ratio ranges, say from $q=10$ to $q=10^4$, it is difficult to perform a detailed comparison between ppBHPT and NR data. Some recent breakthroughs, however, have made it possible to perform  NR simulations for high mass ratio binaries. As we are interested in non-spinning systems, we will consider both SXS NR simulation data for $q=\{15,30\}$~\cite{Yoo:2022erv, Giesler:2022inPrep} and RIT NR simulations for $q=\{15,32\}$ \cite{Lousto:2020tnb}.

\begin{figure*}[htb!]
	\includegraphics[width=1.0\textwidth]{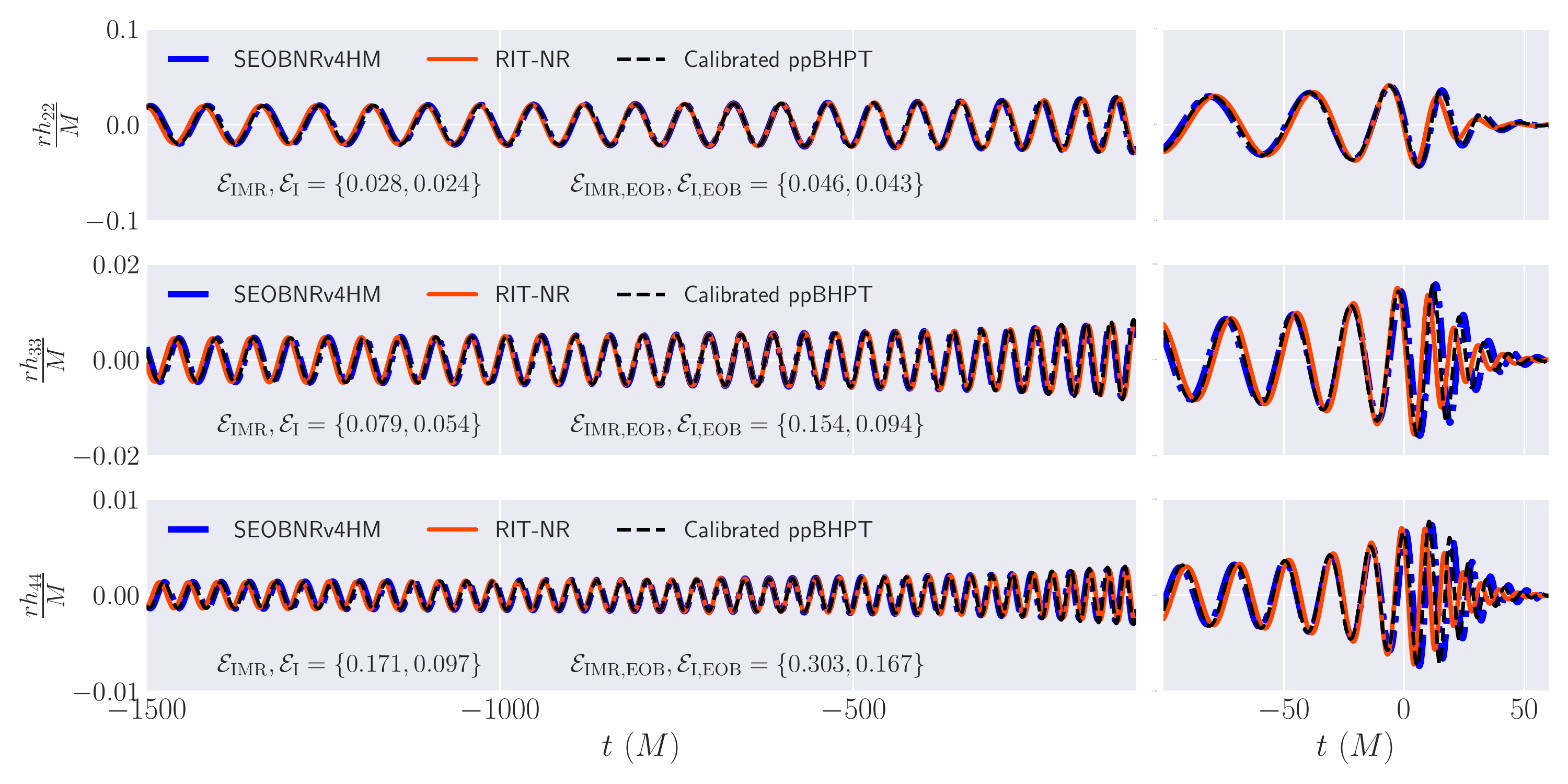}
	\caption{
		Waveform difference between the calibrated ppBHPT waveform (black dashed) and NR data (solid red) from the RIT group (simulation ID RIT:BBH:0792) for $q=32$. We show three different representative modes to demonstrate the efficacy of our model. Errors for the full inspiral-merger-ringdown waveform and only inspiral part is denoted by $\mathcal{E}_{\rm IMR}$ and $\mathcal{E}_{\rm I}$ respectively. For comparison, we also show \texttt{SEOBNRv4HM} waveform modes (solid blue).}		
	\label{Fig:scaled_waveform_q32}
\end{figure*}

\begin{figure*}[htb!]
	\includegraphics[width=1.0\textwidth]{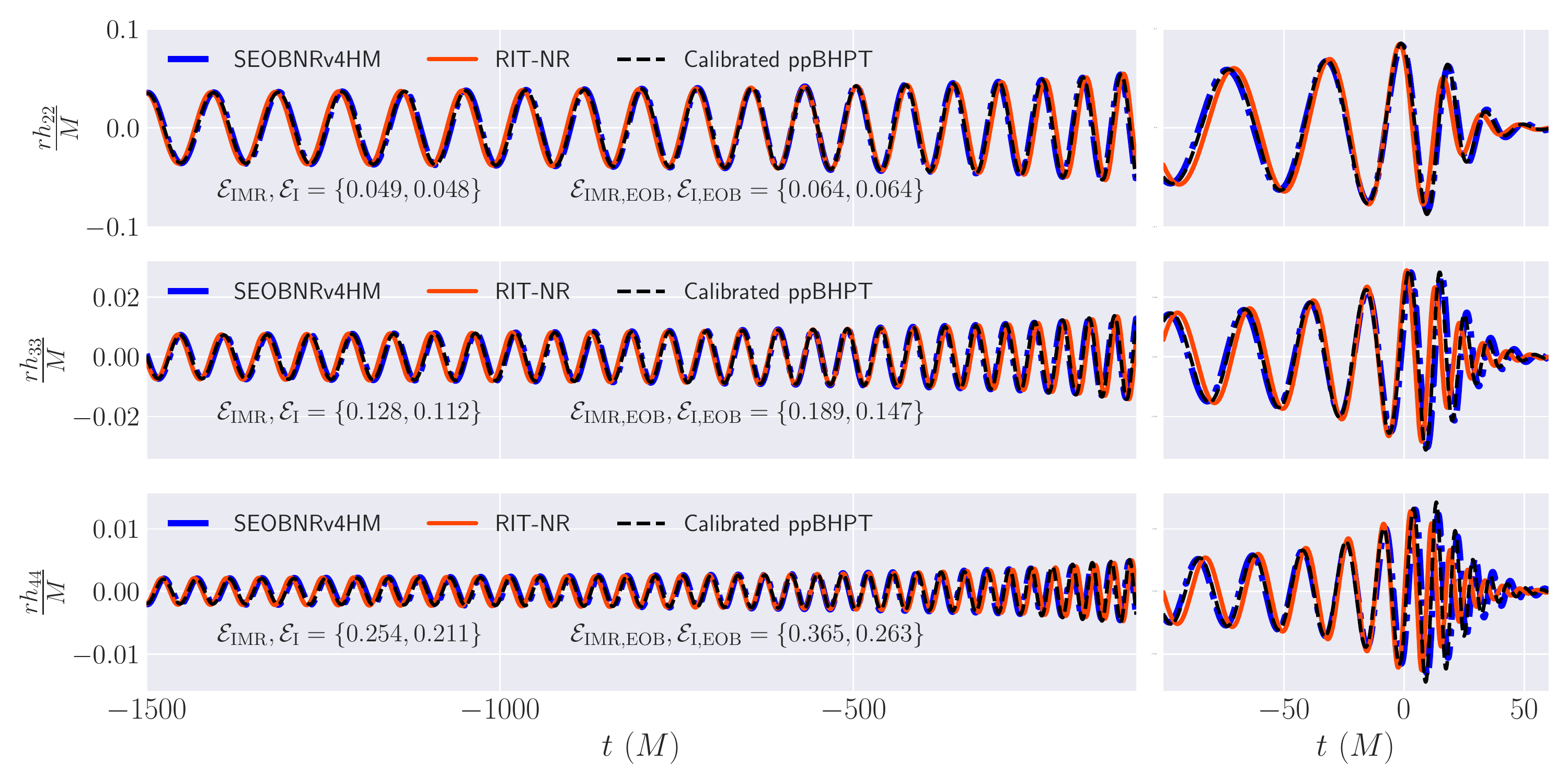}
	\caption{
		Waveform difference between the calibrated ppBHPT waveform (black dashed) and NR data (solid red) from the RIT group (simulation ID RIT:BBH:0373) for $q=15$. We show three different representative modes to demonstrate the efficacy of our model. Errors for the full inspiral-merger-ringdown waveform and only inspiral part is denoted by $\mathcal{E}_{\rm IMR}$ and $\mathcal{E}_{\rm I}$ respectively. For comparison, we also show \texttt{SEOBNRv4HM} waveform modes (solid blue).}	
	\label{Fig:scaled_waveform_q15_rit}
\end{figure*}

\subsubsection{Comparison against SXS NR data at $q=15$ and $q=30$}

We begin with a comparison between the calibrated ppBHPT waveforms, coming from \model{} model, and SXS NR data at mass ratio $q=15$. 
In Fig. \ref{Fig:scaled_waveform_q15}, we plot three representative modes. 
NR data is shown in red solid lines whereas NR-calibrated ppBHPT waveforms are plotted as black dashed lines. 
Note that we do not perform any on-the-fly re-scaling for the ppBHPT waveform
but instead use results from Sec.~\ref{sec:alpha-beta}.
For comparison, we also include the \texttt{SEOBNRv4HM} model. We find both \texttt{SEOBNRv4HM} and \model{} match NR very well at $q=15$. For \model{}, full [inspiral-only] waveform errors for $(2,2)$, $(3,3)$, and $(4,4)$ modes are 0.00076 [0.00068], 0.0021 [0.0020], and 0.0054 [0.0047], respectively. The \texttt{SEOBNRv4HM} shows a noticeable offset from NR around merger and ringdown for the higher modes, although both models deliver good accuracy overall, especially pre-merger. 

When interpreting these errors it is important to note that, from Fig.~\ref{Fig:ppBHPT_error}, it is clear that our model cannot achieve an error $\mathcal{E} \lesssim 0.001$ due to numerical error in the un-calibrated \model{}.   
As this value is consistent with what we see in our comparisons with SXS data, it is not clear if even better agreement with NR could be achieved with higher-accuracy ppBHPT waveform training data.
As an additional check, we also perform an $\{\alpha_{\ell},\beta\}$ optimization between the raw ppBHPT and SXS NR data at $q=15$.
We find that the $\alpha_{\ell}$ and $\beta$ values obtained this way match closely to values obtained from Eq.(\ref{alpha_fit}) and Eq.(\ref{beta_fit}), and negligible improvement in errors observed.
This implies that the calibration carried out in the range $3.0 \leq q \leq 10.0$ continues to work well at mass ratio 15. 

Next, we compare calibrated-ppBHPT waveforms against the highest mass ratio SXS NR data at $q=30$.
Fig. \ref{Fig:scaled_waveform_q30} shows three representative modes for $q=30$ for both
NR (red solid lines) and calibrated-ppBHPT waveforms (black dashed lines). 
We note that at $q=30$, $\beta \approx 0.961$ while $1/(1+1/q) \approx 0.967$, suggesting even an un-calibrated ppBHPT waveform would work reasonably well.
We find for the 22 mode, the errors between calibrated-ppBHPT and NR data are around $\sim 10^{-3}$ implying a good match.
For higher order modes, errors are still around $\sim 0.01$.
Furthermore, when we compare \texttt{SEOBNRv4HM} to NR data, we find the full waveform errors to be similar to the \model{} for higher order modes whereas around one order of magnitude better than the \model{} model for $(2,2)$ mode. 

Just as in the $q = 15$ analysis, as an additional check of our calibration parameters we perform a fresh $\{\alpha_{\ell},\beta\}$ optimization between the raw ppBHPT and SXS NR data at $q=30$.
We find that while the $\alpha_{\ell}$ and $\beta$ values computed this way are quite close to values obtained from Eq.(\ref{alpha_fit}) and Eq.(\ref{beta_fit}), they provide a better match with NR data - with improving the $(2,2)$ mode error by at-least one order of magnitude making it comparable to the $(2,2)$ mode error of \texttt{SEOBNRv4HM}.
This suggests that relatively larger error for calibrated-ppBHPT at $q=30$ when compared to NR data than at $q=15$ is due to the error in extrapolating $\alpha$ and $\beta$ scaling much beyond the mass ratio range ($3\le q \le 10$) it is trained on.
We expect this scaling to become more robust in the future as more NR simulations become available beyond $q=10$. 
We further note that the NR data at $q=30$ is shorter in length (only $\sim1800M$ long).
Careful comparison between ppBHPT and NR data in this regime needs longer NR data.
We leave that study for the future when smoother and longer NR data will be available.
Nonetheless, that a reasonable match between these two types of waveforms is obtained in this regime is a promising sign. 

\subsubsection{Comparison against RIT NR data at $q=15$ and $q=32$}
We now compare calibrated \model{} against the publicly available RIT NR data at mass ratio $q=32$. Fig. \ref{Fig:scaled_waveform_q32} shows three representative modes for both NR (red solid lines), calibrated \model{} (black dashed lines), and \texttt{SEOBNRv4HM} (solid blue) at $q=32$. We find that, for $(2,2)$ mode, scaled ppBHPT and \texttt{SEOBNRv4HM} yields an error of $\sim0.01$, and for subdominant modes the calibrated \model{} matches the NR data a bit more closely than the \texttt{SEOBNRv4HM} model around merger and ringdown. It is interesting to note that at $q=30$, both the calibrated \model{} and \texttt{SEOBNRv4HM} models provide at least one order of magnitude better match when tested against the SXS NR data. This may imply some systematic difference between SXS and RIT NR simulations at high mass ratios. RIT simulations also show evidence of residual eccentricity that introduces additional small modulations in the waveform -- potentially increasing the disagreements. We have also compared the calibrated \model{} ppBHPT and \texttt{SEOBNRv4HM} models with the RIT NR data at $q=15$ (Fig. \ref{Fig:scaled_waveform_q15_rit}), but we encountered issues similar to the $q=32$ case.

\section{Discussion \& Conclusion}
\label{Sec:Conclusion}

In this paper, we have described more fully the methods used to
build \texttt{EMRISur1dq1e4} (which was introduced in a short letter~\cite{Rifat:2019ltp}), a time-domain surrogate model of waveforms obtained through numerically solving the Teukolsky equation sourced by a test-particle with adiabatic-driven inspiral. We apply point-particle black hole perturbation theory (ppBHPT) framework to non-spinning systems with mass ratios from $q=2.5$ to $q=10,000$. While intermediate mass ratio systems ($q > 10$) are targets for our model, we use ppBHPT waveforms in the regime $q < 10$ to (i) carry out comparisons between numerical relativity and ppBHPT and (ii) calibrate the surrogate model to NR thereby vastly improving the model's accuracy.

We have also taken this opportunity to make numerous important improvements to the underlying model. The updated model, \model{}, is $30,500m_1$ in duration, making it suitable to be used in building template banks for LIGO/Virgo data analysis at larger mass-ratios, and also serve as a useful tool for mock data analyses for future observatories. We also employ an improved transition trajectory algorithm between early inspiral and plunge \cite{Lim:2019xrb} -- thereby reducing nonphysical jumps/oscillations in the waveforms (cf. Sec.~\ref{sec:fixer}). The \model{} model includes a total of 50 modes up to $\ell=10$, which is particularly important as subdominant modes are expected to play an important role in intermediate mass ratio systems~\cite{Islam:2021zee, Varma:2016dnf, Varma:2014jxa, Shaik:2019dym}. Please see Table~\ref{Tab:modes} for a complete summary of the \model{}. This model will be made publicly available as part of both the Black Hole Perturbation Toolkit~\cite{BHPToolkit} and GWSurrogate~\cite{gwsurrogate}.

We also perform a detailed comparison between ppBHPT and NR waveforms in the comparable mass ratio regime for  all modes up to $\ell \le 5$. 
We find that after a simple calibration step the ppBHPT waveforms yield remarkable agreement with NR.
The calibrated waveforms are also compared against available SXS NR data at $q=\{15,30\}$ and against RIT NR data at $q=\{15,32\}$, and are found to give good agreement for many of the subdominant modes even up through merger and ringdown.
Furthermore, by construction, the calibration parameters ``turn off" as $q \rightarrow \infty$, so that the correct test-particle behavior is recovered.
Our results suggest that suitably calibrated ppBHPT models may be used to generate accurate late inspiral, merger, and ringdown waveforms in the $q>10$ regime that is especially challenging for NR. Future models should include obvious extensions such as spin, effects of eccentricity, and spin-orbit precession.\\

\begin{acknowledgments}
We thank Keigan Cullen and Nur Rifat for insightful comments and discussions on this project.
The authors acknowledge support from NSF Grants No.~PHY-2106755 (G.K), No. PHY-1806665 (T.I.~and S.F), PHY-1912081 (M.G. and L.E.K), OAC-1931280 (M.G. and L.E.K), and DMS-1912716 (T.I., S.F, and G.K). Part of this work is additionally supported by the Heising-Simons Foundation, the Simons Foundation, and NSF Grants Nos. PHY-1748958.
This project has received funding from the European Union’s Horizon 2020
research and innovation program under the Marie Skłodowska-Curie grant
agreement No.~896869.
Simulations were performed on CARNiE at the Center for Scientific Computing and Visualization Research (CSCVR) of UMassD, which is supported by the ONR/DURIP Grant No.\ N00014181255, the MIT Lincoln Labs {\em SuperCloud} GPU supercomputer supported by the Massachusetts Green High Performance Computing Center (MGHPCC) and ORNL SUMMIT under allocation AST166. 
\end{acknowledgments}  

\bibliography{BHPTNRSur_1D}

\begin{thebibliography}{97}%
\makeatletter
\providecommand \@ifxundefined [1]{%
 \@ifx{#1\undefined}
}%
\providecommand \@ifnum [1]{%
 \ifnum #1\expandafter \@firstoftwo
 \else \expandafter \@secondoftwo
 \fi
}%
\providecommand \@ifx [1]{%
 \ifx #1\expandafter \@firstoftwo
 \else \expandafter \@secondoftwo
 \fi
}%
\providecommand \natexlab [1]{#1}%
\providecommand \enquote  [1]{``#1''}%
\providecommand \bibnamefont  [1]{#1}%
\providecommand \bibfnamefont [1]{#1}%
\providecommand \citenamefont [1]{#1}%
\providecommand \href@noop [0]{\@secondoftwo}%
\providecommand \href [0]{\begingroup \@sanitize@url \@href}%
\providecommand \@href[1]{\@@startlink{#1}\@@href}%
\providecommand \@@href[1]{\endgroup#1\@@endlink}%
\providecommand \@sanitize@url [0]{\catcode `\\12\catcode `\$12\catcode
  `\&12\catcode `\#12\catcode `\^12\catcode `\_12\catcode `\%12\relax}%
\providecommand \@@startlink[1]{}%
\providecommand \@@endlink[0]{}%
\providecommand \url  [0]{\begingroup\@sanitize@url \@url }%
\providecommand \@url [1]{\endgroup\@href {#1}{\urlprefix }}%
\providecommand \urlprefix  [0]{URL }%
\providecommand \Eprint [0]{\href }%
\providecommand \doibase [0]{http://dx.doi.org/}%
\providecommand \selectlanguage [0]{\@gobble}%
\providecommand \bibinfo  [0]{\@secondoftwo}%
\providecommand \bibfield  [0]{\@secondoftwo}%
\providecommand \translation [1]{[#1]}%
\providecommand \BibitemOpen [0]{}%
\providecommand \bibitemStop [0]{}%
\providecommand \bibitemNoStop [0]{.\EOS\space}%
\providecommand \EOS [0]{\spacefactor3000\relax}%
\providecommand \BibitemShut  [1]{\csname bibitem#1\endcsname}%
\let\auto@bib@innerbib\@empty
\bibitem [{\citenamefont {Abbott}\ \emph {et~al.}(2019)\citenamefont {Abbott}
  \emph {et~al.}}]{LIGOScientific:2018mvr}%
  \BibitemOpen
  \bibfield  {author} {\bibinfo {author} {\bibfnamefont {B.~P.}\ \bibnamefont
  {Abbott}} \emph {et~al.} (\bibinfo {collaboration} {LIGO Scientific,
  Virgo}),\ }\bibfield  {title} {\enquote {\bibinfo {title} {{GWTC-1: A
  Gravitational-Wave Transient Catalog of Compact Binary Mergers Observed by
  LIGO and Virgo during the First and Second Observing Runs}},}\ }\href
  {\doibase 10.1103/PhysRevX.9.031040} {\bibfield  {journal} {\bibinfo
  {journal} {Phys. Rev. X}\ }\textbf {\bibinfo {volume} {9}},\ \bibinfo {pages}
  {031040} (\bibinfo {year} {2019})},\ \Eprint
  {http://arxiv.org/abs/1811.12907} {arXiv:1811.12907 [astro-ph.HE]}
  \BibitemShut {NoStop}%
\bibitem [{\citenamefont {Abbott}\ \emph {et~al.}(2020)\citenamefont {Abbott}
  \emph {et~al.}}]{Abbott:2020niy}%
  \BibitemOpen
  \bibfield  {author} {\bibinfo {author} {\bibfnamefont {R.}~\bibnamefont
  {Abbott}} \emph {et~al.} (\bibinfo {collaboration} {LIGO Scientific,
  Virgo}),\ }\bibfield  {title} {\enquote {\bibinfo {title} {{GWTC-2: Compact
  Binary Coalescences Observed by LIGO and Virgo During the First Half of the
  Third Observing Run}},}\ }\href@noop {} {\  (\bibinfo {year} {2020})},\
  \Eprint {http://arxiv.org/abs/2010.14527} {arXiv:2010.14527 [gr-qc]}
  \BibitemShut {NoStop}%
\bibitem [{\citenamefont {Feng}\ and\ \citenamefont
  {Soria}(2011)}]{Feng:2011pc}%
  \BibitemOpen
  \bibfield  {author} {\bibinfo {author} {\bibfnamefont {Hua}\ \bibnamefont
  {Feng}}\ and\ \bibinfo {author} {\bibfnamefont {Roberto}\ \bibnamefont
  {Soria}},\ }\bibfield  {title} {\enquote {\bibinfo {title} {{Ultraluminous
  X-ray Sources in the Chandra and XMM-Newton Era}},}\ }\href {\doibase
  10.1016/j.newar.2011.08.002} {\bibfield  {journal} {\bibinfo  {journal} {New
  Astron. Rev.}\ }\textbf {\bibinfo {volume} {55}},\ \bibinfo {pages}
  {166--183} (\bibinfo {year} {2011})},\ \Eprint
  {http://arxiv.org/abs/1109.1610} {arXiv:1109.1610 [astro-ph.HE]} \BibitemShut
  {NoStop}%
\bibitem [{\citenamefont {Pasham}\ \emph {et~al.}(2014)\citenamefont {Pasham},
  \citenamefont {Strohmayer},\ and\ \citenamefont
  {Mushotzky}}]{Pasham:2015tca}%
  \BibitemOpen
  \bibfield  {author} {\bibinfo {author} {\bibfnamefont {Dheeraj~R.}\
  \bibnamefont {Pasham}}, \bibinfo {author} {\bibfnamefont {Tod~E.}\
  \bibnamefont {Strohmayer}}, \ and\ \bibinfo {author} {\bibfnamefont
  {Richard~F.}\ \bibnamefont {Mushotzky}},\ }\bibfield  {title} {\enquote
  {\bibinfo {title} {{A 400 solar mass black hole in the Ultraluminous X-ray
  source M82 X-1 accreting close to its Eddington limit}},}\ }\href {\doibase
  10.1038/nature13710} {\bibfield  {journal} {\bibinfo  {journal} {Nature}\
  }\textbf {\bibinfo {volume} {513}},\ \bibinfo {pages} {74} (\bibinfo {year}
  {2014})},\ \Eprint {http://arxiv.org/abs/1501.03180} {arXiv:1501.03180
  [astro-ph.HE]} \BibitemShut {NoStop}%
\bibitem [{\citenamefont {Mezcua}(2017)}]{Mezcua:2017npy}%
  \BibitemOpen
  \bibfield  {author} {\bibinfo {author} {\bibfnamefont {Mar}\ \bibnamefont
  {Mezcua}},\ }\bibfield  {title} {\enquote {\bibinfo {title} {{Observational
  evidence for intermediate-mass black holes}},}\ }\href {\doibase
  10.1142/S021827181730021X} {\bibfield  {journal} {\bibinfo  {journal} {Int.
  J. Mod. Phys. D}\ }\textbf {\bibinfo {volume} {26}},\ \bibinfo {pages}
  {1730021} (\bibinfo {year} {2017})},\ \Eprint
  {http://arxiv.org/abs/1705.09667} {arXiv:1705.09667 [astro-ph.GA]}
  \BibitemShut {NoStop}%
\bibitem [{\citenamefont {{Mezcua}}(2017)}]{2017IJMPD..2630021M}%
  \BibitemOpen
  \bibfield  {author} {\bibinfo {author} {\bibfnamefont {Mar}\ \bibnamefont
  {{Mezcua}}},\ }\bibfield  {title} {\enquote {\bibinfo {title} {{Observational
  evidence for intermediate-mass black holes}},}\ }\href {\doibase
  10.1142/S021827181730021X} {\bibfield  {journal} {\bibinfo  {journal}
  {International Journal of Modern Physics D}\ }\textbf {\bibinfo {volume}
  {26}},\ \bibinfo {eid} {1730021} (\bibinfo {year} {2017})},\ \Eprint
  {http://arxiv.org/abs/1705.09667} {arXiv:1705.09667 [astro-ph.GA]}
  \BibitemShut {NoStop}%
\bibitem [{\citenamefont {Leigh}\ \emph {et~al.}(2014)\citenamefont {Leigh},
  \citenamefont {L\"utzgendorf}, \citenamefont {Geller}, \citenamefont
  {Maccarone}, \citenamefont {Heinke},\ and\ \citenamefont
  {Sesana}}]{Leigh:2014oda}%
  \BibitemOpen
  \bibfield  {author} {\bibinfo {author} {\bibfnamefont {Nathan W.~C.}\
  \bibnamefont {Leigh}}, \bibinfo {author} {\bibfnamefont {Nora}\ \bibnamefont
  {L\"utzgendorf}}, \bibinfo {author} {\bibfnamefont {Aaron~M.}\ \bibnamefont
  {Geller}}, \bibinfo {author} {\bibfnamefont {Thomas~J.}\ \bibnamefont
  {Maccarone}}, \bibinfo {author} {\bibfnamefont {Craig}\ \bibnamefont
  {Heinke}}, \ and\ \bibinfo {author} {\bibfnamefont {Alberto}\ \bibnamefont
  {Sesana}},\ }\bibfield  {title} {\enquote {\bibinfo {title} {{On the
  coexistence of stellar-mass and intermediate-mass black holes in globular
  clusters}},}\ }\href {\doibase 10.1093/mnras/stu1437} {\bibfield  {journal}
  {\bibinfo  {journal} {Mon. Not. Roy. Astron. Soc.}\ }\textbf {\bibinfo
  {volume} {444}},\ \bibinfo {pages} {29--42} (\bibinfo {year} {2014})},\
  \Eprint {http://arxiv.org/abs/1407.4459} {arXiv:1407.4459 [astro-ph.SR]}
  \BibitemShut {NoStop}%
\bibitem [{\citenamefont {MacLeod}\ \emph {et~al.}(2016)\citenamefont
  {MacLeod}, \citenamefont {Trenti},\ and\ \citenamefont
  {Ramirez-Ruiz}}]{MacLeod:2015bpa}%
  \BibitemOpen
  \bibfield  {author} {\bibinfo {author} {\bibfnamefont {Morgan}\ \bibnamefont
  {MacLeod}}, \bibinfo {author} {\bibfnamefont {Michele}\ \bibnamefont
  {Trenti}}, \ and\ \bibinfo {author} {\bibfnamefont {Enrico}\ \bibnamefont
  {Ramirez-Ruiz}},\ }\bibfield  {title} {\enquote {\bibinfo {title} {{The Close
  Stellar Companions to Intermediate Mass Black Holes}},}\ }\href {\doibase
  10.3847/0004-637X/819/1/70} {\bibfield  {journal} {\bibinfo  {journal}
  {Astrophys. J.}\ }\textbf {\bibinfo {volume} {819}},\ \bibinfo {pages} {70}
  (\bibinfo {year} {2016})},\ \Eprint {http://arxiv.org/abs/1508.07000}
  {arXiv:1508.07000 [astro-ph.HE]} \BibitemShut {NoStop}%
\bibitem [{\citenamefont {{Amaro-Seoane}}\ \emph {et~al.}(2017)\citenamefont
  {{Amaro-Seoane}}, \citenamefont {{Audley}}, \citenamefont {{Babak}},
  \citenamefont {{Baker}}, \citenamefont {{Barausse}}, \citenamefont
  {{Bender}}, \citenamefont {{Berti}}, \citenamefont {{Binetruy}},
  \citenamefont {{Born}}, \citenamefont {{Bortoluzzi}}, \citenamefont {{Camp}},
  \citenamefont {{Caprini}}, \citenamefont {{Cardoso}}, \citenamefont
  {{Colpi}}, \citenamefont {{Conklin}}, \citenamefont {{Cornish}},
  \citenamefont {{Cutler}}, \citenamefont {{Danzmann}}, \citenamefont
  {{Dolesi}}, \citenamefont {{Ferraioli}}, \citenamefont {{Ferroni}},
  \citenamefont {{Fitzsimons}}, \citenamefont {{Gair}}, \citenamefont {{Gesa
  Bote}}, \citenamefont {{Giardini}}, \citenamefont {{Gibert}}, \citenamefont
  {{Grimani}}, \citenamefont {{Halloin}}, \citenamefont {{Heinzel}},
  \citenamefont {{Hertog}}, \citenamefont {{Hewitson}}, \citenamefont
  {{Holley-Bockelmann}}, \citenamefont {{Hollington}}, \citenamefont
  {{Hueller}}, \citenamefont {{Inchauspe}}, \citenamefont {{Jetzer}},
  \citenamefont {{Karnesis}}, \citenamefont {{Killow}}, \citenamefont
  {{Klein}}, \citenamefont {{Klipstein}}, \citenamefont {{Korsakova}},
  \citenamefont {{Larson}}, \citenamefont {{Livas}}, \citenamefont {{Lloro}},
  \citenamefont {{Man}}, \citenamefont {{Mance}}, \citenamefont {{Martino}},
  \citenamefont {{Mateos}}, \citenamefont {{McKenzie}}, \citenamefont
  {{McWilliams}}, \citenamefont {{Miller}}, \citenamefont {{Mueller}},
  \citenamefont {{Nardini}}, \citenamefont {{Nelemans}}, \citenamefont
  {{Nofrarias}}, \citenamefont {{Petiteau}}, \citenamefont {{Pivato}},
  \citenamefont {{Plagnol}}, \citenamefont {{Porter}}, \citenamefont
  {{Reiche}}, \citenamefont {{Robertson}}, \citenamefont {{Robertson}},
  \citenamefont {{Rossi}}, \citenamefont {{Russano}}, \citenamefont {{Schutz}},
  \citenamefont {{Sesana}}, \citenamefont {{Shoemaker}}, \citenamefont
  {{Slutsky}}, \citenamefont {{Sopuerta}}, \citenamefont {{Sumner}},
  \citenamefont {{Tamanini}}, \citenamefont {{Thorpe}}, \citenamefont
  {{Troebs}}, \citenamefont {{Vallisneri}}, \citenamefont {{Vecchio}},
  \citenamefont {{Vetrugno}}, \citenamefont {{Vitale}}, \citenamefont
  {{Volonteri}}, \citenamefont {{Wanner}}, \citenamefont {{Ward}},
  \citenamefont {{Wass}}, \citenamefont {{Weber}}, \citenamefont {{Ziemer}},\
  and\ \citenamefont {{Zweifel}}}]{2017arXiv170200786A}%
  \BibitemOpen
  \bibfield  {author} {\bibinfo {author} {\bibfnamefont {Pau}\ \bibnamefont
  {{Amaro-Seoane}}}, \bibinfo {author} {\bibfnamefont {Heather}\ \bibnamefont
  {{Audley}}}, \bibinfo {author} {\bibfnamefont {Stanislav}\ \bibnamefont
  {{Babak}}}, \bibinfo {author} {\bibfnamefont {John}\ \bibnamefont {{Baker}}},
  \bibinfo {author} {\bibfnamefont {Enrico}\ \bibnamefont {{Barausse}}},
  \bibinfo {author} {\bibfnamefont {Peter}\ \bibnamefont {{Bender}}}, \bibinfo
  {author} {\bibfnamefont {Emanuele}\ \bibnamefont {{Berti}}}, \bibinfo
  {author} {\bibfnamefont {Pierre}\ \bibnamefont {{Binetruy}}}, \bibinfo
  {author} {\bibfnamefont {Michael}\ \bibnamefont {{Born}}}, \bibinfo {author}
  {\bibfnamefont {Daniele}\ \bibnamefont {{Bortoluzzi}}}, \bibinfo {author}
  {\bibfnamefont {Jordan}\ \bibnamefont {{Camp}}}, \bibinfo {author}
  {\bibfnamefont {Chiara}\ \bibnamefont {{Caprini}}}, \bibinfo {author}
  {\bibfnamefont {Vitor}\ \bibnamefont {{Cardoso}}}, \bibinfo {author}
  {\bibfnamefont {Monica}\ \bibnamefont {{Colpi}}}, \bibinfo {author}
  {\bibfnamefont {John}\ \bibnamefont {{Conklin}}}, \bibinfo {author}
  {\bibfnamefont {Neil}\ \bibnamefont {{Cornish}}}, \bibinfo {author}
  {\bibfnamefont {Curt}\ \bibnamefont {{Cutler}}}, \bibinfo {author}
  {\bibfnamefont {Karsten}\ \bibnamefont {{Danzmann}}}, \bibinfo {author}
  {\bibfnamefont {Rita}\ \bibnamefont {{Dolesi}}}, \bibinfo {author}
  {\bibfnamefont {Luigi}\ \bibnamefont {{Ferraioli}}}, \bibinfo {author}
  {\bibfnamefont {Valerio}\ \bibnamefont {{Ferroni}}}, \bibinfo {author}
  {\bibfnamefont {Ewan}\ \bibnamefont {{Fitzsimons}}}, \bibinfo {author}
  {\bibfnamefont {Jonathan}\ \bibnamefont {{Gair}}}, \bibinfo {author}
  {\bibfnamefont {Lluis}\ \bibnamefont {{Gesa Bote}}}, \bibinfo {author}
  {\bibfnamefont {Domenico}\ \bibnamefont {{Giardini}}}, \bibinfo {author}
  {\bibfnamefont {Ferran}\ \bibnamefont {{Gibert}}}, \bibinfo {author}
  {\bibfnamefont {Catia}\ \bibnamefont {{Grimani}}}, \bibinfo {author}
  {\bibfnamefont {Hubert}\ \bibnamefont {{Halloin}}}, \bibinfo {author}
  {\bibfnamefont {Gerhard}\ \bibnamefont {{Heinzel}}}, \bibinfo {author}
  {\bibfnamefont {Thomas}\ \bibnamefont {{Hertog}}}, \bibinfo {author}
  {\bibfnamefont {Martin}\ \bibnamefont {{Hewitson}}}, \bibinfo {author}
  {\bibfnamefont {Kelly}\ \bibnamefont {{Holley-Bockelmann}}}, \bibinfo
  {author} {\bibfnamefont {Daniel}\ \bibnamefont {{Hollington}}}, \bibinfo
  {author} {\bibfnamefont {Mauro}\ \bibnamefont {{Hueller}}}, \bibinfo {author}
  {\bibfnamefont {Henri}\ \bibnamefont {{Inchauspe}}}, \bibinfo {author}
  {\bibfnamefont {Philippe}\ \bibnamefont {{Jetzer}}}, \bibinfo {author}
  {\bibfnamefont {Nikos}\ \bibnamefont {{Karnesis}}}, \bibinfo {author}
  {\bibfnamefont {Christian}\ \bibnamefont {{Killow}}}, \bibinfo {author}
  {\bibfnamefont {Antoine}\ \bibnamefont {{Klein}}}, \bibinfo {author}
  {\bibfnamefont {Bill}\ \bibnamefont {{Klipstein}}}, \bibinfo {author}
  {\bibfnamefont {Natalia}\ \bibnamefont {{Korsakova}}}, \bibinfo {author}
  {\bibfnamefont {Shane~L}\ \bibnamefont {{Larson}}}, \bibinfo {author}
  {\bibfnamefont {Jeffrey}\ \bibnamefont {{Livas}}}, \bibinfo {author}
  {\bibfnamefont {Ivan}\ \bibnamefont {{Lloro}}}, \bibinfo {author}
  {\bibfnamefont {Nary}\ \bibnamefont {{Man}}}, \bibinfo {author}
  {\bibfnamefont {Davor}\ \bibnamefont {{Mance}}}, \bibinfo {author}
  {\bibfnamefont {Joseph}\ \bibnamefont {{Martino}}}, \bibinfo {author}
  {\bibfnamefont {Ignacio}\ \bibnamefont {{Mateos}}}, \bibinfo {author}
  {\bibfnamefont {Kirk}\ \bibnamefont {{McKenzie}}}, \bibinfo {author}
  {\bibfnamefont {Sean~T}\ \bibnamefont {{McWilliams}}}, \bibinfo {author}
  {\bibfnamefont {Cole}\ \bibnamefont {{Miller}}}, \bibinfo {author}
  {\bibfnamefont {Guido}\ \bibnamefont {{Mueller}}}, \bibinfo {author}
  {\bibfnamefont {Germano}\ \bibnamefont {{Nardini}}}, \bibinfo {author}
  {\bibfnamefont {Gijs}\ \bibnamefont {{Nelemans}}}, \bibinfo {author}
  {\bibfnamefont {Miquel}\ \bibnamefont {{Nofrarias}}}, \bibinfo {author}
  {\bibfnamefont {Antoine}\ \bibnamefont {{Petiteau}}}, \bibinfo {author}
  {\bibfnamefont {Paolo}\ \bibnamefont {{Pivato}}}, \bibinfo {author}
  {\bibfnamefont {Eric}\ \bibnamefont {{Plagnol}}}, \bibinfo {author}
  {\bibfnamefont {Ed}~\bibnamefont {{Porter}}}, \bibinfo {author}
  {\bibfnamefont {Jens}\ \bibnamefont {{Reiche}}}, \bibinfo {author}
  {\bibfnamefont {David}\ \bibnamefont {{Robertson}}}, \bibinfo {author}
  {\bibfnamefont {Norna}\ \bibnamefont {{Robertson}}}, \bibinfo {author}
  {\bibfnamefont {Elena}\ \bibnamefont {{Rossi}}}, \bibinfo {author}
  {\bibfnamefont {Giuliana}\ \bibnamefont {{Russano}}}, \bibinfo {author}
  {\bibfnamefont {Bernard}\ \bibnamefont {{Schutz}}}, \bibinfo {author}
  {\bibfnamefont {Alberto}\ \bibnamefont {{Sesana}}}, \bibinfo {author}
  {\bibfnamefont {David}\ \bibnamefont {{Shoemaker}}}, \bibinfo {author}
  {\bibfnamefont {Jacob}\ \bibnamefont {{Slutsky}}}, \bibinfo {author}
  {\bibfnamefont {Carlos~F.}\ \bibnamefont {{Sopuerta}}}, \bibinfo {author}
  {\bibfnamefont {Tim}\ \bibnamefont {{Sumner}}}, \bibinfo {author}
  {\bibfnamefont {Nicola}\ \bibnamefont {{Tamanini}}}, \bibinfo {author}
  {\bibfnamefont {Ira}\ \bibnamefont {{Thorpe}}}, \bibinfo {author}
  {\bibfnamefont {Michael}\ \bibnamefont {{Troebs}}}, \bibinfo {author}
  {\bibfnamefont {Michele}\ \bibnamefont {{Vallisneri}}}, \bibinfo {author}
  {\bibfnamefont {Alberto}\ \bibnamefont {{Vecchio}}}, \bibinfo {author}
  {\bibfnamefont {Daniele}\ \bibnamefont {{Vetrugno}}}, \bibinfo {author}
  {\bibfnamefont {Stefano}\ \bibnamefont {{Vitale}}}, \bibinfo {author}
  {\bibfnamefont {Marta}\ \bibnamefont {{Volonteri}}}, \bibinfo {author}
  {\bibfnamefont {Gudrun}\ \bibnamefont {{Wanner}}}, \bibinfo {author}
  {\bibfnamefont {Harry}\ \bibnamefont {{Ward}}}, \bibinfo {author}
  {\bibfnamefont {Peter}\ \bibnamefont {{Wass}}}, \bibinfo {author}
  {\bibfnamefont {William}\ \bibnamefont {{Weber}}}, \bibinfo {author}
  {\bibfnamefont {John}\ \bibnamefont {{Ziemer}}}, \ and\ \bibinfo {author}
  {\bibfnamefont {Peter}\ \bibnamefont {{Zweifel}}},\ }\bibfield  {title}
  {\enquote {\bibinfo {title} {{Laser Interferometer Space Antenna}},}\
  }\href@noop {} {\bibfield  {journal} {\bibinfo  {journal} {arXiv e-prints}\
  ,\ \bibinfo {eid} {arXiv:1702.00786}} (\bibinfo {year} {2017})},\ \Eprint
  {http://arxiv.org/abs/1702.00786} {arXiv:1702.00786 [astro-ph.IM]}
  \BibitemShut {NoStop}%
\bibitem [{\citenamefont {Amaro-Seoane}\ \emph {et~al.}(2007)\citenamefont
  {Amaro-Seoane}, \citenamefont {Gair}, \citenamefont {Freitag}, \citenamefont
  {Coleman~Miller}, \citenamefont {Mandel}, \citenamefont {Cutler},\ and\
  \citenamefont {Babak}}]{AmaroSeoane:2007aw}%
  \BibitemOpen
  \bibfield  {author} {\bibinfo {author} {\bibfnamefont {Pau}\ \bibnamefont
  {Amaro-Seoane}}, \bibinfo {author} {\bibfnamefont {Jonathan~R.}\ \bibnamefont
  {Gair}}, \bibinfo {author} {\bibfnamefont {Marc}\ \bibnamefont {Freitag}},
  \bibinfo {author} {\bibfnamefont {M.}~\bibnamefont {Coleman~Miller}},
  \bibinfo {author} {\bibfnamefont {Ilya}\ \bibnamefont {Mandel}}, \bibinfo
  {author} {\bibfnamefont {Curt~J.}\ \bibnamefont {Cutler}}, \ and\ \bibinfo
  {author} {\bibfnamefont {Stanislav}\ \bibnamefont {Babak}},\ }\bibfield
  {title} {\enquote {\bibinfo {title} {{Astrophysics, detection and science
  applications of intermediate- and extreme mass-ratio inspirals}},}\ }\href
  {\doibase 10.1088/0264-9381/24/17/R01} {\bibfield  {journal} {\bibinfo
  {journal} {Class. Quant. Grav.}\ }\textbf {\bibinfo {volume} {24}},\ \bibinfo
  {pages} {R113--R169} (\bibinfo {year} {2007})},\ \Eprint
  {http://arxiv.org/abs/astro-ph/0703495} {arXiv:astro-ph/0703495} \BibitemShut
  {NoStop}%
\bibitem [{\citenamefont {Berry}\ \emph {et~al.}(2019)\citenamefont {Berry},
  \citenamefont {Hughes}, \citenamefont {Sopuerta}, \citenamefont {Chua},
  \citenamefont {Heffernan}, \citenamefont {Holley-Bockelmann}, \citenamefont
  {Mihaylov}, \citenamefont {Miller},\ and\ \citenamefont
  {Sesana}}]{Berry:2019wgg}%
  \BibitemOpen
  \bibfield  {author} {\bibinfo {author} {\bibfnamefont {Christopher P.~L.}\
  \bibnamefont {Berry}}, \bibinfo {author} {\bibfnamefont {Scott~A.}\
  \bibnamefont {Hughes}}, \bibinfo {author} {\bibfnamefont {Carlos~F.}\
  \bibnamefont {Sopuerta}}, \bibinfo {author} {\bibfnamefont {Alvin J.~K.}\
  \bibnamefont {Chua}}, \bibinfo {author} {\bibfnamefont {Anna}\ \bibnamefont
  {Heffernan}}, \bibinfo {author} {\bibfnamefont {Kelly}\ \bibnamefont
  {Holley-Bockelmann}}, \bibinfo {author} {\bibfnamefont {Deyan~P.}\
  \bibnamefont {Mihaylov}}, \bibinfo {author} {\bibfnamefont {M.~Coleman}\
  \bibnamefont {Miller}}, \ and\ \bibinfo {author} {\bibfnamefont {Alberto}\
  \bibnamefont {Sesana}},\ }\bibfield  {title} {\enquote {\bibinfo {title}
  {{The unique potential of extreme mass-ratio inspirals for gravitational-wave
  astronomy}},}\ }\href@noop {} {\  (\bibinfo {year} {2019})},\ \Eprint
  {http://arxiv.org/abs/1903.03686} {arXiv:1903.03686 [astro-ph.HE]}
  \BibitemShut {NoStop}%
\bibitem [{\citenamefont {Amaro-Seoane}(2018)}]{Amaro-Seoane:2018gbb}%
  \BibitemOpen
  \bibfield  {author} {\bibinfo {author} {\bibfnamefont {Pau}\ \bibnamefont
  {Amaro-Seoane}},\ }\bibfield  {title} {\enquote {\bibinfo {title} {{Detecting
  Intermediate-Mass Ratio Inspirals From The Ground And Space}},}\ }\href
  {\doibase 10.1103/PhysRevD.98.063018} {\bibfield  {journal} {\bibinfo
  {journal} {Phys. Rev. D}\ }\textbf {\bibinfo {volume} {98}},\ \bibinfo
  {pages} {063018} (\bibinfo {year} {2018})},\ \Eprint
  {http://arxiv.org/abs/1807.03824} {arXiv:1807.03824 [astro-ph.HE]}
  \BibitemShut {NoStop}%
\bibitem [{\citenamefont {Kawamura}\ \emph {et~al.}(2020)\citenamefont
  {Kawamura} \emph {et~al.}}]{Kawamura:2020pcg}%
  \BibitemOpen
  \bibfield  {author} {\bibinfo {author} {\bibfnamefont {Seiji}\ \bibnamefont
  {Kawamura}} \emph {et~al.},\ }\bibfield  {title} {\enquote {\bibinfo {title}
  {{Current status of space gravitational wave antenna DECIGO and B-DECIGO}},}\
  }\href@noop {} {\  (\bibinfo {year} {2020})},\ \Eprint
  {http://arxiv.org/abs/2006.13545} {arXiv:2006.13545 [gr-qc]} \BibitemShut
  {NoStop}%
\bibitem [{\citenamefont {Sedda}\ \emph {et~al.}(2020)\citenamefont {Sedda}
  \emph {et~al.}}]{Sedda:2019uro}%
  \BibitemOpen
  \bibfield  {author} {\bibinfo {author} {\bibfnamefont {Manuel~Arca}\
  \bibnamefont {Sedda}} \emph {et~al.},\ }\bibfield  {title} {\enquote
  {\bibinfo {title} {{The missing link in gravitational-wave astronomy:
  discoveries waiting in the decihertz range}},}\ }\href {\doibase
  10.1088/1361-6382/abb5c1} {\bibfield  {journal} {\bibinfo  {journal} {Class.
  Quant. Grav.}\ }\textbf {\bibinfo {volume} {37}},\ \bibinfo {pages} {215011}
  (\bibinfo {year} {2020})},\ \Eprint {http://arxiv.org/abs/1908.11375}
  {arXiv:1908.11375 [gr-qc]} \BibitemShut {NoStop}%
\bibitem [{\citenamefont {Gair}\ \emph
  {et~al.}(2011{\natexlab{a}})\citenamefont {Gair}, \citenamefont {Mandel},
  \citenamefont {Miller},\ and\ \citenamefont {Volonteri}}]{Gair:2010dx}%
  \BibitemOpen
  \bibfield  {author} {\bibinfo {author} {\bibfnamefont {Jonathan~R.}\
  \bibnamefont {Gair}}, \bibinfo {author} {\bibfnamefont {Ilya}\ \bibnamefont
  {Mandel}}, \bibinfo {author} {\bibfnamefont {M.~Coleman}\ \bibnamefont
  {Miller}}, \ and\ \bibinfo {author} {\bibfnamefont {Marta}\ \bibnamefont
  {Volonteri}},\ }\bibfield  {title} {\enquote {\bibinfo {title} {{Exploring
  intermediate and massive black-hole binaries with the Einstein Telescope}},}\
  }\href {\doibase 10.1007/s10714-010-1104-3} {\bibfield  {journal} {\bibinfo
  {journal} {Gen. Rel. Grav.}\ }\textbf {\bibinfo {volume} {43}},\ \bibinfo
  {pages} {485--518} (\bibinfo {year} {2011}{\natexlab{a}})},\ \Eprint
  {http://arxiv.org/abs/0907.5450} {arXiv:0907.5450 [astro-ph.CO]} \BibitemShut
  {NoStop}%
\bibitem [{\citenamefont {Huerta}\ and\ \citenamefont
  {Gair}(2011{\natexlab{a}})}]{Huerta:2010un}%
  \BibitemOpen
  \bibfield  {author} {\bibinfo {author} {\bibfnamefont {E.~A.}\ \bibnamefont
  {Huerta}}\ and\ \bibinfo {author} {\bibfnamefont {Jonathan~R.}\ \bibnamefont
  {Gair}},\ }\bibfield  {title} {\enquote {\bibinfo {title}
  {{Intermediate-mass-ratio-inspirals in the Einstein Telescope: I.
  Signal-to-noise ratio calculations}},}\ }\href {\doibase
  10.1103/PhysRevD.83.044020} {\bibfield  {journal} {\bibinfo  {journal} {Phys.
  Rev. D}\ }\textbf {\bibinfo {volume} {83}},\ \bibinfo {pages} {044020}
  (\bibinfo {year} {2011}{\natexlab{a}})},\ \Eprint
  {http://arxiv.org/abs/1009.1985} {arXiv:1009.1985 [gr-qc]} \BibitemShut
  {NoStop}%
\bibitem [{\citenamefont {Huerta}\ and\ \citenamefont
  {Gair}(2011{\natexlab{b}})}]{Huerta:2010tp}%
  \BibitemOpen
  \bibfield  {author} {\bibinfo {author} {\bibfnamefont {E.~A.}\ \bibnamefont
  {Huerta}}\ and\ \bibinfo {author} {\bibfnamefont {Jonathan~R.}\ \bibnamefont
  {Gair}},\ }\bibfield  {title} {\enquote {\bibinfo {title}
  {{Intermediate-mass-ratio-inspirals in the Einstein Telescope. II. Parameter
  estimation errors}},}\ }\href {\doibase 10.1103/PhysRevD.83.044021}
  {\bibfield  {journal} {\bibinfo  {journal} {Phys. Rev. D}\ }\textbf {\bibinfo
  {volume} {83}},\ \bibinfo {pages} {044021} (\bibinfo {year}
  {2011}{\natexlab{b}})},\ \Eprint {http://arxiv.org/abs/1011.0421}
  {arXiv:1011.0421 [gr-qc]} \BibitemShut {NoStop}%
\bibitem [{\citenamefont {Haster}\ \emph {et~al.}(2016)\citenamefont {Haster},
  \citenamefont {Wang}, \citenamefont {Berry}, \citenamefont {Stevenson},
  \citenamefont {Veitch},\ and\ \citenamefont {Mandel}}]{Haster:2015cnn}%
  \BibitemOpen
  \bibfield  {author} {\bibinfo {author} {\bibfnamefont {Carl-Johan}\
  \bibnamefont {Haster}}, \bibinfo {author} {\bibfnamefont {Zhilu}\
  \bibnamefont {Wang}}, \bibinfo {author} {\bibfnamefont {Christopher P.~L.}\
  \bibnamefont {Berry}}, \bibinfo {author} {\bibfnamefont {Simon}\ \bibnamefont
  {Stevenson}}, \bibinfo {author} {\bibfnamefont {John}\ \bibnamefont
  {Veitch}}, \ and\ \bibinfo {author} {\bibfnamefont {Ilya}\ \bibnamefont
  {Mandel}},\ }\bibfield  {title} {\enquote {\bibinfo {title} {{Inference on
  gravitational waves from coalescences of stellar-mass compact objects and
  intermediate-mass black holes}},}\ }\href {\doibase 10.1093/mnras/stw233}
  {\bibfield  {journal} {\bibinfo  {journal} {Mon. Not. Roy. Astron. Soc.}\
  }\textbf {\bibinfo {volume} {457}},\ \bibinfo {pages} {4499--4506} (\bibinfo
  {year} {2016})},\ \Eprint {http://arxiv.org/abs/1511.01431} {arXiv:1511.01431
  [astro-ph.HE]} \BibitemShut {NoStop}%
\bibitem [{\citenamefont {Bellovary}\ \emph {et~al.}(2019)\citenamefont
  {Bellovary}, \citenamefont {Brooks}, \citenamefont {Colpi}, \citenamefont
  {Eracleous}, \citenamefont {Holley-Bockelmann}, \citenamefont
  {Hornschemeier}, \citenamefont {Mayer}, \citenamefont {Natarajan},
  \citenamefont {Slutsky},\ and\ \citenamefont {Tremmel}}]{Bellovary:2019nib}%
  \BibitemOpen
  \bibfield  {author} {\bibinfo {author} {\bibfnamefont {Jillian}\ \bibnamefont
  {Bellovary}}, \bibinfo {author} {\bibfnamefont {Alyson}\ \bibnamefont
  {Brooks}}, \bibinfo {author} {\bibfnamefont {Monica}\ \bibnamefont {Colpi}},
  \bibinfo {author} {\bibfnamefont {Michael}\ \bibnamefont {Eracleous}},
  \bibinfo {author} {\bibfnamefont {Kelly}\ \bibnamefont {Holley-Bockelmann}},
  \bibinfo {author} {\bibfnamefont {Ann}\ \bibnamefont {Hornschemeier}},
  \bibinfo {author} {\bibfnamefont {Lucio}\ \bibnamefont {Mayer}}, \bibinfo
  {author} {\bibfnamefont {Priya}\ \bibnamefont {Natarajan}}, \bibinfo {author}
  {\bibfnamefont {Jacob}\ \bibnamefont {Slutsky}}, \ and\ \bibinfo {author}
  {\bibfnamefont {Michael}\ \bibnamefont {Tremmel}},\ }\bibfield  {title}
  {\enquote {\bibinfo {title} {{Where are the Intermediate Mass Black
  Holes?}}}\ }\href@noop {} {\  (\bibinfo {year} {2019})},\ \Eprint
  {http://arxiv.org/abs/1903.08144} {arXiv:1903.08144 [astro-ph.HE]}
  \BibitemShut {NoStop}%
\bibitem [{\citenamefont {Barausse}\ \emph {et~al.}(2007)\citenamefont
  {Barausse}, \citenamefont {Rezzolla}, \citenamefont {Petroff},\ and\
  \citenamefont {Ansorg}}]{Barausse:2006vt}%
  \BibitemOpen
  \bibfield  {author} {\bibinfo {author} {\bibfnamefont {Enrico}\ \bibnamefont
  {Barausse}}, \bibinfo {author} {\bibfnamefont {Luciano}\ \bibnamefont
  {Rezzolla}}, \bibinfo {author} {\bibfnamefont {David}\ \bibnamefont
  {Petroff}}, \ and\ \bibinfo {author} {\bibfnamefont {Marcus}\ \bibnamefont
  {Ansorg}},\ }\bibfield  {title} {\enquote {\bibinfo {title} {{Gravitational
  waves from Extreme Mass Ratio Inspirals in non-pure Kerr spacetimes}},}\
  }\href {\doibase 10.1103/PhysRevD.75.064026} {\bibfield  {journal} {\bibinfo
  {journal} {Phys. Rev. D}\ }\textbf {\bibinfo {volume} {75}},\ \bibinfo
  {pages} {064026} (\bibinfo {year} {2007})},\ \Eprint
  {http://arxiv.org/abs/gr-qc/0612123} {arXiv:gr-qc/0612123} \BibitemShut
  {NoStop}%
\bibitem [{\citenamefont {Barausse}\ and\ \citenamefont
  {Rezzolla}(2008)}]{Barausse:2007dy}%
  \BibitemOpen
  \bibfield  {author} {\bibinfo {author} {\bibfnamefont {Enrico}\ \bibnamefont
  {Barausse}}\ and\ \bibinfo {author} {\bibfnamefont {Luciano}\ \bibnamefont
  {Rezzolla}},\ }\bibfield  {title} {\enquote {\bibinfo {title} {{The Influence
  of the hydrodynamic drag from an accretion torus on extreme mass-ratio
  inspirals}},}\ }\href {\doibase 10.1103/PhysRevD.77.104027} {\bibfield
  {journal} {\bibinfo  {journal} {Phys. Rev. D}\ }\textbf {\bibinfo {volume}
  {77}},\ \bibinfo {pages} {104027} (\bibinfo {year} {2008})},\ \Eprint
  {http://arxiv.org/abs/0711.4558} {arXiv:0711.4558 [gr-qc]} \BibitemShut
  {NoStop}%
\bibitem [{\citenamefont {Gair}\ \emph
  {et~al.}(2011{\natexlab{b}})\citenamefont {Gair}, \citenamefont {Flanagan},
  \citenamefont {Drasco}, \citenamefont {Hinderer},\ and\ \citenamefont
  {Babak}}]{Gair:2010iv}%
  \BibitemOpen
  \bibfield  {author} {\bibinfo {author} {\bibfnamefont {Jonathan~R.}\
  \bibnamefont {Gair}}, \bibinfo {author} {\bibfnamefont {Eanna~E.}\
  \bibnamefont {Flanagan}}, \bibinfo {author} {\bibfnamefont {Steve}\
  \bibnamefont {Drasco}}, \bibinfo {author} {\bibfnamefont {Tanja}\
  \bibnamefont {Hinderer}}, \ and\ \bibinfo {author} {\bibfnamefont
  {Stanislav}\ \bibnamefont {Babak}},\ }\bibfield  {title} {\enquote {\bibinfo
  {title} {{Forced motion near black holes}},}\ }\href {\doibase
  10.1103/PhysRevD.83.044037} {\bibfield  {journal} {\bibinfo  {journal} {Phys.
  Rev. D}\ }\textbf {\bibinfo {volume} {83}},\ \bibinfo {pages} {044037}
  (\bibinfo {year} {2011}{\natexlab{b}})},\ \Eprint
  {http://arxiv.org/abs/1012.5111} {arXiv:1012.5111 [gr-qc]} \BibitemShut
  {NoStop}%
\bibitem [{\citenamefont {Yunes}\ \emph {et~al.}(2011)\citenamefont {Yunes},
  \citenamefont {Kocsis}, \citenamefont {Loeb},\ and\ \citenamefont
  {Haiman}}]{Yunes:2011ws}%
  \BibitemOpen
  \bibfield  {author} {\bibinfo {author} {\bibfnamefont {Nicolas}\ \bibnamefont
  {Yunes}}, \bibinfo {author} {\bibfnamefont {Bence}\ \bibnamefont {Kocsis}},
  \bibinfo {author} {\bibfnamefont {Abraham}\ \bibnamefont {Loeb}}, \ and\
  \bibinfo {author} {\bibfnamefont {Zoltan}\ \bibnamefont {Haiman}},\
  }\bibfield  {title} {\enquote {\bibinfo {title} {{Imprint of Accretion
  Disk-Induced Migration on Gravitational Waves from Extreme Mass Ratio
  Inspirals}},}\ }\href {\doibase 10.1103/PhysRevLett.107.171103} {\bibfield
  {journal} {\bibinfo  {journal} {Phys. Rev. Lett.}\ }\textbf {\bibinfo
  {volume} {107}},\ \bibinfo {pages} {171103} (\bibinfo {year} {2011})},\
  \Eprint {http://arxiv.org/abs/1103.4609} {arXiv:1103.4609 [astro-ph.CO]}
  \BibitemShut {NoStop}%
\bibitem [{\citenamefont {Barausse}\ \emph {et~al.}(2014)\citenamefont
  {Barausse}, \citenamefont {Cardoso},\ and\ \citenamefont
  {Pani}}]{Barausse:2014tra}%
  \BibitemOpen
  \bibfield  {author} {\bibinfo {author} {\bibfnamefont {Enrico}\ \bibnamefont
  {Barausse}}, \bibinfo {author} {\bibfnamefont {Vitor}\ \bibnamefont
  {Cardoso}}, \ and\ \bibinfo {author} {\bibfnamefont {Paolo}\ \bibnamefont
  {Pani}},\ }\bibfield  {title} {\enquote {\bibinfo {title} {{Can environmental
  effects spoil precision gravitational-wave astrophysics?}}}\ }\href {\doibase
  10.1103/PhysRevD.89.104059} {\bibfield  {journal} {\bibinfo  {journal} {Phys.
  Rev. D}\ }\textbf {\bibinfo {volume} {89}},\ \bibinfo {pages} {104059}
  (\bibinfo {year} {2014})},\ \Eprint {http://arxiv.org/abs/1404.7149}
  {arXiv:1404.7149 [gr-qc]} \BibitemShut {NoStop}%
\bibitem [{\citenamefont {Barausse}\ \emph {et~al.}(2015)\citenamefont
  {Barausse}, \citenamefont {Cardoso},\ and\ \citenamefont
  {Pani}}]{Barausse:2014pra}%
  \BibitemOpen
  \bibfield  {author} {\bibinfo {author} {\bibfnamefont {Enrico}\ \bibnamefont
  {Barausse}}, \bibinfo {author} {\bibfnamefont {Vitor}\ \bibnamefont
  {Cardoso}}, \ and\ \bibinfo {author} {\bibfnamefont {Paolo}\ \bibnamefont
  {Pani}},\ }\bibfield  {title} {\enquote {\bibinfo {title} {{Environmental
  Effects for Gravitational-wave Astrophysics}},}\ }\href {\doibase
  10.1088/1742-6596/610/1/012044} {\bibfield  {journal} {\bibinfo  {journal}
  {J. Phys. Conf. Ser.}\ }\textbf {\bibinfo {volume} {610}},\ \bibinfo {pages}
  {012044} (\bibinfo {year} {2015})},\ \Eprint {http://arxiv.org/abs/1404.7140}
  {arXiv:1404.7140 [astro-ph.CO]} \BibitemShut {NoStop}%
\bibitem [{\citenamefont {Derdzinski}\ \emph {et~al.}(2020)\citenamefont
  {Derdzinski}, \citenamefont {D'Orazio}, \citenamefont {Duffell},
  \citenamefont {Haiman},\ and\ \citenamefont
  {Macfadyen}}]{Derdzinski:2020wlw}%
  \BibitemOpen
  \bibfield  {author} {\bibinfo {author} {\bibfnamefont {A.}~\bibnamefont
  {Derdzinski}}, \bibinfo {author} {\bibfnamefont {D.}~\bibnamefont
  {D'Orazio}}, \bibinfo {author} {\bibfnamefont {P.}~\bibnamefont {Duffell}},
  \bibinfo {author} {\bibfnamefont {Z.}~\bibnamefont {Haiman}}, \ and\ \bibinfo
  {author} {\bibfnamefont {A.}~\bibnamefont {Macfadyen}},\ }\bibfield  {title}
  {\enquote {\bibinfo {title} {{Evolution of gas disc-embedded intermediate
  mass ratio inspirals in the LISA band}},}\ }\href {\doibase
  10.1093/mnras/staa3976} {\  (\bibinfo {year} {2020}),\
  10.1093/mnras/staa3976},\ \Eprint {http://arxiv.org/abs/2005.11333}
  {arXiv:2005.11333 [astro-ph.HE]} \BibitemShut {NoStop}%
\bibitem [{\citenamefont {Gair}\ \emph {et~al.}(2013)\citenamefont {Gair},
  \citenamefont {Vallisneri}, \citenamefont {Larson},\ and\ \citenamefont
  {Baker}}]{Gair:2012nm}%
  \BibitemOpen
  \bibfield  {author} {\bibinfo {author} {\bibfnamefont {Jonathan~R.}\
  \bibnamefont {Gair}}, \bibinfo {author} {\bibfnamefont {Michele}\
  \bibnamefont {Vallisneri}}, \bibinfo {author} {\bibfnamefont {Shane~L.}\
  \bibnamefont {Larson}}, \ and\ \bibinfo {author} {\bibfnamefont {John~G.}\
  \bibnamefont {Baker}},\ }\bibfield  {title} {\enquote {\bibinfo {title}
  {{Testing General Relativity with Low-Frequency, Space-Based
  Gravitational-Wave Detectors}},}\ }\href {\doibase 10.12942/lrr-2013-7}
  {\bibfield  {journal} {\bibinfo  {journal} {Living Rev. Rel.}\ }\textbf
  {\bibinfo {volume} {16}},\ \bibinfo {pages} {7} (\bibinfo {year} {2013})},\
  \Eprint {http://arxiv.org/abs/1212.5575} {arXiv:1212.5575 [gr-qc]}
  \BibitemShut {NoStop}%
\bibitem [{\citenamefont {Piovano}\ \emph {et~al.}(2020)\citenamefont
  {Piovano}, \citenamefont {Maselli},\ and\ \citenamefont
  {Pani}}]{Piovano:2020ooe}%
  \BibitemOpen
  \bibfield  {author} {\bibinfo {author} {\bibfnamefont {Gabriel~Andres}\
  \bibnamefont {Piovano}}, \bibinfo {author} {\bibfnamefont {Andrea}\
  \bibnamefont {Maselli}}, \ and\ \bibinfo {author} {\bibfnamefont {Paolo}\
  \bibnamefont {Pani}},\ }\bibfield  {title} {\enquote {\bibinfo {title}
  {{Model independent tests of the Kerr bound with extreme mass ratio
  inspirals}},}\ }\href {\doibase 10.1016/j.physletb.2020.135860} {\bibfield
  {journal} {\bibinfo  {journal} {Phys. Lett. B}\ }\textbf {\bibinfo {volume}
  {811}},\ \bibinfo {pages} {135860} (\bibinfo {year} {2020})},\ \Eprint
  {http://arxiv.org/abs/2003.08448} {arXiv:2003.08448 [gr-qc]} \BibitemShut
  {NoStop}%
\bibitem [{\citenamefont {Yunes}\ and\ \citenamefont
  {Sopuerta}(2010)}]{Yunes:2009ry}%
  \BibitemOpen
  \bibfield  {author} {\bibinfo {author} {\bibfnamefont {Nicolas}\ \bibnamefont
  {Yunes}}\ and\ \bibinfo {author} {\bibfnamefont {C.~F.}\ \bibnamefont
  {Sopuerta}},\ }\bibfield  {title} {\enquote {\bibinfo {title} {{Testing
  Effective Quantum Gravity with Gravitational Waves from Extreme Mass Ratio
  Inspirals}},}\ }\href {\doibase 10.1088/1742-6596/228/1/012051} {\bibfield
  {journal} {\bibinfo  {journal} {J. Phys. Conf. Ser.}\ }\textbf {\bibinfo
  {volume} {228}},\ \bibinfo {pages} {012051} (\bibinfo {year} {2010})},\
  \Eprint {http://arxiv.org/abs/0909.3636} {arXiv:0909.3636 [gr-qc]}
  \BibitemShut {NoStop}%
\bibitem [{\citenamefont {Canizares}\ \emph
  {et~al.}(2012{\natexlab{a}})\citenamefont {Canizares}, \citenamefont {Gair},\
  and\ \citenamefont {Sopuerta}}]{Canizares:2012ji}%
  \BibitemOpen
  \bibfield  {author} {\bibinfo {author} {\bibfnamefont {P.}~\bibnamefont
  {Canizares}}, \bibinfo {author} {\bibfnamefont {J.~R.}\ \bibnamefont {Gair}},
  \ and\ \bibinfo {author} {\bibfnamefont {C.~F.}\ \bibnamefont {Sopuerta}},\
  }\bibfield  {title} {\enquote {\bibinfo {title} {{Testing Chern-Simons
  modified gravity with observations of extreme-mass-ratio binaries}},}\ }\href
  {\doibase 10.1088/1742-6596/363/1/012019} {\bibfield  {journal} {\bibinfo
  {journal} {J. Phys. Conf. Ser.}\ }\textbf {\bibinfo {volume} {363}},\
  \bibinfo {pages} {012019} (\bibinfo {year} {2012}{\natexlab{a}})},\ \Eprint
  {http://arxiv.org/abs/1206.0322} {arXiv:1206.0322 [gr-qc]} \BibitemShut
  {NoStop}%
\bibitem [{\citenamefont {Canizares}\ \emph
  {et~al.}(2012{\natexlab{b}})\citenamefont {Canizares}, \citenamefont {Gair},\
  and\ \citenamefont {Sopuerta}}]{Canizares:2012is}%
  \BibitemOpen
  \bibfield  {author} {\bibinfo {author} {\bibfnamefont {Priscilla}\
  \bibnamefont {Canizares}}, \bibinfo {author} {\bibfnamefont {Jonathan~R.}\
  \bibnamefont {Gair}}, \ and\ \bibinfo {author} {\bibfnamefont {Carlos~F.}\
  \bibnamefont {Sopuerta}},\ }\bibfield  {title} {\enquote {\bibinfo {title}
  {{Testing Chern-Simons Modified Gravity with Gravitational-Wave Detections of
  Extreme-Mass-Ratio Binaries}},}\ }\href {\doibase 10.1103/PhysRevD.86.044010}
  {\bibfield  {journal} {\bibinfo  {journal} {Phys. Rev. D}\ }\textbf {\bibinfo
  {volume} {86}},\ \bibinfo {pages} {044010} (\bibinfo {year}
  {2012}{\natexlab{b}})},\ \Eprint {http://arxiv.org/abs/1205.1253}
  {arXiv:1205.1253 [gr-qc]} \BibitemShut {NoStop}%
\bibitem [{\citenamefont {Rodriguez}\ \emph {et~al.}(2012)\citenamefont
  {Rodriguez}, \citenamefont {Mandel},\ and\ \citenamefont
  {Gair}}]{Rodriguez:2011aa}%
  \BibitemOpen
  \bibfield  {author} {\bibinfo {author} {\bibfnamefont {Carl~L.}\ \bibnamefont
  {Rodriguez}}, \bibinfo {author} {\bibfnamefont {Ilya}\ \bibnamefont
  {Mandel}}, \ and\ \bibinfo {author} {\bibfnamefont {Jonathan~R.}\
  \bibnamefont {Gair}},\ }\bibfield  {title} {\enquote {\bibinfo {title}
  {{Verifying the no-hair property of massive compact objects with
  intermediate-mass-ratio inspirals in advanced gravitational-wave
  detectors}},}\ }\href {\doibase 10.1103/PhysRevD.85.062002} {\bibfield
  {journal} {\bibinfo  {journal} {Phys. Rev. D}\ }\textbf {\bibinfo {volume}
  {85}},\ \bibinfo {pages} {062002} (\bibinfo {year} {2012})},\ \Eprint
  {http://arxiv.org/abs/1112.1404} {arXiv:1112.1404 [astro-ph.HE]} \BibitemShut
  {NoStop}%
\bibitem [{\citenamefont {Chua}\ \emph {et~al.}(2018)\citenamefont {Chua},
  \citenamefont {Hee}, \citenamefont {Handley}, \citenamefont {Higson},
  \citenamefont {Moore}, \citenamefont {Gair}, \citenamefont {Hobson},\ and\
  \citenamefont {Lasenby}}]{Chua:2018yng}%
  \BibitemOpen
  \bibfield  {author} {\bibinfo {author} {\bibfnamefont {Alvin J.~K.}\
  \bibnamefont {Chua}}, \bibinfo {author} {\bibfnamefont {Sonke}\ \bibnamefont
  {Hee}}, \bibinfo {author} {\bibfnamefont {Will~J.}\ \bibnamefont {Handley}},
  \bibinfo {author} {\bibfnamefont {Edward}\ \bibnamefont {Higson}}, \bibinfo
  {author} {\bibfnamefont {Christopher~J.}\ \bibnamefont {Moore}}, \bibinfo
  {author} {\bibfnamefont {Jonathan~R.}\ \bibnamefont {Gair}}, \bibinfo
  {author} {\bibfnamefont {Michael~P.}\ \bibnamefont {Hobson}}, \ and\ \bibinfo
  {author} {\bibfnamefont {Anthony~N.}\ \bibnamefont {Lasenby}},\ }\bibfield
  {title} {\enquote {\bibinfo {title} {{Towards a framework for testing general
  relativity with extreme-mass-ratio-inspiral observations}},}\ }\href
  {\doibase 10.1093/mnras/sty1079} {\bibfield  {journal} {\bibinfo  {journal}
  {Mon. Not. Roy. Astron. Soc.}\ }\textbf {\bibinfo {volume} {478}},\ \bibinfo
  {pages} {28--40} (\bibinfo {year} {2018})},\ \Eprint
  {http://arxiv.org/abs/1803.10210} {arXiv:1803.10210 [gr-qc]} \BibitemShut
  {NoStop}%
\bibitem [{\citenamefont {Abbott}\ \emph {et~al.}(2021)\citenamefont {Abbott}
  \emph {et~al.}}]{LIGOScientific:2020tif}%
  \BibitemOpen
  \bibfield  {author} {\bibinfo {author} {\bibfnamefont {R.}~\bibnamefont
  {Abbott}} \emph {et~al.} (\bibinfo {collaboration} {LIGO Scientific,
  Virgo}),\ }\bibfield  {title} {\enquote {\bibinfo {title} {{Tests of general
  relativity with binary black holes from the second LIGO-Virgo
  gravitational-wave transient catalog}},}\ }\href {\doibase
  10.1103/PhysRevD.103.122002} {\bibfield  {journal} {\bibinfo  {journal}
  {Phys. Rev. D}\ }\textbf {\bibinfo {volume} {103}},\ \bibinfo {pages}
  {122002} (\bibinfo {year} {2021})},\ \Eprint
  {http://arxiv.org/abs/2010.14529} {arXiv:2010.14529 [gr-qc]} \BibitemShut
  {NoStop}%
\bibitem [{\citenamefont {Islam}\ \emph
  {et~al.}(2021{\natexlab{a}})\citenamefont {Islam}, \citenamefont {Field},
  \citenamefont {Haster},\ and\ \citenamefont {Smith}}]{Islam:2021zee}%
  \BibitemOpen
  \bibfield  {author} {\bibinfo {author} {\bibfnamefont {Tousif}\ \bibnamefont
  {Islam}}, \bibinfo {author} {\bibfnamefont {Scott~E.}\ \bibnamefont {Field}},
  \bibinfo {author} {\bibfnamefont {Carl-Johan}\ \bibnamefont {Haster}}, \ and\
  \bibinfo {author} {\bibfnamefont {Rory}\ \bibnamefont {Smith}},\ }\bibfield
  {title} {\enquote {\bibinfo {title} {{High precision source characterization
  of intermediate mass-ratio black hole coalescences with gravitational waves:
  The importance of higher order multipoles}},}\ }\href {\doibase
  10.1103/PhysRevD.104.084068} {\bibfield  {journal} {\bibinfo  {journal}
  {Phys. Rev. D}\ }\textbf {\bibinfo {volume} {104}},\ \bibinfo {pages}
  {084068} (\bibinfo {year} {2021}{\natexlab{a}})},\ \Eprint
  {http://arxiv.org/abs/2105.04422} {arXiv:2105.04422 [gr-qc]} \BibitemShut
  {NoStop}%
\bibitem [{\citenamefont {Blackman}\ \emph {et~al.}(2015)\citenamefont
  {Blackman}, \citenamefont {Field}, \citenamefont {Galley}, \citenamefont
  {Szil\'agyi}, \citenamefont {Scheel}, \citenamefont {Tiglio},\ and\
  \citenamefont {Hemberger}}]{Blackman:2015pia}%
  \BibitemOpen
  \bibfield  {author} {\bibinfo {author} {\bibfnamefont {Jonathan}\
  \bibnamefont {Blackman}}, \bibinfo {author} {\bibfnamefont {Scott~E.}\
  \bibnamefont {Field}}, \bibinfo {author} {\bibfnamefont {Chad~R.}\
  \bibnamefont {Galley}}, \bibinfo {author} {\bibfnamefont {B\'ela}\
  \bibnamefont {Szil\'agyi}}, \bibinfo {author} {\bibfnamefont {Mark~A.}\
  \bibnamefont {Scheel}}, \bibinfo {author} {\bibfnamefont {Manuel}\
  \bibnamefont {Tiglio}}, \ and\ \bibinfo {author} {\bibfnamefont {Daniel~A.}\
  \bibnamefont {Hemberger}},\ }\bibfield  {title} {\enquote {\bibinfo {title}
  {{Fast and Accurate Prediction of Numerical Relativity Waveforms from Binary
  Black Hole Coalescences Using Surrogate Models}},}\ }\href {\doibase
  10.1103/PhysRevLett.115.121102} {\bibfield  {journal} {\bibinfo  {journal}
  {Phys. Rev. Lett.}\ }\textbf {\bibinfo {volume} {115}},\ \bibinfo {pages}
  {121102} (\bibinfo {year} {2015})},\ \Eprint
  {http://arxiv.org/abs/1502.07758} {arXiv:1502.07758 [gr-qc]} \BibitemShut
  {NoStop}%
\bibitem [{\citenamefont {Blackman}\ \emph
  {et~al.}(2017{\natexlab{a}})\citenamefont {Blackman}, \citenamefont {Field},
  \citenamefont {Scheel}, \citenamefont {Galley}, \citenamefont {Ott},
  \citenamefont {Boyle}, \citenamefont {Kidder}, \citenamefont {Pfeiffer},\
  and\ \citenamefont {Szilágyi}}]{Blackman:2017pcm}%
  \BibitemOpen
  \bibfield  {author} {\bibinfo {author} {\bibfnamefont {Jonathan}\
  \bibnamefont {Blackman}}, \bibinfo {author} {\bibfnamefont {Scott~E.}\
  \bibnamefont {Field}}, \bibinfo {author} {\bibfnamefont {Mark~A.}\
  \bibnamefont {Scheel}}, \bibinfo {author} {\bibfnamefont {Chad~R.}\
  \bibnamefont {Galley}}, \bibinfo {author} {\bibfnamefont {Christian~D.}\
  \bibnamefont {Ott}}, \bibinfo {author} {\bibfnamefont {Michael}\ \bibnamefont
  {Boyle}}, \bibinfo {author} {\bibfnamefont {Lawrence~E.}\ \bibnamefont
  {Kidder}}, \bibinfo {author} {\bibfnamefont {Harald~P.}\ \bibnamefont
  {Pfeiffer}}, \ and\ \bibinfo {author} {\bibfnamefont {Béla}\ \bibnamefont
  {Szilágyi}},\ }\bibfield  {title} {\enquote {\bibinfo {title} {{Numerical
  relativity waveform surrogate model for generically precessing binary black
  hole mergers}},}\ }\href {\doibase 10.1103/PhysRevD.96.024058} {\bibfield
  {journal} {\bibinfo  {journal} {Phys. Rev.}\ }\textbf {\bibinfo {volume}
  {D96}},\ \bibinfo {pages} {024058} (\bibinfo {year} {2017}{\natexlab{a}})},\
  \Eprint {http://arxiv.org/abs/1705.07089} {arXiv:1705.07089 [gr-qc]}
  \BibitemShut {NoStop}%
\bibitem [{\citenamefont {Blackman}\ \emph
  {et~al.}(2017{\natexlab{b}})\citenamefont {Blackman}, \citenamefont {Field},
  \citenamefont {Scheel}, \citenamefont {Galley}, \citenamefont {Hemberger},
  \citenamefont {Schmidt},\ and\ \citenamefont {Smith}}]{Blackman:2017dfb}%
  \BibitemOpen
  \bibfield  {author} {\bibinfo {author} {\bibfnamefont {Jonathan}\
  \bibnamefont {Blackman}}, \bibinfo {author} {\bibfnamefont {Scott~E.}\
  \bibnamefont {Field}}, \bibinfo {author} {\bibfnamefont {Mark~A.}\
  \bibnamefont {Scheel}}, \bibinfo {author} {\bibfnamefont {Chad~R.}\
  \bibnamefont {Galley}}, \bibinfo {author} {\bibfnamefont {Daniel~A.}\
  \bibnamefont {Hemberger}}, \bibinfo {author} {\bibfnamefont {Patricia}\
  \bibnamefont {Schmidt}}, \ and\ \bibinfo {author} {\bibfnamefont {Rory}\
  \bibnamefont {Smith}},\ }\bibfield  {title} {\enquote {\bibinfo {title} {{A
  Surrogate Model of Gravitational Waveforms from Numerical Relativity
  Simulations of Precessing Binary Black Hole Mergers}},}\ }\href {\doibase
  10.1103/PhysRevD.95.104023} {\bibfield  {journal} {\bibinfo  {journal} {Phys.
  Rev. D}\ }\textbf {\bibinfo {volume} {95}},\ \bibinfo {pages} {104023}
  (\bibinfo {year} {2017}{\natexlab{b}})},\ \Eprint
  {http://arxiv.org/abs/1701.00550} {arXiv:1701.00550 [gr-qc]} \BibitemShut
  {NoStop}%
\bibitem [{\citenamefont {Varma}\ \emph
  {et~al.}(2019{\natexlab{a}})\citenamefont {Varma}, \citenamefont {Field},
  \citenamefont {Scheel}, \citenamefont {Blackman}, \citenamefont {Kidder},\
  and\ \citenamefont {Pfeiffer}}]{Varma:2018mmi}%
  \BibitemOpen
  \bibfield  {author} {\bibinfo {author} {\bibfnamefont {Vijay}\ \bibnamefont
  {Varma}}, \bibinfo {author} {\bibfnamefont {Scott~E.}\ \bibnamefont {Field}},
  \bibinfo {author} {\bibfnamefont {Mark~A.}\ \bibnamefont {Scheel}}, \bibinfo
  {author} {\bibfnamefont {Jonathan}\ \bibnamefont {Blackman}}, \bibinfo
  {author} {\bibfnamefont {Lawrence~E.}\ \bibnamefont {Kidder}}, \ and\
  \bibinfo {author} {\bibfnamefont {Harald~P.}\ \bibnamefont {Pfeiffer}},\
  }\bibfield  {title} {\enquote {\bibinfo {title} {{Surrogate model of
  hybridized numerical relativity binary black hole waveforms}},}\ }\href
  {\doibase 10.1103/PhysRevD.99.064045} {\bibfield  {journal} {\bibinfo
  {journal} {Phys. Rev.}\ }\textbf {\bibinfo {volume} {D99}},\ \bibinfo {pages}
  {064045} (\bibinfo {year} {2019}{\natexlab{a}})},\ \Eprint
  {http://arxiv.org/abs/1812.07865} {arXiv:1812.07865 [gr-qc]} \BibitemShut
  {NoStop}%
\bibitem [{\citenamefont {Varma}\ \emph
  {et~al.}(2019{\natexlab{b}})\citenamefont {Varma}, \citenamefont {Field},
  \citenamefont {Scheel}, \citenamefont {Blackman}, \citenamefont {Gerosa},
  \citenamefont {Stein}, \citenamefont {Kidder},\ and\ \citenamefont
  {Pfeiffer}}]{Varma:2019csw}%
  \BibitemOpen
  \bibfield  {author} {\bibinfo {author} {\bibfnamefont {Vijay}\ \bibnamefont
  {Varma}}, \bibinfo {author} {\bibfnamefont {Scott~E.}\ \bibnamefont {Field}},
  \bibinfo {author} {\bibfnamefont {Mark~A.}\ \bibnamefont {Scheel}}, \bibinfo
  {author} {\bibfnamefont {Jonathan}\ \bibnamefont {Blackman}}, \bibinfo
  {author} {\bibfnamefont {Davide}\ \bibnamefont {Gerosa}}, \bibinfo {author}
  {\bibfnamefont {Leo~C.}\ \bibnamefont {Stein}}, \bibinfo {author}
  {\bibfnamefont {Lawrence~E.}\ \bibnamefont {Kidder}}, \ and\ \bibinfo
  {author} {\bibfnamefont {Harald~P.}\ \bibnamefont {Pfeiffer}},\ }\bibfield
  {title} {\enquote {\bibinfo {title} {{Surrogate models for precessing binary
  black hole simulations with unequal masses}},}\ }\href {\doibase
  10.1103/PhysRevResearch.1.033015} {\bibfield  {journal} {\bibinfo  {journal}
  {Phys. Rev. Research.}\ }\textbf {\bibinfo {volume} {1}},\ \bibinfo {pages}
  {033015} (\bibinfo {year} {2019}{\natexlab{b}})},\ \Eprint
  {http://arxiv.org/abs/1905.09300} {arXiv:1905.09300 [gr-qc]} \BibitemShut
  {NoStop}%
\bibitem [{\citenamefont {Islam}\ \emph
  {et~al.}(2021{\natexlab{b}})\citenamefont {Islam}, \citenamefont {Varma},
  \citenamefont {Lodman}, \citenamefont {Field}, \citenamefont {Khanna},
  \citenamefont {Scheel}, \citenamefont {Pfeiffer}, \citenamefont {Gerosa},\
  and\ \citenamefont {Kidder}}]{Islam:2021mha}%
  \BibitemOpen
  \bibfield  {author} {\bibinfo {author} {\bibfnamefont {Tousif}\ \bibnamefont
  {Islam}}, \bibinfo {author} {\bibfnamefont {Vijay}\ \bibnamefont {Varma}},
  \bibinfo {author} {\bibfnamefont {Jackie}\ \bibnamefont {Lodman}}, \bibinfo
  {author} {\bibfnamefont {Scott~E.}\ \bibnamefont {Field}}, \bibinfo {author}
  {\bibfnamefont {Gaurav}\ \bibnamefont {Khanna}}, \bibinfo {author}
  {\bibfnamefont {Mark~A.}\ \bibnamefont {Scheel}}, \bibinfo {author}
  {\bibfnamefont {Harald~P.}\ \bibnamefont {Pfeiffer}}, \bibinfo {author}
  {\bibfnamefont {Davide}\ \bibnamefont {Gerosa}}, \ and\ \bibinfo {author}
  {\bibfnamefont {Lawrence~E.}\ \bibnamefont {Kidder}},\ }\bibfield  {title}
  {\enquote {\bibinfo {title} {{Eccentric binary black hole surrogate models
  for the gravitational waveform and remnant properties: comparable mass,
  nonspinning case}},}\ }\href {\doibase 10.1103/PhysRevD.103.064022}
  {\bibfield  {journal} {\bibinfo  {journal} {Phys. Rev. D}\ }\textbf {\bibinfo
  {volume} {103}},\ \bibinfo {pages} {064022} (\bibinfo {year}
  {2021}{\natexlab{b}})},\ \Eprint {http://arxiv.org/abs/2101.11798}
  {arXiv:2101.11798 [gr-qc]} \BibitemShut {NoStop}%
\bibitem [{\citenamefont {Boh\'e}\ \emph
  {et~al.}(2017{\natexlab{a}})\citenamefont {Boh\'e} \emph
  {et~al.}}]{bohe2017improved}%
  \BibitemOpen
  \bibfield  {author} {\bibinfo {author} {\bibfnamefont {Alejandro}\
  \bibnamefont {Boh\'e}} \emph {et~al.},\ }\bibfield  {title} {\enquote
  {\bibinfo {title} {{Improved effective-one-body model of spinning,
  nonprecessing binary black holes for the era of gravitational-wave
  astrophysics with advanced detectors}},}\ }\href {\doibase
  10.1103/PhysRevD.95.044028} {\bibfield  {journal} {\bibinfo  {journal} {Phys.
  Rev. D}\ }\textbf {\bibinfo {volume} {95}},\ \bibinfo {pages} {044028}
  (\bibinfo {year} {2017}{\natexlab{a}})},\ \Eprint
  {http://arxiv.org/abs/1611.03703} {arXiv:1611.03703 [gr-qc]} \BibitemShut
  {NoStop}%
\bibitem [{\citenamefont {Cotesta}\ \emph {et~al.}(2018)\citenamefont
  {Cotesta}, \citenamefont {Buonanno}, \citenamefont {Boh\'e}, \citenamefont
  {Taracchini}, \citenamefont {Hinder},\ and\ \citenamefont
  {Ossokine}}]{cotesta2018enriching}%
  \BibitemOpen
  \bibfield  {author} {\bibinfo {author} {\bibfnamefont {Roberto}\ \bibnamefont
  {Cotesta}}, \bibinfo {author} {\bibfnamefont {Alessandra}\ \bibnamefont
  {Buonanno}}, \bibinfo {author} {\bibfnamefont {Alejandro}\ \bibnamefont
  {Boh\'e}}, \bibinfo {author} {\bibfnamefont {Andrea}\ \bibnamefont
  {Taracchini}}, \bibinfo {author} {\bibfnamefont {Ian}\ \bibnamefont
  {Hinder}}, \ and\ \bibinfo {author} {\bibfnamefont {Serguei}\ \bibnamefont
  {Ossokine}},\ }\bibfield  {title} {\enquote {\bibinfo {title} {{Enriching the
  Symphony of Gravitational Waves from Binary Black Holes by Tuning Higher
  Harmonics}},}\ }\href {\doibase 10.1103/PhysRevD.98.084028} {\bibfield
  {journal} {\bibinfo  {journal} {Phys. Rev. D}\ }\textbf {\bibinfo {volume}
  {98}},\ \bibinfo {pages} {084028} (\bibinfo {year} {2018})},\ \Eprint
  {http://arxiv.org/abs/1803.10701} {arXiv:1803.10701 [gr-qc]} \BibitemShut
  {NoStop}%
\bibitem [{\citenamefont {Cotesta}\ \emph {et~al.}(2020)\citenamefont
  {Cotesta}, \citenamefont {Marsat},\ and\ \citenamefont
  {P{\"u}rrer}}]{cotesta2020frequency}%
  \BibitemOpen
  \bibfield  {author} {\bibinfo {author} {\bibfnamefont {Roberto}\ \bibnamefont
  {Cotesta}}, \bibinfo {author} {\bibfnamefont {Sylvain}\ \bibnamefont
  {Marsat}}, \ and\ \bibinfo {author} {\bibfnamefont {Michael}\ \bibnamefont
  {P{\"u}rrer}},\ }\bibfield  {title} {\enquote {\bibinfo {title} {{Frequency
  domain reduced order model of aligned-spin effective-one-body waveforms with
  higher-order modes}},}\ }\href {\doibase 10.1103/PhysRevD.101.124040}
  {\bibfield  {journal} {\bibinfo  {journal} {Phys. Rev. D}\ }\textbf {\bibinfo
  {volume} {101}},\ \bibinfo {pages} {124040} (\bibinfo {year} {2020})},\
  \Eprint {http://arxiv.org/abs/2003.12079} {arXiv:2003.12079 [gr-qc]}
  \BibitemShut {NoStop}%
\bibitem [{\citenamefont {Pan}\ \emph {et~al.}(2014)\citenamefont {Pan},
  \citenamefont {Buonanno}, \citenamefont {Taracchini}, \citenamefont {Kidder},
  \citenamefont {Mrou\'e}, \citenamefont {Pfeiffer}, \citenamefont {Scheel},\
  and\ \citenamefont {Szil\'agyi}}]{pan2014inspiral}%
  \BibitemOpen
  \bibfield  {author} {\bibinfo {author} {\bibfnamefont {Yi}~\bibnamefont
  {Pan}}, \bibinfo {author} {\bibfnamefont {Alessandra}\ \bibnamefont
  {Buonanno}}, \bibinfo {author} {\bibfnamefont {Andrea}\ \bibnamefont
  {Taracchini}}, \bibinfo {author} {\bibfnamefont {Lawrence~E.}\ \bibnamefont
  {Kidder}}, \bibinfo {author} {\bibfnamefont {Abdul~H.}\ \bibnamefont
  {Mrou\'e}}, \bibinfo {author} {\bibfnamefont {Harald~P.}\ \bibnamefont
  {Pfeiffer}}, \bibinfo {author} {\bibfnamefont {Mark~A.}\ \bibnamefont
  {Scheel}}, \ and\ \bibinfo {author} {\bibfnamefont {B\'ela}\ \bibnamefont
  {Szil\'agyi}},\ }\bibfield  {title} {\enquote {\bibinfo {title}
  {{Inspiral-merger-ringdown waveforms of spinning, precessing black-hole
  binaries in the effective-one-body formalism}},}\ }\href {\doibase
  10.1103/PhysRevD.89.084006} {\bibfield  {journal} {\bibinfo  {journal} {Phys.
  Rev. D}\ }\textbf {\bibinfo {volume} {89}},\ \bibinfo {pages} {084006}
  (\bibinfo {year} {2014})},\ \Eprint {http://arxiv.org/abs/1307.6232}
  {arXiv:1307.6232 [gr-qc]} \BibitemShut {NoStop}%
\bibitem [{\citenamefont {Babak}\ \emph {et~al.}(2017)\citenamefont {Babak},
  \citenamefont {Taracchini},\ and\ \citenamefont
  {Buonanno}}]{babak2017validating}%
  \BibitemOpen
  \bibfield  {author} {\bibinfo {author} {\bibfnamefont {Stanislav}\
  \bibnamefont {Babak}}, \bibinfo {author} {\bibfnamefont {Andrea}\
  \bibnamefont {Taracchini}}, \ and\ \bibinfo {author} {\bibfnamefont
  {Alessandra}\ \bibnamefont {Buonanno}},\ }\bibfield  {title} {\enquote
  {\bibinfo {title} {{Validating the effective-one-body model of spinning,
  precessing binary black holes against numerical relativity}},}\ }\href
  {\doibase 10.1103/PhysRevD.95.024010} {\bibfield  {journal} {\bibinfo
  {journal} {Phys. Rev. D}\ }\textbf {\bibinfo {volume} {95}},\ \bibinfo
  {pages} {024010} (\bibinfo {year} {2017})},\ \Eprint
  {http://arxiv.org/abs/1607.05661} {arXiv:1607.05661 [gr-qc]} \BibitemShut
  {NoStop}%
\bibitem [{\citenamefont {Husa}\ \emph {et~al.}(2016)\citenamefont {Husa},
  \citenamefont {Khan}, \citenamefont {Hannam}, \citenamefont {P\"urrer},
  \citenamefont {Ohme}, \citenamefont {Jim\'enez~Forteza},\ and\ \citenamefont
  {Boh\'e}}]{husa2016frequency}%
  \BibitemOpen
  \bibfield  {author} {\bibinfo {author} {\bibfnamefont {Sascha}\ \bibnamefont
  {Husa}}, \bibinfo {author} {\bibfnamefont {Sebastian}\ \bibnamefont {Khan}},
  \bibinfo {author} {\bibfnamefont {Mark}\ \bibnamefont {Hannam}}, \bibinfo
  {author} {\bibfnamefont {Michael}\ \bibnamefont {P\"urrer}}, \bibinfo
  {author} {\bibfnamefont {Frank}\ \bibnamefont {Ohme}}, \bibinfo {author}
  {\bibfnamefont {Xisco}\ \bibnamefont {Jim\'enez~Forteza}}, \ and\ \bibinfo
  {author} {\bibfnamefont {Alejandro}\ \bibnamefont {Boh\'e}},\ }\bibfield
  {title} {\enquote {\bibinfo {title} {{Frequency-domain gravitational waves
  from nonprecessing black-hole binaries. I. New numerical waveforms and
  anatomy of the signal}},}\ }\href {\doibase 10.1103/PhysRevD.93.044006}
  {\bibfield  {journal} {\bibinfo  {journal} {Phys. Rev. D}\ }\textbf {\bibinfo
  {volume} {93}},\ \bibinfo {pages} {044006} (\bibinfo {year} {2016})},\
  \Eprint {http://arxiv.org/abs/1508.07250} {arXiv:1508.07250 [gr-qc]}
  \BibitemShut {NoStop}%
\bibitem [{\citenamefont {Khan}\ \emph {et~al.}(2016)\citenamefont {Khan},
  \citenamefont {Husa}, \citenamefont {Hannam}, \citenamefont {Ohme},
  \citenamefont {P\"urrer}, \citenamefont {Jim\'enez~Forteza},\ and\
  \citenamefont {Boh\'e}}]{khan2016frequency}%
  \BibitemOpen
  \bibfield  {author} {\bibinfo {author} {\bibfnamefont {Sebastian}\
  \bibnamefont {Khan}}, \bibinfo {author} {\bibfnamefont {Sascha}\ \bibnamefont
  {Husa}}, \bibinfo {author} {\bibfnamefont {Mark}\ \bibnamefont {Hannam}},
  \bibinfo {author} {\bibfnamefont {Frank}\ \bibnamefont {Ohme}}, \bibinfo
  {author} {\bibfnamefont {Michael}\ \bibnamefont {P\"urrer}}, \bibinfo
  {author} {\bibfnamefont {Xisco}\ \bibnamefont {Jim\'enez~Forteza}}, \ and\
  \bibinfo {author} {\bibfnamefont {Alejandro}\ \bibnamefont {Boh\'e}},\
  }\bibfield  {title} {\enquote {\bibinfo {title} {{Frequency-domain
  gravitational waves from nonprecessing black-hole binaries. II. A
  phenomenological model for the advanced detector era}},}\ }\href {\doibase
  10.1103/PhysRevD.93.044007} {\bibfield  {journal} {\bibinfo  {journal} {Phys.
  Rev. D}\ }\textbf {\bibinfo {volume} {93}},\ \bibinfo {pages} {044007}
  (\bibinfo {year} {2016})},\ \Eprint {http://arxiv.org/abs/1508.07253}
  {arXiv:1508.07253 [gr-qc]} \BibitemShut {NoStop}%
\bibitem [{\citenamefont {London}\ \emph {et~al.}(2018)\citenamefont {London},
  \citenamefont {Khan}, \citenamefont {Fauchon-Jones}, \citenamefont
  {Garc\'\i{}a}, \citenamefont {Hannam}, \citenamefont {Husa}, \citenamefont
  {Jim\'enez-Forteza}, \citenamefont {Kalaghatgi}, \citenamefont {Ohme},\ and\
  \citenamefont {Pannarale}}]{london2018first}%
  \BibitemOpen
  \bibfield  {author} {\bibinfo {author} {\bibfnamefont {Lionel}\ \bibnamefont
  {London}}, \bibinfo {author} {\bibfnamefont {Sebastian}\ \bibnamefont
  {Khan}}, \bibinfo {author} {\bibfnamefont {Edward}\ \bibnamefont
  {Fauchon-Jones}}, \bibinfo {author} {\bibfnamefont {Cecilio}\ \bibnamefont
  {Garc\'\i{}a}}, \bibinfo {author} {\bibfnamefont {Mark}\ \bibnamefont
  {Hannam}}, \bibinfo {author} {\bibfnamefont {Sascha}\ \bibnamefont {Husa}},
  \bibinfo {author} {\bibfnamefont {Xisco}\ \bibnamefont {Jim\'enez-Forteza}},
  \bibinfo {author} {\bibfnamefont {Chinmay}\ \bibnamefont {Kalaghatgi}},
  \bibinfo {author} {\bibfnamefont {Frank}\ \bibnamefont {Ohme}}, \ and\
  \bibinfo {author} {\bibfnamefont {Francesco}\ \bibnamefont {Pannarale}},\
  }\bibfield  {title} {\enquote {\bibinfo {title} {{First higher-multipole
  model of gravitational waves from spinning and coalescing black-hole
  binaries}},}\ }\href {\doibase 10.1103/PhysRevLett.120.161102} {\bibfield
  {journal} {\bibinfo  {journal} {Phys. Rev. Lett.}\ }\textbf {\bibinfo
  {volume} {120}},\ \bibinfo {pages} {161102} (\bibinfo {year} {2018})},\
  \Eprint {http://arxiv.org/abs/1708.00404} {arXiv:1708.00404 [gr-qc]}
  \BibitemShut {NoStop}%
\bibitem [{\citenamefont {Khan}\ \emph {et~al.}(2019)\citenamefont {Khan},
  \citenamefont {Chatziioannou}, \citenamefont {Hannam},\ and\ \citenamefont
  {Ohme}}]{khan2019phenomenological}%
  \BibitemOpen
  \bibfield  {author} {\bibinfo {author} {\bibfnamefont {Sebastian}\
  \bibnamefont {Khan}}, \bibinfo {author} {\bibfnamefont {Katerina}\
  \bibnamefont {Chatziioannou}}, \bibinfo {author} {\bibfnamefont {Mark}\
  \bibnamefont {Hannam}}, \ and\ \bibinfo {author} {\bibfnamefont {Frank}\
  \bibnamefont {Ohme}},\ }\bibfield  {title} {\enquote {\bibinfo {title}
  {{Phenomenological model for the gravitational-wave signal from precessing
  binary black holes with two-spin effects}},}\ }\href {\doibase
  10.1103/PhysRevD.100.024059} {\bibfield  {journal} {\bibinfo  {journal}
  {Phys. Rev. D}\ }\textbf {\bibinfo {volume} {100}},\ \bibinfo {pages}
  {024059} (\bibinfo {year} {2019})},\ \Eprint
  {http://arxiv.org/abs/1809.10113} {arXiv:1809.10113 [gr-qc]} \BibitemShut
  {NoStop}%
\bibitem [{\citenamefont {Nagar}\ \emph {et~al.}(2022)\citenamefont {Nagar},
  \citenamefont {Healy}, \citenamefont {Lousto}, \citenamefont {Bernuzzi},\
  and\ \citenamefont {Albertini}}]{Nagar:2022icd}%
  \BibitemOpen
  \bibfield  {author} {\bibinfo {author} {\bibfnamefont {Alessandro}\
  \bibnamefont {Nagar}}, \bibinfo {author} {\bibfnamefont {James}\ \bibnamefont
  {Healy}}, \bibinfo {author} {\bibfnamefont {Carlos~O.}\ \bibnamefont
  {Lousto}}, \bibinfo {author} {\bibfnamefont {Sebastiano}\ \bibnamefont
  {Bernuzzi}}, \ and\ \bibinfo {author} {\bibfnamefont {Angelica}\ \bibnamefont
  {Albertini}},\ }\bibfield  {title} {\enquote {\bibinfo {title}
  {{Numerical-relativity validation of effective-one-body waveforms in the
  intermediate-mass-ratio regime}},}\ }\href@noop {} {\  (\bibinfo {year}
  {2022})},\ \Eprint {http://arxiv.org/abs/2202.05643} {arXiv:2202.05643
  [gr-qc]} \BibitemShut {NoStop}%
\bibitem [{\citenamefont {Lousto}\ and\ \citenamefont
  {Healy}(2020)}]{Lousto:2020tnb}%
  \BibitemOpen
  \bibfield  {author} {\bibinfo {author} {\bibfnamefont {Carlos~O.}\
  \bibnamefont {Lousto}}\ and\ \bibinfo {author} {\bibfnamefont {James}\
  \bibnamefont {Healy}},\ }\bibfield  {title} {\enquote {\bibinfo {title}
  {{Exploring the Small Mass Ratio Binary Black Hole Merger via
  Zeno\textquoteright{}s Dichotomy Approach}},}\ }\href {\doibase
  10.1103/PhysRevLett.125.191102} {\bibfield  {journal} {\bibinfo  {journal}
  {Phys. Rev. Lett.}\ }\textbf {\bibinfo {volume} {125}},\ \bibinfo {pages}
  {191102} (\bibinfo {year} {2020})},\ \Eprint
  {http://arxiv.org/abs/2006.04818} {arXiv:2006.04818 [gr-qc]} \BibitemShut
  {NoStop}%
\bibitem [{\citenamefont {Lousto}\ and\ \citenamefont
  {Healy}(2022)}]{Lousto:2022hoq}%
  \BibitemOpen
  \bibfield  {author} {\bibinfo {author} {\bibfnamefont {Carlos~O.}\
  \bibnamefont {Lousto}}\ and\ \bibinfo {author} {\bibfnamefont {James}\
  \bibnamefont {Healy}},\ }\bibfield  {title} {\enquote {\bibinfo {title}
  {{Study of the Intermediate Mass Ratio Black Hole Binary Merger up to 1000:1
  with Numerical Relativity}},}\ }\href@noop {} {\  (\bibinfo {year} {2022})},\
  \Eprint {http://arxiv.org/abs/2203.08831} {arXiv:2203.08831 [gr-qc]}
  \BibitemShut {NoStop}%
\bibitem [{\citenamefont {Yoo}\ \emph {et~al.}(2022)\citenamefont {Yoo},
  \citenamefont {Varma}, \citenamefont {Giesler}, \citenamefont {Scheel},
  \citenamefont {Haster}, \citenamefont {Pfeiffer}, \citenamefont {Kidder},\
  and\ \citenamefont {Boyle}}]{Yoo:2022erv}%
  \BibitemOpen
  \bibfield  {author} {\bibinfo {author} {\bibfnamefont {Jooheon}\ \bibnamefont
  {Yoo}}, \bibinfo {author} {\bibfnamefont {Vijay}\ \bibnamefont {Varma}},
  \bibinfo {author} {\bibfnamefont {Matthew}\ \bibnamefont {Giesler}}, \bibinfo
  {author} {\bibfnamefont {Mark~A.}\ \bibnamefont {Scheel}}, \bibinfo {author}
  {\bibfnamefont {Carl-Johan}\ \bibnamefont {Haster}}, \bibinfo {author}
  {\bibfnamefont {Harald~P.}\ \bibnamefont {Pfeiffer}}, \bibinfo {author}
  {\bibfnamefont {Lawrence~E.}\ \bibnamefont {Kidder}}, \ and\ \bibinfo
  {author} {\bibfnamefont {Michael}\ \bibnamefont {Boyle}},\ }\bibfield
  {title} {\enquote {\bibinfo {title} {{Targeted large mass ratio numerical
  relativity surrogate waveform model for GW190814}},}\ }\href@noop {} {\
  (\bibinfo {year} {2022})},\ \Eprint {http://arxiv.org/abs/2203.10109}
  {arXiv:2203.10109 [gr-qc]} \BibitemShut {NoStop}%
\bibitem [{\citenamefont {Giesler}\ \emph {et~al.}(2022)\citenamefont
  {Giesler}, \citenamefont {Scheel},\ and\ \citenamefont
  {Teukolsky}}]{Giesler:2022inPrep}%
  \BibitemOpen
  \bibfield  {author} {\bibinfo {author} {\bibfnamefont {Matthew}\ \bibnamefont
  {Giesler}}, \bibinfo {author} {\bibfnamefont {Mark~A.}\ \bibnamefont
  {Scheel}}, \ and\ \bibinfo {author} {\bibfnamefont {Saul~A.}\ \bibnamefont
  {Teukolsky}},\ }\href@noop {} {\enquote {\bibinfo {title} {{Numerical
  simulations of extreme mass ratio binary black holes}},}\ } (\bibinfo {year}
  {2022}),\ \bibinfo {note} {in preparation}\BibitemShut {NoStop}%
\bibitem [{\citenamefont {Boh\'e}\ \emph
  {et~al.}(2017{\natexlab{b}})\citenamefont {Boh\'e} \emph
  {et~al.}}]{Bohe:2016gbl}%
  \BibitemOpen
  \bibfield  {author} {\bibinfo {author} {\bibfnamefont {Alejandro}\
  \bibnamefont {Boh\'e}} \emph {et~al.},\ }\bibfield  {title} {\enquote
  {\bibinfo {title} {{Improved effective-one-body model of spinning,
  nonprecessing binary black holes for the era of gravitational-wave
  astrophysics with advanced detectors}},}\ }\href {\doibase
  10.1103/PhysRevD.95.044028} {\bibfield  {journal} {\bibinfo  {journal} {Phys.
  Rev. D}\ }\textbf {\bibinfo {volume} {95}},\ \bibinfo {pages} {044028}
  (\bibinfo {year} {2017}{\natexlab{b}})},\ \Eprint
  {http://arxiv.org/abs/1611.03703} {arXiv:1611.03703 [gr-qc]} \BibitemShut
  {NoStop}%
\bibitem [{\citenamefont {Barack}\ and\ \citenamefont
  {Cutler}(2004)}]{Barack:2003fp}%
  \BibitemOpen
  \bibfield  {author} {\bibinfo {author} {\bibfnamefont {Leor}\ \bibnamefont
  {Barack}}\ and\ \bibinfo {author} {\bibfnamefont {Curt}\ \bibnamefont
  {Cutler}},\ }\bibfield  {title} {\enquote {\bibinfo {title} {{LISA capture
  sources: Approximate waveforms, signal-to-noise ratios, and parameter
  estimation accuracy}},}\ }\href {\doibase 10.1103/PhysRevD.69.082005}
  {\bibfield  {journal} {\bibinfo  {journal} {Phys. Rev. D}\ }\textbf {\bibinfo
  {volume} {69}},\ \bibinfo {pages} {082005} (\bibinfo {year} {2004})},\
  \Eprint {http://arxiv.org/abs/gr-qc/0310125} {arXiv:gr-qc/0310125}
  \BibitemShut {NoStop}%
\bibitem [{\citenamefont {Babak}\ \emph {et~al.}(2007)\citenamefont {Babak},
  \citenamefont {Fang}, \citenamefont {Gair}, \citenamefont {Glampedakis},\
  and\ \citenamefont {Hughes}}]{Babak:2006uv}%
  \BibitemOpen
  \bibfield  {author} {\bibinfo {author} {\bibfnamefont {Stanislav}\
  \bibnamefont {Babak}}, \bibinfo {author} {\bibfnamefont {Hua}\ \bibnamefont
  {Fang}}, \bibinfo {author} {\bibfnamefont {Jonathan~R.}\ \bibnamefont
  {Gair}}, \bibinfo {author} {\bibfnamefont {Kostas}\ \bibnamefont
  {Glampedakis}}, \ and\ \bibinfo {author} {\bibfnamefont {Scott~A.}\
  \bibnamefont {Hughes}},\ }\bibfield  {title} {\enquote {\bibinfo {title}
  {{'Kludge' gravitational waveforms for a test-body orbiting a Kerr black
  hole}},}\ }\href {\doibase 10.1103/PhysRevD.75.024005} {\bibfield  {journal}
  {\bibinfo  {journal} {Phys. Rev. D}\ }\textbf {\bibinfo {volume} {75}},\
  \bibinfo {pages} {024005} (\bibinfo {year} {2007})},\ \bibinfo {note}
  {[Erratum: Phys.Rev.D 77, 04990 (2008)]},\ \Eprint
  {http://arxiv.org/abs/gr-qc/0607007} {arXiv:gr-qc/0607007} \BibitemShut
  {NoStop}%
\bibitem [{\citenamefont {Chua}\ \emph {et~al.}(2017)\citenamefont {Chua},
  \citenamefont {Moore},\ and\ \citenamefont {Gair}}]{Chua:2017ujo}%
  \BibitemOpen
  \bibfield  {author} {\bibinfo {author} {\bibfnamefont {Alvin J.~K.}\
  \bibnamefont {Chua}}, \bibinfo {author} {\bibfnamefont {Christopher~J.}\
  \bibnamefont {Moore}}, \ and\ \bibinfo {author} {\bibfnamefont {Jonathan~R.}\
  \bibnamefont {Gair}},\ }\bibfield  {title} {\enquote {\bibinfo {title}
  {{Augmented kludge waveforms for detecting extreme-mass-ratio inspirals}},}\
  }\href {\doibase 10.1103/PhysRevD.96.044005} {\bibfield  {journal} {\bibinfo
  {journal} {Phys. Rev. D}\ }\textbf {\bibinfo {volume} {96}},\ \bibinfo
  {pages} {044005} (\bibinfo {year} {2017})},\ \Eprint
  {http://arxiv.org/abs/1705.04259} {arXiv:1705.04259 [gr-qc]} \BibitemShut
  {NoStop}%
\bibitem [{\citenamefont {Chua}\ and\ \citenamefont
  {Gair}(2015)}]{Chua:2015mua}%
  \BibitemOpen
  \bibfield  {author} {\bibinfo {author} {\bibfnamefont {Alvin J.~K.}\
  \bibnamefont {Chua}}\ and\ \bibinfo {author} {\bibfnamefont {Jonathan~R.}\
  \bibnamefont {Gair}},\ }\bibfield  {title} {\enquote {\bibinfo {title}
  {{Improved analytic extreme-mass-ratio inspiral model for scoping out eLISA
  data analysis}},}\ }\href {\doibase 10.1088/0264-9381/32/23/232002}
  {\bibfield  {journal} {\bibinfo  {journal} {Class. Quant. Grav.}\ }\textbf
  {\bibinfo {volume} {32}},\ \bibinfo {pages} {232002} (\bibinfo {year}
  {2015})},\ \Eprint {http://arxiv.org/abs/1510.06245} {arXiv:1510.06245
  [gr-qc]} \BibitemShut {NoStop}%
\bibitem [{\citenamefont {Chua}\ \emph {et~al.}(2021)\citenamefont {Chua},
  \citenamefont {Katz}, \citenamefont {Warburton},\ and\ \citenamefont
  {Hughes}}]{Chua:2020stf}%
  \BibitemOpen
  \bibfield  {author} {\bibinfo {author} {\bibfnamefont {Alvin J.~K.}\
  \bibnamefont {Chua}}, \bibinfo {author} {\bibfnamefont {Michael~L.}\
  \bibnamefont {Katz}}, \bibinfo {author} {\bibfnamefont {Niels}\ \bibnamefont
  {Warburton}}, \ and\ \bibinfo {author} {\bibfnamefont {Scott~A.}\
  \bibnamefont {Hughes}},\ }\bibfield  {title} {\enquote {\bibinfo {title}
  {{Rapid generation of fully relativistic extreme-mass-ratio-inspiral waveform
  templates for LISA data analysis}},}\ }\href {\doibase
  10.1103/PhysRevLett.126.051102} {\bibfield  {journal} {\bibinfo  {journal}
  {Phys. Rev. Lett.}\ }\textbf {\bibinfo {volume} {126}},\ \bibinfo {pages}
  {051102} (\bibinfo {year} {2021})},\ \Eprint
  {http://arxiv.org/abs/2008.06071} {arXiv:2008.06071 [gr-qc]} \BibitemShut
  {NoStop}%
\bibitem [{\citenamefont {Katz}\ \emph {et~al.}(2021)\citenamefont {Katz},
  \citenamefont {Chua}, \citenamefont {Speri}, \citenamefont {Warburton},\ and\
  \citenamefont {Hughes}}]{Katz:2021yft}%
  \BibitemOpen
  \bibfield  {author} {\bibinfo {author} {\bibfnamefont {Michael~L.}\
  \bibnamefont {Katz}}, \bibinfo {author} {\bibfnamefont {Alvin J.~K.}\
  \bibnamefont {Chua}}, \bibinfo {author} {\bibfnamefont {Lorenzo}\
  \bibnamefont {Speri}}, \bibinfo {author} {\bibfnamefont {Niels}\ \bibnamefont
  {Warburton}}, \ and\ \bibinfo {author} {\bibfnamefont {Scott~A.}\
  \bibnamefont {Hughes}},\ }\bibfield  {title} {\enquote {\bibinfo {title}
  {{Fast extreme-mass-ratio-inspiral waveforms: New tools for millihertz
  gravitational-wave data analysis}},}\ }\href {\doibase
  10.1103/PhysRevD.104.064047} {\bibfield  {journal} {\bibinfo  {journal}
  {Phys. Rev. D}\ }\textbf {\bibinfo {volume} {104}},\ \bibinfo {pages}
  {064047} (\bibinfo {year} {2021})},\ \Eprint
  {http://arxiv.org/abs/2104.04582} {arXiv:2104.04582 [gr-qc]} \BibitemShut
  {NoStop}%
\bibitem [{\citenamefont {Wardell}\ \emph {et~al.}(2021)\citenamefont
  {Wardell}, \citenamefont {Pound}, \citenamefont {Warburton}, \citenamefont
  {Miller}, \citenamefont {Durkan},\ and\ \citenamefont
  {Le~Tiec}}]{Wardell:2021fyy}%
  \BibitemOpen
  \bibfield  {author} {\bibinfo {author} {\bibfnamefont {Barry}\ \bibnamefont
  {Wardell}}, \bibinfo {author} {\bibfnamefont {Adam}\ \bibnamefont {Pound}},
  \bibinfo {author} {\bibfnamefont {Niels}\ \bibnamefont {Warburton}}, \bibinfo
  {author} {\bibfnamefont {Jeremy}\ \bibnamefont {Miller}}, \bibinfo {author}
  {\bibfnamefont {Leanne}\ \bibnamefont {Durkan}}, \ and\ \bibinfo {author}
  {\bibfnamefont {Alexandre}\ \bibnamefont {Le~Tiec}},\ }\bibfield  {title}
  {\enquote {\bibinfo {title} {{Gravitational waveforms for compact binaries
  from second-order self-force theory}},}\ }\href@noop {} {\  (\bibinfo {year}
  {2021})},\ \Eprint {http://arxiv.org/abs/2112.12265} {arXiv:2112.12265
  [gr-qc]} \BibitemShut {NoStop}%
\bibitem [{\citenamefont {Rifat}\ \emph {et~al.}(2020)\citenamefont {Rifat},
  \citenamefont {Field}, \citenamefont {Khanna},\ and\ \citenamefont
  {Varma}}]{Rifat:2019ltp}%
  \BibitemOpen
  \bibfield  {author} {\bibinfo {author} {\bibfnamefont {Nur E.~M.}\
  \bibnamefont {Rifat}}, \bibinfo {author} {\bibfnamefont {Scott~E.}\
  \bibnamefont {Field}}, \bibinfo {author} {\bibfnamefont {Gaurav}\
  \bibnamefont {Khanna}}, \ and\ \bibinfo {author} {\bibfnamefont {Vijay}\
  \bibnamefont {Varma}},\ }\bibfield  {title} {\enquote {\bibinfo {title}
  {{Surrogate model for gravitational wave signals from comparable and
  large-mass-ratio black hole binaries}},}\ }\href {\doibase
  10.1103/PhysRevD.101.081502} {\bibfield  {journal} {\bibinfo  {journal}
  {Phys. Rev. D}\ }\textbf {\bibinfo {volume} {101}},\ \bibinfo {pages}
  {081502} (\bibinfo {year} {2020})},\ \Eprint
  {http://arxiv.org/abs/1910.10473} {arXiv:1910.10473 [gr-qc]} \BibitemShut
  {NoStop}%
\bibitem [{\citenamefont {Sundararajan}\ \emph {et~al.}(2007)\citenamefont
  {Sundararajan}, \citenamefont {Khanna},\ and\ \citenamefont
  {Hughes}}]{Sundararajan:2007jg}%
  \BibitemOpen
  \bibfield  {author} {\bibinfo {author} {\bibfnamefont {Pranesh~A.}\
  \bibnamefont {Sundararajan}}, \bibinfo {author} {\bibfnamefont {Gaurav}\
  \bibnamefont {Khanna}}, \ and\ \bibinfo {author} {\bibfnamefont {Scott~A.}\
  \bibnamefont {Hughes}},\ }\bibfield  {title} {\enquote {\bibinfo {title}
  {{Towards adiabatic waveforms for inspiral into Kerr black holes. I. A New
  model of the source for the time domain perturbation equation}},}\ }\href
  {\doibase 10.1103/PhysRevD.76.104005} {\bibfield  {journal} {\bibinfo
  {journal} {Phys. Rev. D}\ }\textbf {\bibinfo {volume} {76}},\ \bibinfo
  {pages} {104005} (\bibinfo {year} {2007})},\ \Eprint
  {http://arxiv.org/abs/gr-qc/0703028} {arXiv:gr-qc/0703028} \BibitemShut
  {NoStop}%
\bibitem [{\citenamefont {Sundararajan}\ \emph {et~al.}(2008)\citenamefont
  {Sundararajan}, \citenamefont {Khanna}, \citenamefont {Hughes},\ and\
  \citenamefont {Drasco}}]{Sundararajan:2008zm}%
  \BibitemOpen
  \bibfield  {author} {\bibinfo {author} {\bibfnamefont {Pranesh~A.}\
  \bibnamefont {Sundararajan}}, \bibinfo {author} {\bibfnamefont {Gaurav}\
  \bibnamefont {Khanna}}, \bibinfo {author} {\bibfnamefont {Scott~A.}\
  \bibnamefont {Hughes}}, \ and\ \bibinfo {author} {\bibfnamefont {Steve}\
  \bibnamefont {Drasco}},\ }\bibfield  {title} {\enquote {\bibinfo {title}
  {{Towards adiabatic waveforms for inspiral into Kerr black holes: II.
  Dynamical sources and generic orbits}},}\ }\href {\doibase
  10.1103/PhysRevD.78.024022} {\bibfield  {journal} {\bibinfo  {journal} {Phys.
  Rev. D}\ }\textbf {\bibinfo {volume} {78}},\ \bibinfo {pages} {024022}
  (\bibinfo {year} {2008})},\ \Eprint {http://arxiv.org/abs/0803.0317}
  {arXiv:0803.0317 [gr-qc]} \BibitemShut {NoStop}%
\bibitem [{\citenamefont {Sundararajan}\ \emph {et~al.}(2010)\citenamefont
  {Sundararajan}, \citenamefont {Khanna},\ and\ \citenamefont
  {Hughes}}]{Sundararajan:2010sr}%
  \BibitemOpen
  \bibfield  {author} {\bibinfo {author} {\bibfnamefont {Pranesh~A.}\
  \bibnamefont {Sundararajan}}, \bibinfo {author} {\bibfnamefont {Gaurav}\
  \bibnamefont {Khanna}}, \ and\ \bibinfo {author} {\bibfnamefont {Scott~A.}\
  \bibnamefont {Hughes}},\ }\bibfield  {title} {\enquote {\bibinfo {title}
  {{Binary black hole merger gravitational waves and recoil in the large mass
  ratio limit}},}\ }\href {\doibase 10.1103/PhysRevD.81.104009} {\bibfield
  {journal} {\bibinfo  {journal} {Phys. Rev. D}\ }\textbf {\bibinfo {volume}
  {81}},\ \bibinfo {pages} {104009} (\bibinfo {year} {2010})},\ \Eprint
  {http://arxiv.org/abs/1003.0485} {arXiv:1003.0485 [gr-qc]} \BibitemShut
  {NoStop}%
\bibitem [{\citenamefont {Zenginoglu}\ and\ \citenamefont
  {Khanna}(2011)}]{Zenginoglu:2011zz}%
  \BibitemOpen
  \bibfield  {author} {\bibinfo {author} {\bibfnamefont {Anil}\ \bibnamefont
  {Zenginoglu}}\ and\ \bibinfo {author} {\bibfnamefont {Gaurav}\ \bibnamefont
  {Khanna}},\ }\bibfield  {title} {\enquote {\bibinfo {title} {{Null infinity
  waveforms from extreme-mass-ratio inspirals in Kerr spacetime}},}\ }\href
  {\doibase 10.1103/PhysRevX.1.021017} {\bibfield  {journal} {\bibinfo
  {journal} {Phys. Rev. X}\ }\textbf {\bibinfo {volume} {1}},\ \bibinfo {pages}
  {021017} (\bibinfo {year} {2011})},\ \Eprint {http://arxiv.org/abs/1108.1816}
  {arXiv:1108.1816 [gr-qc]} \BibitemShut {NoStop}%
\bibitem [{\citenamefont {Field}\ \emph {et~al.}(2021)\citenamefont {Field},
  \citenamefont {Gottlieb}, \citenamefont {Grant}, \citenamefont {Isherwood},\
  and\ \citenamefont {Khanna}}]{Field:2021}%
  \BibitemOpen
  \bibfield  {author} {\bibinfo {author} {\bibfnamefont {S.~E.}\ \bibnamefont
  {Field}}, \bibinfo {author} {\bibfnamefont {S.}~\bibnamefont {Gottlieb}},
  \bibinfo {author} {\bibfnamefont {Z.~J.}\ \bibnamefont {Grant}}, \bibinfo
  {author} {\bibfnamefont {L.~F.}\ \bibnamefont {Isherwood}}, \ and\ \bibinfo
  {author} {\bibfnamefont {G.}~\bibnamefont {Khanna}},\ }\bibfield  {title}
  {\enquote {\bibinfo {title} {{A GPU-Accelerated Mixed-Precision WENO Method
  for Extremal Black Hole and Gravitational Wave Physics Computations}},}\
  }\href {\doibase 10.1007/s42967-021-00129-2} {\bibfield  {journal} {\bibinfo
  {journal} {Commun. Appl. Math. Comput.}\ } (\bibinfo {year} {2021}),\
  10.1007/s42967-021-00129-2},\ \Eprint {http://arxiv.org/abs/2010.04760}
  {arXiv:2010.04760 [math.NA]} \BibitemShut {NoStop}%
\bibitem [{\citenamefont {McKennon}\ \emph {et~al.}(2012)\citenamefont
  {McKennon}, \citenamefont {Forrester},\ and\ \citenamefont
  {Khanna}}]{McKennon:2012iq}%
  \BibitemOpen
  \bibfield  {author} {\bibinfo {author} {\bibfnamefont {Justin}\ \bibnamefont
  {McKennon}}, \bibinfo {author} {\bibfnamefont {Gary}\ \bibnamefont
  {Forrester}}, \ and\ \bibinfo {author} {\bibfnamefont {Gaurav}\ \bibnamefont
  {Khanna}},\ }\bibfield  {title} {\enquote {\bibinfo {title} {{High Accuracy
  Gravitational Waveforms from Black Hole Binary Inspirals Using OpenCL}},}\
  }\href@noop {} {\  (\bibinfo {year} {2012})},\ \Eprint
  {http://arxiv.org/abs/1206.0270} {arXiv:1206.0270 [gr-qc]} \BibitemShut
  {NoStop}%
\bibitem [{\citenamefont {Fujita}\ and\ \citenamefont
  {Tagoshi}(2004)}]{Fujita:2004rb}%
  \BibitemOpen
  \bibfield  {author} {\bibinfo {author} {\bibfnamefont {Ryuichi}\ \bibnamefont
  {Fujita}}\ and\ \bibinfo {author} {\bibfnamefont {Hideyuki}\ \bibnamefont
  {Tagoshi}},\ }\bibfield  {title} {\enquote {\bibinfo {title} {{New numerical
  methods to evaluate homogeneous solutions of the Teukolsky equation}},}\
  }\href {\doibase 10.1143/PTP.112.415} {\bibfield  {journal} {\bibinfo
  {journal} {Prog. Theor. Phys.}\ }\textbf {\bibinfo {volume} {112}},\ \bibinfo
  {pages} {415--450} (\bibinfo {year} {2004})},\ \Eprint
  {http://arxiv.org/abs/gr-qc/0410018} {arXiv:gr-qc/0410018} \BibitemShut
  {NoStop}%
\bibitem [{\citenamefont {Fujita}\ and\ \citenamefont
  {Tagoshi}(2005)}]{Fujita:2005kng}%
  \BibitemOpen
  \bibfield  {author} {\bibinfo {author} {\bibfnamefont {Ryuichi}\ \bibnamefont
  {Fujita}}\ and\ \bibinfo {author} {\bibfnamefont {Hideyuki}\ \bibnamefont
  {Tagoshi}},\ }\bibfield  {title} {\enquote {\bibinfo {title} {{New Numerical
  Methods to Evaluate Homogeneous Solutions of the Teukolsky Equation II.
  Solutions of the Continued Fraction Equation}},}\ }\href {\doibase
  10.1143/PTP.113.1165} {\bibfield  {journal} {\bibinfo  {journal} {Prog.
  Theor. Phys.}\ }\textbf {\bibinfo {volume} {113}},\ \bibinfo {pages}
  {1165--1182} (\bibinfo {year} {2005})},\ \Eprint
  {http://arxiv.org/abs/0904.3818} {arXiv:0904.3818 [gr-qc]} \BibitemShut
  {NoStop}%
\bibitem [{\citenamefont {Mano}\ \emph {et~al.}(1996)\citenamefont {Mano},
  \citenamefont {Suzuki},\ and\ \citenamefont {Takasugi}}]{Mano:1996vt}%
  \BibitemOpen
  \bibfield  {author} {\bibinfo {author} {\bibfnamefont {Shuhei}\ \bibnamefont
  {Mano}}, \bibinfo {author} {\bibfnamefont {Hisao}\ \bibnamefont {Suzuki}}, \
  and\ \bibinfo {author} {\bibfnamefont {Eiichi}\ \bibnamefont {Takasugi}},\
  }\bibfield  {title} {\enquote {\bibinfo {title} {{Analytic solutions of the
  Teukolsky equation and their low frequency expansions}},}\ }\href {\doibase
  10.1143/PTP.95.1079} {\bibfield  {journal} {\bibinfo  {journal} {Prog. Theor.
  Phys.}\ }\textbf {\bibinfo {volume} {95}},\ \bibinfo {pages} {1079--1096}
  (\bibinfo {year} {1996})},\ \Eprint {http://arxiv.org/abs/gr-qc/9603020}
  {arXiv:gr-qc/9603020} \BibitemShut {NoStop}%
\bibitem [{\citenamefont {Throwe}(2010)}]{throwe2010high}%
  \BibitemOpen
  \bibfield  {author} {\bibinfo {author} {\bibfnamefont {William
  William~Thomas}\ \bibnamefont {Throwe}},\ }\emph {\bibinfo {title} {High
  precision calculation of generic extreme mass ratio inspirals}},\ \href@noop
  {} {Ph.D. thesis},\ \bibinfo  {school} {Massachusetts Institute of
  Technology} (\bibinfo {year} {2010})\BibitemShut {NoStop}%
\bibitem [{\citenamefont {O'Sullivan}\ and\ \citenamefont
  {Hughes}(2014)}]{OSullivan:2014ywd}%
  \BibitemOpen
  \bibfield  {author} {\bibinfo {author} {\bibfnamefont {Stephen}\ \bibnamefont
  {O'Sullivan}}\ and\ \bibinfo {author} {\bibfnamefont {Scott~A.}\ \bibnamefont
  {Hughes}},\ }\bibfield  {title} {\enquote {\bibinfo {title} {{Strong-field
  tidal distortions of rotating black holes: Formalism and results for
  circular, equatorial orbits}},}\ }\href {\doibase 10.1103/PhysRevD.91.109901}
  {\bibfield  {journal} {\bibinfo  {journal} {Phys. Rev. D}\ }\textbf {\bibinfo
  {volume} {90}},\ \bibinfo {pages} {124039} (\bibinfo {year} {2014})},\
  \bibinfo {note} {[Erratum: Phys.Rev.D 91, 109901 (2015)]},\ \Eprint
  {http://arxiv.org/abs/1407.6983} {arXiv:1407.6983 [gr-qc]} \BibitemShut
  {NoStop}%
\bibitem [{\citenamefont {Drasco}\ and\ \citenamefont
  {Hughes}(2006)}]{Drasco:2005kz}%
  \BibitemOpen
  \bibfield  {author} {\bibinfo {author} {\bibfnamefont {Steve}\ \bibnamefont
  {Drasco}}\ and\ \bibinfo {author} {\bibfnamefont {Scott~A.}\ \bibnamefont
  {Hughes}},\ }\bibfield  {title} {\enquote {\bibinfo {title} {{Gravitational
  wave snapshots of generic extreme mass ratio inspirals}},}\ }\href {\doibase
  10.1103/PhysRevD.73.024027} {\bibfield  {journal} {\bibinfo  {journal} {Phys.
  Rev. D}\ }\textbf {\bibinfo {volume} {73}},\ \bibinfo {pages} {024027}
  (\bibinfo {year} {2006})},\ \bibinfo {note} {[Erratum: Phys.Rev.D 88, 109905
  (2013), Erratum: Phys.Rev.D 90, 109905 (2014)]},\ \Eprint
  {http://arxiv.org/abs/gr-qc/0509101} {arXiv:gr-qc/0509101} \BibitemShut
  {NoStop}%
\bibitem [{BHP()}]{BHPToolkit}%
  \BibitemOpen
  \href@noop {} {\enquote {\bibinfo {title} {{Black Hole Perturbation
  Toolkit}},}\ }\bibinfo {howpublished}
  {(\href{http://bhptoolkit.org/}{bhptoolkit.org})}\BibitemShut {NoStop}%
\bibitem [{\citenamefont {Ori}\ and\ \citenamefont
  {Thorne}(2000)}]{Ori:2000zn}%
  \BibitemOpen
  \bibfield  {author} {\bibinfo {author} {\bibfnamefont {Amos}\ \bibnamefont
  {Ori}}\ and\ \bibinfo {author} {\bibfnamefont {Kip~S.}\ \bibnamefont
  {Thorne}},\ }\bibfield  {title} {\enquote {\bibinfo {title} {{The Transition
  from inspiral to plunge for a compact body in a circular equatorial orbit
  around a massive, spinning black hole}},}\ }\href {\doibase
  10.1103/PhysRevD.62.124022} {\bibfield  {journal} {\bibinfo  {journal} {Phys.
  Rev. D}\ }\textbf {\bibinfo {volume} {62}},\ \bibinfo {pages} {124022}
  (\bibinfo {year} {2000})},\ \Eprint {http://arxiv.org/abs/gr-qc/0003032}
  {arXiv:gr-qc/0003032} \BibitemShut {NoStop}%
\bibitem [{\citenamefont {Hughes}\ \emph {et~al.}(2019)\citenamefont {Hughes},
  \citenamefont {Apte}, \citenamefont {Khanna},\ and\ \citenamefont
  {Lim}}]{Hughes:2019zmt}%
  \BibitemOpen
  \bibfield  {author} {\bibinfo {author} {\bibfnamefont {Scott~A.}\
  \bibnamefont {Hughes}}, \bibinfo {author} {\bibfnamefont {Anuj}\ \bibnamefont
  {Apte}}, \bibinfo {author} {\bibfnamefont {Gaurav}\ \bibnamefont {Khanna}}, \
  and\ \bibinfo {author} {\bibfnamefont {Halston}\ \bibnamefont {Lim}},\
  }\bibfield  {title} {\enquote {\bibinfo {title} {{Learning about black hole
  binaries from their ringdown spectra}},}\ }\href {\doibase
  10.1103/PhysRevLett.123.161101} {\bibfield  {journal} {\bibinfo  {journal}
  {Phys. Rev. Lett.}\ }\textbf {\bibinfo {volume} {123}},\ \bibinfo {pages}
  {161101} (\bibinfo {year} {2019})},\ \Eprint
  {http://arxiv.org/abs/1901.05900} {arXiv:1901.05900 [gr-qc]} \BibitemShut
  {NoStop}%
\bibitem [{\citenamefont {Apte}\ and\ \citenamefont
  {Hughes}(2019)}]{Apte:2019txp}%
  \BibitemOpen
  \bibfield  {author} {\bibinfo {author} {\bibfnamefont {Anuj}\ \bibnamefont
  {Apte}}\ and\ \bibinfo {author} {\bibfnamefont {Scott~A.}\ \bibnamefont
  {Hughes}},\ }\bibfield  {title} {\enquote {\bibinfo {title} {{Exciting black
  hole modes via misaligned coalescences: I. Inspiral, transition, and plunge
  trajectories using a generalized Ori-Thorne procedure}},}\ }\href {\doibase
  10.1103/PhysRevD.100.084031} {\bibfield  {journal} {\bibinfo  {journal}
  {Phys. Rev. D}\ }\textbf {\bibinfo {volume} {100}},\ \bibinfo {pages}
  {084031} (\bibinfo {year} {2019})},\ \Eprint
  {http://arxiv.org/abs/1901.05901} {arXiv:1901.05901 [gr-qc]} \BibitemShut
  {NoStop}%
\bibitem [{\citenamefont {Hinderer}\ and\ \citenamefont
  {Flanagan}(2008)}]{Hinderer:2008dm}%
  \BibitemOpen
  \bibfield  {author} {\bibinfo {author} {\bibfnamefont {Tanja}\ \bibnamefont
  {Hinderer}}\ and\ \bibinfo {author} {\bibfnamefont {Eanna~E.}\ \bibnamefont
  {Flanagan}},\ }\bibfield  {title} {\enquote {\bibinfo {title} {{Two timescale
  analysis of extreme mass ratio inspirals in Kerr. I. Orbital Motion}},}\
  }\href {\doibase 10.1103/PhysRevD.78.064028} {\bibfield  {journal} {\bibinfo
  {journal} {Phys. Rev. D}\ }\textbf {\bibinfo {volume} {78}},\ \bibinfo
  {pages} {064028} (\bibinfo {year} {2008})},\ \Eprint
  {http://arxiv.org/abs/0805.3337} {arXiv:0805.3337 [gr-qc]} \BibitemShut
  {NoStop}%
\bibitem [{\citenamefont {Field}\ \emph {et~al.}(2014)\citenamefont {Field},
  \citenamefont {Galley}, \citenamefont {Hesthaven}, \citenamefont {Kaye},\
  and\ \citenamefont {Tiglio}}]{Field:2013cfa}%
  \BibitemOpen
  \bibfield  {author} {\bibinfo {author} {\bibfnamefont {Scott~E.}\
  \bibnamefont {Field}}, \bibinfo {author} {\bibfnamefont {Chad~R.}\
  \bibnamefont {Galley}}, \bibinfo {author} {\bibfnamefont {Jan~S.}\
  \bibnamefont {Hesthaven}}, \bibinfo {author} {\bibfnamefont {Jason}\
  \bibnamefont {Kaye}}, \ and\ \bibinfo {author} {\bibfnamefont {Manuel}\
  \bibnamefont {Tiglio}},\ }\bibfield  {title} {\enquote {\bibinfo {title}
  {{Fast prediction and evaluation of gravitational waveforms using surrogate
  models}},}\ }\href {\doibase 10.1103/PhysRevX.4.031006} {\bibfield  {journal}
  {\bibinfo  {journal} {Phys. Rev. X}\ }\textbf {\bibinfo {volume} {4}},\
  \bibinfo {pages} {031006} (\bibinfo {year} {2014})},\ \Eprint
  {http://arxiv.org/abs/1308.3565} {arXiv:1308.3565 [gr-qc]} \BibitemShut
  {NoStop}%
\bibitem [{\citenamefont {P\"urrer}(2014)}]{Purrer:2014fza}%
  \BibitemOpen
  \bibfield  {author} {\bibinfo {author} {\bibfnamefont {Michael}\ \bibnamefont
  {P\"urrer}},\ }\bibfield  {title} {\enquote {\bibinfo {title} {{Frequency
  domain reduced order models for gravitational waves from aligned-spin compact
  binaries}},}\ }\href {\doibase 10.1088/0264-9381/31/19/195010} {\bibfield
  {journal} {\bibinfo  {journal} {Class. Quant. Grav.}\ }\textbf {\bibinfo
  {volume} {31}},\ \bibinfo {pages} {195010} (\bibinfo {year} {2014})},\
  \Eprint {http://arxiv.org/abs/1402.4146} {arXiv:1402.4146 [gr-qc]}
  \BibitemShut {NoStop}%
\bibitem [{\citenamefont {Maday}\ \emph {et~al.}(2009)\citenamefont {Maday},
  \citenamefont {Nguyen}, \citenamefont {Patera},\ and\ \citenamefont
  {Pau}}]{Maday:2009}%
  \BibitemOpen
  \bibfield  {author} {\bibinfo {author} {\bibfnamefont {Y.}~\bibnamefont
  {Maday}}, \bibinfo {author} {\bibfnamefont {N.~.C}\ \bibnamefont {Nguyen}},
  \bibinfo {author} {\bibfnamefont {A.~T.}\ \bibnamefont {Patera}}, \ and\
  \bibinfo {author} {\bibfnamefont {S.~H.}\ \bibnamefont {Pau}},\ }\bibfield
  {title} {\enquote {\bibinfo {title} {A general multipurpose interpolation
  procedure: the magic points},}\ }\href {\doibase 10.3934/cpaa.2009.8.383}
  {\bibfield  {journal} {\bibinfo  {journal} {Communications on Pure and
  Applied Analysis}\ }\textbf {\bibinfo {volume} {8}},\ \bibinfo {pages}
  {383--404} (\bibinfo {year} {2009})}\BibitemShut {NoStop}%
\bibitem [{\citenamefont {Chaturantabut}\ and\ \citenamefont
  {Sorensen}(2010)}]{chaturantabut2010nonlinear}%
  \BibitemOpen
  \bibfield  {author} {\bibinfo {author} {\bibfnamefont {Saifon}\ \bibnamefont
  {Chaturantabut}}\ and\ \bibinfo {author} {\bibfnamefont {Danny~C}\
  \bibnamefont {Sorensen}},\ }\bibfield  {title} {\enquote {\bibinfo {title}
  {Nonlinear model reduction via discrete empirical interpolation},}\
  }\href@noop {} {\bibfield  {journal} {\bibinfo  {journal} {SIAM Journal on
  Scientific Computing}\ }\textbf {\bibinfo {volume} {32}},\ \bibinfo {pages}
  {2737--2764} (\bibinfo {year} {2010})}\BibitemShut {NoStop}%
\bibitem [{\citenamefont {Canizares}\ \emph {et~al.}(2015)\citenamefont
  {Canizares}, \citenamefont {Field}, \citenamefont {Gair}, \citenamefont
  {Raymond}, \citenamefont {Smith},\ and\ \citenamefont
  {Tiglio}}]{Canizares:2014fya}%
  \BibitemOpen
  \bibfield  {author} {\bibinfo {author} {\bibfnamefont {Priscilla}\
  \bibnamefont {Canizares}}, \bibinfo {author} {\bibfnamefont {Scott~E.}\
  \bibnamefont {Field}}, \bibinfo {author} {\bibfnamefont {Jonathan}\
  \bibnamefont {Gair}}, \bibinfo {author} {\bibfnamefont {Vivien}\ \bibnamefont
  {Raymond}}, \bibinfo {author} {\bibfnamefont {Rory}\ \bibnamefont {Smith}}, \
  and\ \bibinfo {author} {\bibfnamefont {Manuel}\ \bibnamefont {Tiglio}},\
  }\bibfield  {title} {\enquote {\bibinfo {title} {{Accelerated
  gravitational-wave parameter estimation with reduced order modeling}},}\
  }\href {\doibase 10.1103/PhysRevLett.114.071104} {\bibfield  {journal}
  {\bibinfo  {journal} {Phys. Rev. Lett.}\ }\textbf {\bibinfo {volume} {114}},\
  \bibinfo {pages} {071104} (\bibinfo {year} {2015})},\ \Eprint
  {http://arxiv.org/abs/1404.6284} {arXiv:1404.6284 [gr-qc]} \BibitemShut
  {NoStop}%
\bibitem [{\citenamefont {Varma}\ \emph
  {et~al.}(2019{\natexlab{c}})\citenamefont {Varma}, \citenamefont {Gerosa},
  \citenamefont {Stein}, \citenamefont {H\'ebert},\ and\ \citenamefont
  {Zhang}}]{Varma:2018aht}%
  \BibitemOpen
  \bibfield  {author} {\bibinfo {author} {\bibfnamefont {Vijay}\ \bibnamefont
  {Varma}}, \bibinfo {author} {\bibfnamefont {Davide}\ \bibnamefont {Gerosa}},
  \bibinfo {author} {\bibfnamefont {Leo~C.}\ \bibnamefont {Stein}}, \bibinfo
  {author} {\bibfnamefont {Fran\c{c}ois}\ \bibnamefont {H\'ebert}}, \ and\
  \bibinfo {author} {\bibfnamefont {Hao}\ \bibnamefont {Zhang}},\ }\bibfield
  {title} {\enquote {\bibinfo {title} {{High-accuracy mass, spin, and recoil
  predictions of generic black-hole merger remnants}},}\ }\href {\doibase
  10.1103/PhysRevLett.122.011101} {\bibfield  {journal} {\bibinfo  {journal}
  {Phys. Rev. Lett.}\ }\textbf {\bibinfo {volume} {122}},\ \bibinfo {pages}
  {011101} (\bibinfo {year} {2019}{\natexlab{c}})},\ \Eprint
  {http://arxiv.org/abs/1809.09125} {arXiv:1809.09125 [gr-qc]} \BibitemShut
  {NoStop}%
\bibitem [{\citenamefont {Boyle}\ \emph {et~al.}(2007)\citenamefont {Boyle},
  \citenamefont {Brown}, \citenamefont {Kidder}, \citenamefont {Mroue},
  \citenamefont {Pfeiffer}, \citenamefont {Scheel}, \citenamefont {Cook},\ and\
  \citenamefont {Teukolsky}}]{Boyle:2007ft}%
  \BibitemOpen
  \bibfield  {author} {\bibinfo {author} {\bibfnamefont {Michael}\ \bibnamefont
  {Boyle}}, \bibinfo {author} {\bibfnamefont {Duncan~A.}\ \bibnamefont
  {Brown}}, \bibinfo {author} {\bibfnamefont {Lawrence~E.}\ \bibnamefont
  {Kidder}}, \bibinfo {author} {\bibfnamefont {Abdul~H.}\ \bibnamefont
  {Mroue}}, \bibinfo {author} {\bibfnamefont {Harald~P.}\ \bibnamefont
  {Pfeiffer}}, \bibinfo {author} {\bibfnamefont {Mark~A.}\ \bibnamefont
  {Scheel}}, \bibinfo {author} {\bibfnamefont {Gregory~B.}\ \bibnamefont
  {Cook}}, \ and\ \bibinfo {author} {\bibfnamefont {Saul~A.}\ \bibnamefont
  {Teukolsky}},\ }\bibfield  {title} {\enquote {\bibinfo {title}
  {{High-accuracy comparison of numerical relativity simulations with
  post-Newtonian expansions}},}\ }\href {\doibase 10.1103/PhysRevD.76.124038}
  {\bibfield  {journal} {\bibinfo  {journal} {Phys. Rev. D}\ }\textbf {\bibinfo
  {volume} {76}},\ \bibinfo {pages} {124038} (\bibinfo {year} {2007})},\
  \Eprint {http://arxiv.org/abs/0710.0158} {arXiv:0710.0158 [gr-qc]}
  \BibitemShut {NoStop}%
\bibitem [{\citenamefont {Boyle}\ \emph {et~al.}(2019)\citenamefont {Boyle}
  \emph {et~al.}}]{Boyle:2019kee}%
  \BibitemOpen
  \bibfield  {author} {\bibinfo {author} {\bibfnamefont {Michael}\ \bibnamefont
  {Boyle}} \emph {et~al.},\ }\bibfield  {title} {\enquote {\bibinfo {title}
  {{The SXS Collaboration catalog of binary black hole simulations}},}\ }\href
  {\doibase 10.1088/1361-6382/ab34e2} {\bibfield  {journal} {\bibinfo
  {journal} {Class. Quant. Grav.}\ }\textbf {\bibinfo {volume} {36}},\ \bibinfo
  {pages} {195006} (\bibinfo {year} {2019})},\ \Eprint
  {http://arxiv.org/abs/1904.04831} {arXiv:1904.04831 [gr-qc]} \BibitemShut
  {NoStop}%
\bibitem [{\citenamefont {{SXS Collaboration}}()}]{SXSCatalog}%
  \BibitemOpen
  \bibfield  {author} {\bibinfo {author} {\bibnamefont {{SXS Collaboration}}},\
  }\href@noop {} {\enquote {\bibinfo {title} {The {SXS} collaboration catalog
  of gravitational waveforms},}\ }\bibinfo {note}
  {\url{http://www.black-holes.org/waveforms}}\BibitemShut {NoStop}%
\bibitem [{\citenamefont {Mroue}\ \emph {et~al.}(2013)\citenamefont {Mroue}
  \emph {et~al.}}]{Mroue:2013xna}%
  \BibitemOpen
  \bibfield  {author} {\bibinfo {author} {\bibfnamefont {Abdul~H.}\
  \bibnamefont {Mroue}} \emph {et~al.},\ }\bibfield  {title} {\enquote
  {\bibinfo {title} {{Catalog of 174 Binary Black Hole Simulations for
  Gravitational Wave Astronomy}},}\ }\href {\doibase
  10.1103/PhysRevLett.111.241104} {\bibfield  {journal} {\bibinfo  {journal}
  {Phys. Rev. Lett.}\ }\textbf {\bibinfo {volume} {111}},\ \bibinfo {pages}
  {241104} (\bibinfo {year} {2013})},\ \Eprint {http://arxiv.org/abs/1304.6077}
  {arXiv:1304.6077 [gr-qc]} \BibitemShut {NoStop}%
\bibitem [{\citenamefont {McKechan}\ \emph {et~al.}(2010)\citenamefont
  {McKechan}, \citenamefont {Robinson},\ and\ \citenamefont
  {Sathyaprakash}}]{McKechan:2010kp}%
  \BibitemOpen
  \bibfield  {author} {\bibinfo {author} {\bibfnamefont {D.~J.~A.}\
  \bibnamefont {McKechan}}, \bibinfo {author} {\bibfnamefont {C.}~\bibnamefont
  {Robinson}}, \ and\ \bibinfo {author} {\bibfnamefont {B.~S.}\ \bibnamefont
  {Sathyaprakash}},\ }\bibfield  {title} {\enquote {\bibinfo {title} {{A
  tapering window for time-domain templates and simulated signals in the
  detection of gravitational waves from coalescing compact binaries}},}\ }\href
  {\doibase 10.1088/0264-9381/27/8/084020} {\bibfield  {journal} {\bibinfo
  {journal} {Class. Quant. Grav.}\ }\textbf {\bibinfo {volume} {27}},\ \bibinfo
  {pages} {084020} (\bibinfo {year} {2010})},\ \Eprint
  {http://arxiv.org/abs/1003.2939} {arXiv:1003.2939 [gr-qc]} \BibitemShut
  {NoStop}%
\bibitem [{\citenamefont {Lim}\ \emph {et~al.}(2019)\citenamefont {Lim},
  \citenamefont {Khanna}, \citenamefont {Apte},\ and\ \citenamefont
  {Hughes}}]{Lim:2019xrb}%
  \BibitemOpen
  \bibfield  {author} {\bibinfo {author} {\bibfnamefont {Halston}\ \bibnamefont
  {Lim}}, \bibinfo {author} {\bibfnamefont {Gaurav}\ \bibnamefont {Khanna}},
  \bibinfo {author} {\bibfnamefont {Anuj}\ \bibnamefont {Apte}}, \ and\
  \bibinfo {author} {\bibfnamefont {Scott~A.}\ \bibnamefont {Hughes}},\
  }\bibfield  {title} {\enquote {\bibinfo {title} {{Exciting black hole modes
  via misaligned coalescences: II. The mode content of late-time coalescence
  waveforms}},}\ }\href {\doibase 10.1103/PhysRevD.100.084032} {\bibfield
  {journal} {\bibinfo  {journal} {Phys. Rev. D}\ }\textbf {\bibinfo {volume}
  {100}},\ \bibinfo {pages} {084032} (\bibinfo {year} {2019})},\ \Eprint
  {http://arxiv.org/abs/1901.05902} {arXiv:1901.05902 [gr-qc]} \BibitemShut
  {NoStop}%
\bibitem [{\citenamefont {Varma}\ and\ \citenamefont
  {Ajith}(2017)}]{Varma:2016dnf}%
  \BibitemOpen
  \bibfield  {author} {\bibinfo {author} {\bibfnamefont {Vijay}\ \bibnamefont
  {Varma}}\ and\ \bibinfo {author} {\bibfnamefont {Parameswaran}\ \bibnamefont
  {Ajith}},\ }\bibfield  {title} {\enquote {\bibinfo {title} {{Effects of
  nonquadrupole modes in the detection and parameter estimation of black hole
  binaries with nonprecessing spins}},}\ }\href {\doibase
  10.1103/PhysRevD.96.124024} {\bibfield  {journal} {\bibinfo  {journal} {Phys.
  Rev. D}\ }\textbf {\bibinfo {volume} {96}},\ \bibinfo {pages} {124024}
  (\bibinfo {year} {2017})},\ \Eprint {http://arxiv.org/abs/1612.05608}
  {arXiv:1612.05608 [gr-qc]} \BibitemShut {NoStop}%
\bibitem [{\citenamefont {Varma}\ \emph {et~al.}(2014)\citenamefont {Varma},
  \citenamefont {Ajith}, \citenamefont {Husa}, \citenamefont {Bustillo},
  \citenamefont {Hannam},\ and\ \citenamefont {P\"urrer}}]{Varma:2014jxa}%
  \BibitemOpen
  \bibfield  {author} {\bibinfo {author} {\bibfnamefont {Vijay}\ \bibnamefont
  {Varma}}, \bibinfo {author} {\bibfnamefont {Parameswaran}\ \bibnamefont
  {Ajith}}, \bibinfo {author} {\bibfnamefont {Sascha}\ \bibnamefont {Husa}},
  \bibinfo {author} {\bibfnamefont {Juan~Calderon}\ \bibnamefont {Bustillo}},
  \bibinfo {author} {\bibfnamefont {Mark}\ \bibnamefont {Hannam}}, \ and\
  \bibinfo {author} {\bibfnamefont {Michael}\ \bibnamefont {P\"urrer}},\
  }\bibfield  {title} {\enquote {\bibinfo {title} {{Gravitational-wave
  observations of binary black holes: Effect of nonquadrupole modes}},}\ }\href
  {\doibase 10.1103/PhysRevD.90.124004} {\bibfield  {journal} {\bibinfo
  {journal} {Phys. Rev. D}\ }\textbf {\bibinfo {volume} {90}},\ \bibinfo
  {pages} {124004} (\bibinfo {year} {2014})},\ \Eprint
  {http://arxiv.org/abs/1409.2349} {arXiv:1409.2349 [gr-qc]} \BibitemShut
  {NoStop}%
\bibitem [{\citenamefont {Shaik}\ \emph {et~al.}(2020)\citenamefont {Shaik},
  \citenamefont {Lange}, \citenamefont {Field}, \citenamefont {O'Shaughnessy},
  \citenamefont {Varma}, \citenamefont {Kidder}, \citenamefont {Pfeiffer},\
  and\ \citenamefont {Wysocki}}]{Shaik:2019dym}%
  \BibitemOpen
  \bibfield  {author} {\bibinfo {author} {\bibfnamefont {Feroz~H.}\
  \bibnamefont {Shaik}}, \bibinfo {author} {\bibfnamefont {Jacob}\ \bibnamefont
  {Lange}}, \bibinfo {author} {\bibfnamefont {Scott~E.}\ \bibnamefont {Field}},
  \bibinfo {author} {\bibfnamefont {Richard}\ \bibnamefont {O'Shaughnessy}},
  \bibinfo {author} {\bibfnamefont {Vijay}\ \bibnamefont {Varma}}, \bibinfo
  {author} {\bibfnamefont {Lawrence~E.}\ \bibnamefont {Kidder}}, \bibinfo
  {author} {\bibfnamefont {Harald~P.}\ \bibnamefont {Pfeiffer}}, \ and\
  \bibinfo {author} {\bibfnamefont {Daniel}\ \bibnamefont {Wysocki}},\
  }\bibfield  {title} {\enquote {\bibinfo {title} {{Impact of subdominant modes
  on the interpretation of gravitational-wave signals from heavy binary black
  hole systems}},}\ }\href {\doibase 10.1103/PhysRevD.101.124054} {\bibfield
  {journal} {\bibinfo  {journal} {Phys. Rev. D}\ }\textbf {\bibinfo {volume}
  {101}},\ \bibinfo {pages} {124054} (\bibinfo {year} {2020})},\ \Eprint
  {http://arxiv.org/abs/1911.02693} {arXiv:1911.02693 [gr-qc]} \BibitemShut
  {NoStop}%
\bibitem [{\citenamefont {Blackman}\ \emph {et~al.}()\citenamefont {Blackman},
  \citenamefont {Field}, \citenamefont {Galley},\ and\ \citenamefont
  {Varma}}]{gwsurrogate}%
  \BibitemOpen
  \bibfield  {author} {\bibinfo {author} {\bibfnamefont {Jonathan}\
  \bibnamefont {Blackman}}, \bibinfo {author} {\bibfnamefont {Scott}\
  \bibnamefont {Field}}, \bibinfo {author} {\bibfnamefont {Chad}\ \bibnamefont
  {Galley}}, \ and\ \bibinfo {author} {\bibfnamefont {Vijay}\ \bibnamefont
  {Varma}},\ }\href@noop {} {\enquote {\bibinfo {title} {gwsurrogate},}\
  }\bibinfo {note}
  {\url{https://pypi.python.org/pypi/gwsurrogate/}}\BibitemShut {NoStop}%
\end{thebibliography}%

\end{document}